%% file: swn.tex
\documentclass[pre,twocolumn,showpacs,groupaddress,preprintnumbers,floatfix]{revtex4-1}

\usepackage{etex}
\usepackage[mathscr]{eucal}
\usepackage{ifpdf}
\usepackage[pagebackref]{hyperref}
\usepackage{dcolumn}
\usepackage{url}
\usepackage{amsmath}
\usepackage{amscd}
\usepackage{amsfonts}
\usepackage{amssymb}
\usepackage{amsthm}
\usepackage{bigints}
\usepackage{xfrac}
\usepackage{braket}
\usepackage{bm}   
\usepackage{bbm}
\usepackage{verbatim}
\usepackage{stmaryrd}
\usepackage{xcolor}
\usepackage{setspace}
\usepackage{tikz}
\usetikzlibrary{arrows}
\usetikzlibrary{automata}
\usetikzlibrary{positioning}
\usetikzlibrary{plotmarks}
\usepackage{pgfplots}
\pgfplotsset{compat=newest}

\usepackage{epigraph}
\usepackage{qcircuit}

\usepackage[abs]{overpic}   
\usepackage{graphicx}
\usepackage{subfigure}
\usepackage{hyphenat}
\tikzstyle{vaucanson}=[
  node distance=2.5cm,
  bend angle=15,
  auto,
  every loop/.style={},
  every edge/.style={->,draw=black,line width=1.2,>=latex,shorten <=1pt,shorten >=1pt},
  every state/.style={draw=black,line width=2,font=\large},
  loop right/.style={right,out=22,in=-22,loop},
  loop above/.style={above,out=112,in=68,loop},
  loop left/.style={left,out=202,in=158,loop},
  loop below/.style={below,out=292,in=248,loop},
  loop above right/.style={right,out=67,in=23,loop},
  loop above left/.style={left,out=157,in=113,loop},
  loop below left/.style={left,out=247,in=203,loop},
  loop below right/.style={right,out=337,in=293,loop},
]

\DeclareMathAlphabet{\mathpzc}{OT1}{pzc}{m}{it}

\newcommand{\realtime}{ \mathpzc{t} }

\newcommand{\DHD}{doubly harmonic diminution}   

\newcommand{\FonezeroA}{
\hspace{-1.75em}
\raisebox{-0.4em}{
\scalebox{0.5}{
\begin{tikzpicture}[style=vaucanson, rounded corners=4pt]
    \filldraw[fill=yellow!20] (-0.25,-0.25)   rectangle (1.25,0.75);
    \node (F10_0i) at (0,0) {$0$};
    \node (F10_1i) at (0,.5) {$1$};    
    \node (F10_0f) at (1,0) {$0$};
    \path
        (F10_0i) edge (F10_0f)
        (F10_1i) edge (F10_0f)
    ;
\end{tikzpicture}    
}
}
\hspace{-1.25em}
}

\newcommand{\Fonezero}{
$\Bigl\{ $
\FonezeroA
$ \Bigr\}$
}

\newcommand{\FonezeroAapp}{\hspace{0.15em} \FonezeroA \hspace{0.25em}}
\newcommand{\Fonezeroapp}{$\Bigl\{ $  \FonezeroAapp  $ \Bigr\}$}

\newcommand{\FoneoneA}{
\hspace{-1.75em}
\raisebox{-0.4em}{
\scalebox{0.5}{
\begin{tikzpicture}[style=vaucanson, rounded corners=4pt]
    \filldraw[fill=yellow!20] (-0.25,-0.25)   rectangle (1.25,0.75);
    \node (F11_0i) at (0,0)  {$0$};
    \node (F11_1i) at (0,0.5) {$1$};    
    \node (F11_0f) at (1,0)  {$0$};
    \node (F11_1f) at (1,0.5) {$1$};
    \path
        (F11_0i) edge (F11_0f)
        (F11_1i) edge (F11_1f)
    ;
\end{tikzpicture}    
}
}
\hspace{-1.25em}
}

\newcommand{\FoneoneB}{
\hspace{-1.75em}
\raisebox{-0.4em}{
\scalebox{0.5}{
\begin{tikzpicture}[style=vaucanson, rounded corners=4pt]
    \filldraw[fill=yellow!20] (-0.25,-0.25)   rectangle (1.25,0.75);
    \node (F11_0i) at (0,0)  {$0$};
    \node (F11_1i) at (0,0.5) {$1$};    
    \node (F11_0f) at (1,0)  {$0$};
    \node (F11_1f) at (1,0.5) {$1$};
    \path
        (0.2,0.05) edge (0.8,0.5)
        (0.2,0.45) edge  (0.8,0)
    ;
\end{tikzpicture}    
}
}
\hspace{-1.25em}
}

\newcommand{\Foneone}{
$\Bigl\{ $
\FoneoneA , \FoneoneB
$ \Bigr\}$
}

\newcommand{\FoneoneAapp}{\hspace{0.15em} \FoneoneA \hspace{0.25em}}
\newcommand{\FoneoneBapp}{\hspace{0.15em} \FoneoneB \hspace{0.25em}}
\newcommand{\Foneoneapp}{$\Bigl\{ $  \FoneoneAapp , \FoneoneBapp  $ \Bigr\}$}

\theoremstyle{plain}    \newtheorem{Lem}{Lemma}
\theoremstyle{plain}    \newtheorem*{ProLem}{Proof}
\theoremstyle{plain}    \newtheorem{Cor}{Corollary}
\theoremstyle{plain}    \newtheorem*{ProCor}{Proof}
\theoremstyle{plain}    \newtheorem{The}{Theorem}
\theoremstyle{plain}    \newtheorem*{ProThe}{Proof}
\theoremstyle{plain}    
\theoremstyle{plain}    
\theoremstyle{plain}    
\theoremstyle{plain}    
\theoremstyle{plain}    \newtheorem{Def}{Definition}
\theoremstyle{plain}    
\theoremstyle{plain}

\input{cmechabbrev}

\renewcommand{\H}{\operatorname{H}}
\renewcommand{\I}{\operatorname{I}}

\colorlet {R_color}    {blue}
\colorlet {k_color}    {black!30!green}

\def\clap#1{\hbox to 0pt{\hss#1\hss}}

\begin{document}

\title{Fraudulent White Noise:\\
\vspace{0.05in}
Flat power spectra belie arbitrarily complex processes}

\author{Paul M. Riechers}
\email{pmriechers@gmail.com}

\affiliation{Complexity Institute\\
Nanyang Technological University\\
639798 Singapore, Singapore}

\affiliation{School of Physical and Mathematical Sciences\\
Nanyang Technological University\\
637371 Singapore, Singapore}

\author{James P. Crutchfield}
\email{chaos@ucdavis.edu}

\affiliation{Complexity Sciences Center\\
Department of Physics\\
University of California at Davis\\
One Shields Avenue, Davis, CA 95616}

\date{\today}
\bibliographystyle{unsrt}

\begin{abstract}
Power spectral densities are a common, convenient, and powerful way to analyze
signals. So much so that they are now broadly deployed across the sciences and
engineering---from quantum physics to cosmology, and
from crystallography to neuroscience to speech recognition.
The features they reveal not only identify prominent
signal-frequencies but also hint at mechanisms that generate correlation and
lead to resonance. Despite their near-centuries-long run of successes in signal
analysis, here we show that flat power spectra can be generated by highly
complex processes, effectively hiding all inherent structure in complex
signals. Historically, this circumstance has been widely misinterpreted, being
taken as the renowned signature of ``structureless'' \emph{white noise}---the
benchmark of randomness. We argue, in contrast, to the extent that most
real-world complex systems exhibit correlations beyond pairwise statistics
their structures evade power spectra and other pairwise statistical measures.
{ \color{blue} 
As concrete physical examples, we demonstrate that
fraudulent white noise hides the predictable structure of both 
entangled quantum systems and chaotic crystals.
}
To make these words of warning operational, we present constructive results
that explore how this situation comes about and the high toll it takes in
understanding complex mechanisms. First, we give the closed-form solution for
the power spectrum of a very broad class of structurally-complex signal
generators. Second, we demonstrate the close relationship between eigen-spectra
of evolution operators and power spectra. Third, we characterize the minimal
generative structure implied by \emph{any} power spectrum. Fourth, we show how
to construct arbitrarily complex processes with flat power spectra. Finally,
leveraging this diagnosis of the problem, we point the way to developing more
incisive tools for discovering structure in complex signals.
\end{abstract}



\preprint{arxiv.org:1908.11405}

\maketitle



\setstretch{1.1}

\newcommand{\pdf}{\text{p}}
\newcommand{\Abet}{\ProcessAlphabet}
\newcommand{\MS}{\MeasSymbol}
\newcommand{\ms}{\meassymbol}
\newcommand{\SSet}{\CausalStateSet}
\newcommand{\St}{\CausalState}
\newcommand{\st}{s}

\newcommand{\distr}{\boldsymbol{\mu}}
\newcommand{\rate}{G}
\newcommand{\stationary}{\boldsymbol{\pi}}

\newcommand{\kB}{k_\text{B}}
 
\newcommand{\enum}{\boldsymbol{\#}}


\section{Introduction}

Innovative science probes the unknown. Success in discovering the mechanisms
that underlie the systems we seek to understand, though, requires
distinguishing structure from noise. Often, this distinction falls to
discretion: structure is that part of a signal we can predict, while noise
stands in as a catch-all for everything else. This conundrum holds especially
in the analysis of signals from truly complex systems, as when analyzing data
from multi-electrode arrays in brain tissue or social experiments. These
systems are often said to be `noisy' even though the so-called noise may be
entirely functionally relevant, but in an unknown way. Such descriptions fall
far short of a principled approach that explains all trends and correlational
structure, which would claim success only when all that remains unexplained in
the signal is structureless white noise. Even this principled approach
ultimately begs the central question, though: how do we test if an apparently
random signal is truly white noise?

The challenge of discovering structure in noisy signals is compounded manifold,
as we demonstrate in the following, when our chosen observables hide arbitrary
amounts of in-principle-predictable structure behind a familiar signature
of white noise---the flat power spectrum. Said simply, observables can be
completely devoid of pairwise correlation, while still
embodying structure in higher-order correlations. More precisely, structure can
be hidden beyond any arbitrarily-large order-$N$ correlation---that not
appearing in pairwise, three-way, nor any $n$-way statistics, up to some
arbitrarily large $N$. Moreover, the hidden structure
can be arbitrarily sophisticated. It can be used, for example, to embed
messages while shifting (and so hiding) the messages' content beyond $N$-way
correlation. Here, we explore the structures conveyed and hidden by power
spectra, revealing a novel perspective on the interplay between structure and
noise in Fourier analysis.

Section~\ref{sec:TemporalStructure} discusses temporal structure and provides
closed-form expressions for the power spectra from autonomous signal
generators. It highlights the intimate connection between power spectra and
eigen-spectra of a system's time-evolution generator.
Section~\ref{sec:HiddenStructure} then introduces a suite of results on
structure that is \emph{hidden} by power spectra. Notably, it introduces a
general condition for \emph{fraudulent white noise} processes---structured
processes with a flat power spectrum---which applies very broadly, including to
input-dependent processes with nonstationary high-order statistics. 
{ \color{blue} 
Section \ref{sec:FWN_in_PhysicalSystems} 
demonstrates that fraudulent white noise indeed
arises in the observation of important physical systems.
We show that fraudulent white noise
arises in measurements of entangled quantum systems.
We also show that flat diffraction patterns belie
the predictable structure of chaotic crystals.
}
Taken altogether the results emphasize the power spectrum's shortcomings for the task
of structure detection. In response, Sec.~\ref{sec:SeekingStructureInNoise}
considers more sophisticated measures of structure.  
{ \color{blue} 
We give closed-form expressions
for polyspectra---which are often advocated as the natural next step for detecting higher-order structure---but 
show that these too have severe blind spots.  
}
This motivates us to introduce the
\emph{dependence function} which identifies the presence of novel finite-range
dependencies that contribute to total correlation. Section~\ref{sec:Conclusion}
concludes the development. Appendices present detailed derivations, as well as
several generalizations, of the main results.

\section{Structure in Space and Time?}
\label{sec:TemporalStructure}

Pairwise correlations are encountered throughout the sciences and engineering,
especially in statistical physics. They are assumed, estimated, relied on,
designed with, and used for interpretation widely. The following explores
several specific examples of pairwise correlation that arise in different
fields. These will set the context for our development, particularly for
experts in the associated fields. However, our general results should be
accessible and relevant across disciplines, as they rely primarily on basic
probability theory and linear algebra.

A well-studied lesson from statistical physics is that diverging correlation
length heralds the emergence of new types of order. Remarkably,
mechanistically-distinct physical systems share many universal behaviors near a
critical point of emergent order, including the scaling of spatial pairwise
correlation length~\cite{Path11}. More broadly, pairwise correlations are
indicators of fundamental physical processes. For example, the
fluctuation--dissipation theorem says that pairwise temporal correlations in
equilibrium determine the friction encountered in transport processes. The
Green--Kubo relations~\cite{Kubo66} make this explicit. Far from equilibrium,
say in computing devices and biological systems composed of excitable media,
temporal correlations are signatures of richly coordinated state-trajectories.

Pairwise correlations are directly viewed in the frequency domain via power
spectral densities. Indeed, power spectra are employed as a basic data analysis
tool in many scientific domains and have been key to major scientific
discoveries. For example, comparing alternative theoretical predictions for
power spectra of incident electromagnetic radiation from locally-thermalized
bodies, a unexpected discrepancy---the ultraviolet catastrophe---led to the
acceptance of Planck's theory of quantized energies and the subsequent birth of
quantum theory~\cite{Plan00aa,Eins05,Klei61}. A contemporary example of the
prominent role of power spectra is seen in the exquisitely detailed map of the
cosmic microwave background (CMB)---a snapshot of the early universe's spatial
correlations. In fact, models of the early universe are now benchmarked against
their ability to replicate the CMB power spectrum~\cite{Akra18}.

In applied mathematics, power spectra played a key role in highlighting the
defining features of the strange attractors of dynamical systems
theory~\cite{Holm77,Farm80}. This led to the discovery of Ruelle--Pollicott
resonances, where mixing and the decay of correlations in chaotic systems were
related to the point spectrum of the Ruelle--Perron--Frobenius
operator~\cite{Poll85, Ruel86, Gasp05}. Indeed, the power spectra of chaotic
systems are still actively used to analyze the behavior of everything from open
quantum systems~\cite{Panc00, Garc04} to climate models~\cite{Chek14}.

{ \color{blue} 
The famous $1/f$ decay of power spectra found in many complex systems has
received considerable attention throughout many decades \cite{Pres78a, Mand99,
Grave17}---sometimes being attributed to self-organized
criticality~\cite{Bak87}; almost always being taken as a signature of truly
complex systems. More recently, the value of $\alpha$ in $1/{f^\alpha}$
noise---and deviations from this mean behavior---are used to interpret particle
tracking experiments~\cite{Krap18, Krap19}. Related advances have enabled
extraction of physical properties from power spectral analyses of nonstationary
processes~\cite{Sade14, Dech15, Leib15}.
}

Power spectra are regularly used to discover structure in materials science and
biology, too. X-ray diffraction patterns---used to identify crystalline and
molecular organization and central to discovering DNA's
double-helix~\cite{Wats53a,Wilk53,Fran53,Wats53b}---are power spectra of
scatterer densities, as we explain in App.~\ref{sec:DPsAsPSs}. 
Power spectra have been used to identify temporal correlations in single-neuron
spike trains, refuting the common Poissonian white-noise assumption common 
in theoretical and computational neuroscience~\cite{Bair94, Fark09, Dumm14, Pena18}.  
This allows the possibility that 
temporal correlations in the spike train---rather than just the firing rate---can 
play an important role in the neural code~\cite{Chac05, Goll08}.
On a much larger (mean-field) scale, brain wave
activity in different frequency bands gives signatures of normal brain
functioning, as well as pathological conditions. Rhythmic brain-wave activity
is clinically assessed through real-time power spectra of
electroencephalography (EEG) signals~\cite{Mell07,Claa13,Tatu14}.

From the smallest to the largest scales in the universe, when probing both the
inanimate and the animate, power spectra are a central diagnostic tool for
structure and validating scientific models. Their use is so important that
special-purpose spectral and network analyzers are standard laboratory test
equipment; they can be readily purchased from dozens of major manufacturers.

Power spectra report pairwise correlations in a signal. But how much of a
system's structure is faithfully represented by pairwise correlation? Are there
important types of order that evade power spectra completely? To answer these
questions, we first consider the problem of hidden structure through
the lens of autocorrelation and power spectra. Only then, once the strengths
and weakness of power spectra are clear, do we move on along to more
sophisticated measures of structure. Along the way we trace a path that begins
to reveal what one can mean by ``statistical dependency'', ``correlation'', and
``structure''.

\subsection{Correlation and Power Spectra}

To provide a common ground, consider discrete-time processes described by an
interdependent sequence $\dots X_0 X_1 X_2 \dots$ of random variables $X_t$
that take on values $x \in \Abet$ within an alphabet assumed (for now) to be a
subset of the complex numbers: $\Abet \subset \mathbb{C}$.  (For concreteness
here, we interpret $t$ as indexing time { \color{blue} $\realtime = t \tau_0$,
where $\tau_0$ is the duration of each time-step}.  For other kinds of
stochastic process, $t$ may represent spatial or angular coordinates.) An
observed process may have a discrete domain, as with a classical discrete-time
communication channel or a series of quantum measurements or, otherwise, may be
a regularly-sampled process evolving in continuous time.

A signal's \emph{power spectrum} or, more properly, its \emph{power spectral
density} quantifies how its power is distributed across
frequency~\cite{Blac58b,Stoi05}. For a discrete-domain process it is:
\begin{align}
P(\omega) = \lim_{N \to \infty} \tfrac{1}{f_0 N }
 \Braket{ \Bigl| \sum_{t=1}^{N} X_t e^{-i \omega t} \Bigr|^2 } ~,
\label{eq:PSDdef}
\end{align}
where the angle brackets denote the expected value over the random variable
chain $X_1 X_2 X_3 \ldots X_N$,
{\color{blue} 
$\omega = 2 \pi f / f_0$ is the angular frequency, and $f$ is the frequency and
$f_0 = 1 /\tau_0$ is the fundamental frequency. We set $f_0$ to
unity in the discrete-time case. In the continuous-time limit where $\tau_0 = d
\realtime \to 0$, the power spectrum becomes:
\begin{align*}
P(f) = \lim_{L \to \infty} \frac{1}{L}
  \Braket{ \bigl| \int_0^L X_{\realtime}
  e^{- i 2 \pi f \realtime} \, d \realtime \bigr|^2 }
  ~,
\end{align*}
where we use the fact that $\omega t = 2 \pi f \realtime$. In either discrete
or continuous time, integrating over any band of frequencies gives the power in
that band.}

For \emph{wide-sense stationary} stochastic processes the \emph{autocorrelation
function}: 
\begin{align}
\gamma(\tau) = \Braket{ \, \overline{X}_t  X_{t + \tau}}
  ~,
\label{eq:AutocorrDef}  
\end{align}
is independent of the global time shift $t$ and depends only on the relative
time-separation $\tau$ between observables \footnote{For wide-sense stationary
processes that are also ergodic, the expected value over realizations
$\gamma(\tau) = \Braket{ \, \overline{X_t} X_{t + \tau}}_{\pdf(X_t ,
X_{t+\tau})}$ coincides with the time-average $\Braket{ \, \overline{x}_t
x_{t+\tau}}_t$  over a single realization. In practice, scientists and
engineers typically perform this latter time-average to obtain the
autocorrelation function $\gamma(\tau) = \Braket{ \, \overline{X_t} X_{t +
\tau}}_{\pdf(X_t , X_{t+\tau})} = \Braket{ \, \overline{x}_t x_{t+\tau}}_t$
from experimental data.}. The bar above $X_t$ denotes its complex conjugate.
Equation~\eqref{eq:AutocorrDef} makes plain the connection between pairwise
statistics and the pairwise correlation function. For wide-sense stationary
stochastic processes, the power spectrum is also determined by the 
signal's autocorrelation function $\gamma(\tau)$:
\begin{align}
P(\omega) = \lim_{N \to \infty} \tfrac{1}{f_0 N} \sum_{\tau=-N}^{N} \bigl( N - \left| \tau \right| \bigr) \gamma(\tau) e^{-i \omega \tau}
  ~.
\label{eq:PSDfromACF}
\end{align}
The windowing function $N - \left| \tau \right|$ appearing in
Eq.~\eqref{eq:PSDfromACF} is a direct consequence of Eq.~\eqref{eq:PSDdef};  it
is not imposed externally, as is common practice in signal analysis. (This
factor is important for controlling convergence in our subsequent
derivations.)

Equation~\eqref{eq:PSDfromACF} suggests that the power spectrum is very \emph{nearly}
the Fourier transform of the autocorrelation function, except for the $N-|\tau|$ term.
In fact, the Wiener--Khinchin theorem proves that 
the power spectrum 
is indeed \emph{equal} to the Fourier transform of the autocorrelation function for wide-sense stationary processes~\cite{Wien30,Khin34}.  
Note, too, that the pairwise
correlation function $\gamma(\tau)$ can be obtained via the inverse Fourier
transform of the power spectrum $P(\omega)$.

\subsection{Temporal Structurelessness}

Our goal is to understand temporal structure and to identify it in stochastic
processes. To detect structure, even when hidden, we first must establish
a baseline reference process that has no temporal structure: genuine white
noise.

\emph{White noise processes},
{\color{blue} if we remove their mean value, have zero autocorrelation for all
$\tau > 0$. Colloquially, white noise is often taken as a synonym for any
completely random process with no statistical dependencies whatsoever.  To be
precise, we will define \emph{genuine white noise}} as those processes for
which each random variable $X_t$ is statistically independent of all others
$X_{t^\prime \neq t}$, and each is identically distributed according to the
same probability density function (PDF) over the alphabet. That is, the random
variables in the sequence are \emph{independent and identically distributed}
(IID).

Familiar examples include a sequence of coin flips or the sequence of sums when
rolling a pair of dice. As an example from contemporary physics, consider the
(classical) process that results from observing a sequence of Bell-pair quantum
states \cite{Niel11a}. For each Bell pair, one of the entangled particles is
sent to Alice and the other sent to Bob. Alice makes a sequence of measurements
(along any measurement axis). The measurement output sequence she observes is
pure white noise, with each measurement outcome having equal and independent
probability of being up or down along the measurement axis. In fact, more
sophisticated deployments of Bell pairs are being developed to provide
certifiable random number generation~\cite{Acin16}. Experiments now concentrate
on increasing the rate of generating white noise~\cite{Shen18,Liu18}.

The most recognizable feature of all white noise processes is their flat power
spectrum. For any IID process, it follows directly from
Eq.~\eqref{eq:AutocorrDef} that $\gamma(0) = \braket{ | X_t |^2 }$, whereas
$\gamma(\tau) = | \braket{ X_t } |^2$ for $\tau \neq 0$. From
Eq.~\eqref{eq:PSDfromACF}, this immediately yields the familiar flat power
spectrum of white noise, together with a $\delta$-function at zero frequency,
corresponding to the signal's constant offset. For real-valued IID processes
with zero mean (and $f_0=1$), this simplifies further to $\gamma( \tau ) =
\sigma^2 \, \delta_{0, \tau} $ and so $P(\omega) = \sigma^2$. In fact, the flat
power spectrum has height equal to the variance $\sigma^2 = \braket{X_t^2} -
\braket{X_t}^2$ of the white noise for any real-valued IID process. The flat
power spectrum for IID processes indicates that any temporal structure in the
generating source has such short memory that it vanishes within the short
sampling time $\tau_0$ between each observation.

Gaussian white noises tend to be the most commonly employed white noise
processes and, usually, for good reason. By the central limit theorem, Gaussian
white noise arises generically in systems whenever \emph{many} events---with
amplitude of finite variance and with rapidly decaying correlation (compared to
the timescale between observations)---contribute additively to each individual
observation. Suppose, for example, that the expected number of these
contributions to each new observation is simply proportional to the time since
the last observation. When sampled at interval $d \realtime = \tau_0$, 
the central limit
theorem then tells us that each observation of the accumulated noise is IID and
Gaussian distributed with variance $\sigma_\eta^2 \propto d \realtime$. 
This immediately leads to the familiar standard deviation
$\sigma_\eta \propto  \sqrt{d \realtime} $  
 of the additive noise $\eta(t)$ that appears when
numerically integrating stochastic differential equations (e.g., Langevin
equations); this, in turn, produces the trajectories of slower random
variables~\footnote{The noise standard deviation being proportional to the
square root of the timestep---$\sigma_\eta \propto \sqrt{ d \realtime }$---rather
than proportional to the timestep itself is familiar in numerically integrating
Langevin and other stochastic differential equations. Though sometimes met with
confusion, the dependence is a direct consequence of the central limit theorem
since many independent noise contributions accumulate within each timestep and 
the variance of the accumulated value is proportional to the number of contributions. 
The simplest stochastic integral---integrating Gaussian white noise---produces the
famous Wiener process $W_{\realtime}$ which is the canonical model for Brownian motion.
Conversely, regularly sampling changes in the Wiener state produces Gaussian
white noise: 
$W_{\realtime+ \tau_0} - W_{\realtime} \sim \mathcal{N}(0, \tau_0)$.}.

\begin{figure}[t]
\begin{center}
\includegraphics[width=0.9\columnwidth]{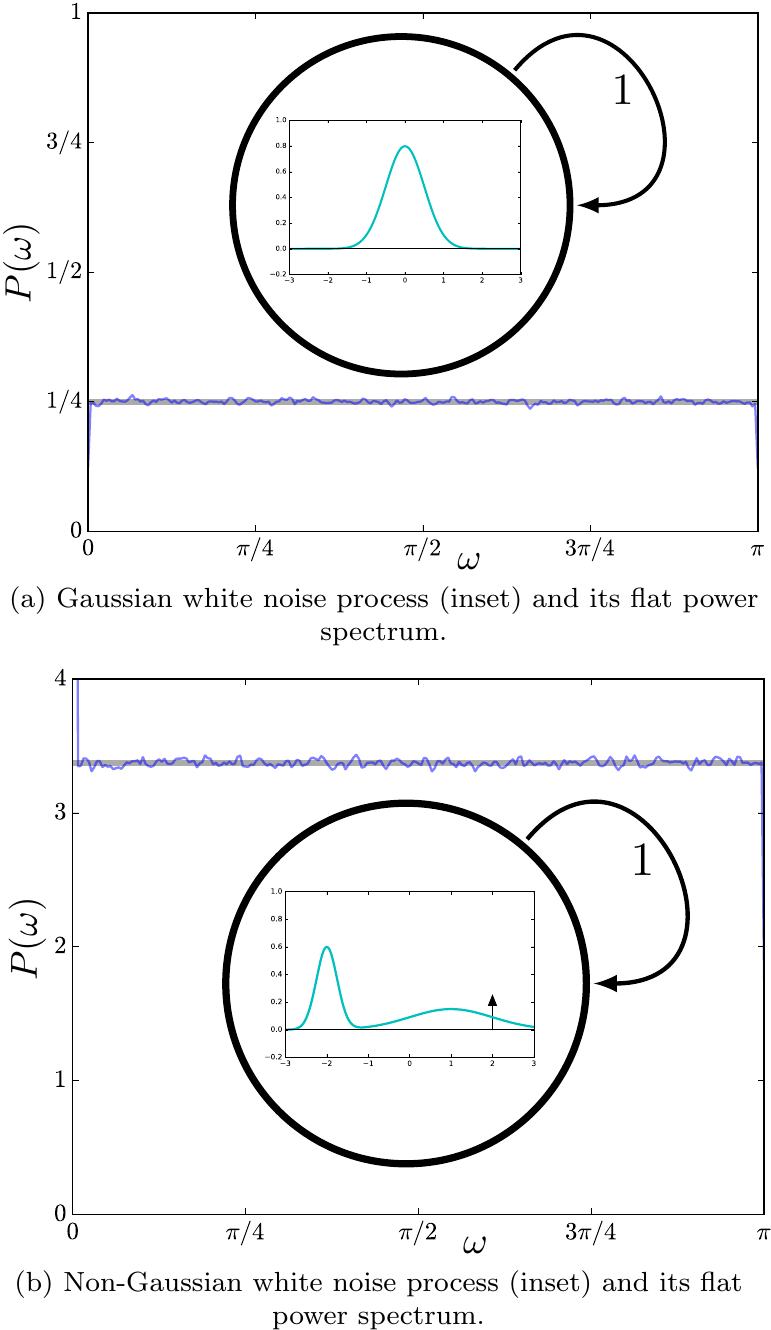}
\end{center}
\caption{Genuine white noise processes have no memory: Represented structurally
	by a state machine with a single state that is repeatedly visited with each
	observation. The same probability density function, inscribed in the state,
	is sampled at each timestep. (a) Gaussian white noise memoryless stochastic
	process. (b) Another white noise process, although non-Gaussian. For each
	(a) and (b), the flat power spectrum is given theoretically (thick gray),
	with height equal to the variance of the probability density function. We
	also display the numerically-obtained power spectrum (thin blue) for each.
	The class of all possible (not-necessarily Gaussian) memoryless white
	noises is identical with the class of processes generated by single-state
	machines. This class, in turn, is identical to that of all IID processes
	(spanning all possible probability density functions). These temporally
	structureless processes constitute all possible varieties of genuine white
	noise.
	}
\label{fig:WhiteNoises}
\end{figure}

The memoryless nature of repetitive sampling from a distribution is apparent in
the state machine shown in Fig.~\ref{fig:WhiteNoises}(a). The same Gaussian
distribution is repeatedly sampled with probability $1$ (as depicted by the
self-transition probability there) for each observation, regardless of what
happened previously.~\footnote{Gaussian genuine white noise is a particular
case of a \emph{Gaussian process}. Gaussian processes, often considered in
machine learning, are defined as those processes with a multivariate Gaussian
distribution over all finite collections of observable random variables $\{ X_t
\}_{t \in \mathcal{T}}$, where $\mathcal{T}$ is any finite set of
times~\cite[Ch.~6.4]{Bish06a}\cite{Gort19a}. There is only one Gaussian process
consistent with each choice of second-order statistics. Therefore, Gaussian
genuine white noise is the \emph{only} Gaussian process which is a white noise.
(Infinitely many other processes can also have low-order multivariate Gaussian
distributions but, if they are not multivariate Gaussian distributed to all
orders, they are not ``Gaussian processes'' in the technical sense.)}

Other ``structureless'' white noises are also possible. In fact, any of an
uncountably infinite set of different IID processes---Gaussian, Poisson,
Bernoulli, or any process that resamples a particular distribution at each
timestep---all yield the flat power spectrum or white noise. Non-Gaussian noise
can emerge from repetitive sampling of a system's (non-Gaussian) stationary
distribution when the relaxation timescales are far shorter than the time
elapsed between samples. Alternatively, non-Gaussian white noise can arise when
only a few physical events contribute to each observation, in which case the
non-Gaussianity may reveal features of the physical generative mechanism.
Nevertheless, these processes possess no temporal structure on the timescale of
observation and, in particular, generate absolutely no correlations in the
sequence of observations.

The hallmark of this structural paucity is the single state for the hidden
Markov model (HMM) that describes all of these IID processes, as depicted in
Fig.~\ref{fig:WhiteNoises}(b) \footnote{We choose to ignore the power
spectrum's $\delta$-function at zero frequency when classifying white noise,
since it merely corresponds to a constant (DC) offset of the signal. Naturally,
we make an exception for \emph{completely} deterministic DC signals (with no
noise at all), since these should not be considered among the white noise
processes.}. The single state means that no influences from the past can affect
the next or future samples. These are the genuine white noises.

In sharp contrast, we will consider stochastic processes with arbitrarily
sophisticated temporal structure on the timescale of observation. The much more
general class we next consider allows for a thorough investigation of
temporally structured stochastic processes. One surprising feature is that
these very structured processes, described by arbitrarily complicated transition dynamics within
memoryful collections of internal states, can have the flat power spectrum of white
noise.
{ \color{blue} 
These are the \emph{fraudulent white noise processes}: white noise processes
with a flat power spectrum that are nevertheless not \emph{genuine} white
noise. Fraudulent white noise contains statistical dependencies---predictable
structure completely veiled by common measures of correlation.
}

\subsection{Models of Temporal Structure}

Structure arises over time from the interdependence between observables. To
explicitly address structure in a broad class of temporally structured
processes, we use \emph{hidden Markov models} (HMMs) as our preferred
representation for autonomous signal generators
\cite{Paz71a,Rabi86a,Rabi89a,Elli95a,Ephr02a,Kell12a,Bech15a}. Later sections
introduce yet more sophisticated models with input dependence.

Despite Markovian state-to-state transitions, HMMs can generate
temporally-structured non-Markovian stochastic processes---those with infinite
history dependence (infinite Markov order). Processes generated by even
finite-state HMMs, in fact, \emph{typically} have infinite-range statistical
dependencies between observables since simple state-transition motifs guarantee
this feature \cite{Jame10a}. In addition to this richness and their ability to
compactly generate the exact temporal statistics of nonlinear dynamical
systems, HMMs are attractive since they are amenable to linear operator
techniques \cite{Koop31,Crut83a,Jaeg00,Fras08,Riec18a,Riec18b,Riec18c}.

{ \color{blue}
Section \ref{sec:FWN_in_PhysicalSystems} employs HMMs to represent (i)
sequential measurements of entangled quantum systems, (ii) scattering factors
of disordered materials, and (iii) ion transport through biomolecular channels.
But, to get there, we must first introduce the general properties of HMMs.
}

Let the $4$-tuple $\mathcal{M} = \bigl( \SSet , \Abet , \mathcal{P}, T  \bigr)$
be a discrete-time HMM that generates the stationary stochastic process $\dots
X_{-2} X_{-1} X_0 X_1 X_2 \dots$ according to the following. $\SSet$ is the
(finite) set of states of the internal Markov chain and $\Abet \subseteq
\mathbb{C}$ is the observable alphabet. $\St_t$ is the random variable for the
hidden state at time $t$ that takes on values $s \in \SSet$. $X_t$ is the
random variable for the observation at time $t$ that takes on values $x \in
\Abet$.

Given the hidden state at time $t$, the possible observations are distributed
according to the conditional probability density functions: $\mathcal{P} =
\bigl\{ \pdf(X_t | \St_t = s) \bigr\}_{s \in \SSet}$. For each $s \in \SSet$,
$\pdf(X_t | \St_t = s)$ may be abbreviated as $\pdf(X|s)$ since the probability
density function in each state is assumed to not change over $t$. Similarly, we
will write $\pdf(x|s)$ for $\pdf(X_t = x | \St_t = s)$. Finally, the
hidden-state-to-state stochastic transition matrix $T$ has elements $T_{s,s'} =
\Pr(\St_{t+1} = s' | \St_t = s)$, which give the probability of transitioning
from hidden state $s$ to $s'$ given that the system is in state $s$, where $s,
s' \in \SSet$. It is important for subsequent developments that
$\Pr(\cdot)$ denotes a probability in contrast to $\pdf(\cdot)$
which denotes a probability \emph{density}. 

\begin{figure}[t]
\begin{center}
\includegraphics[width=0.78\columnwidth]{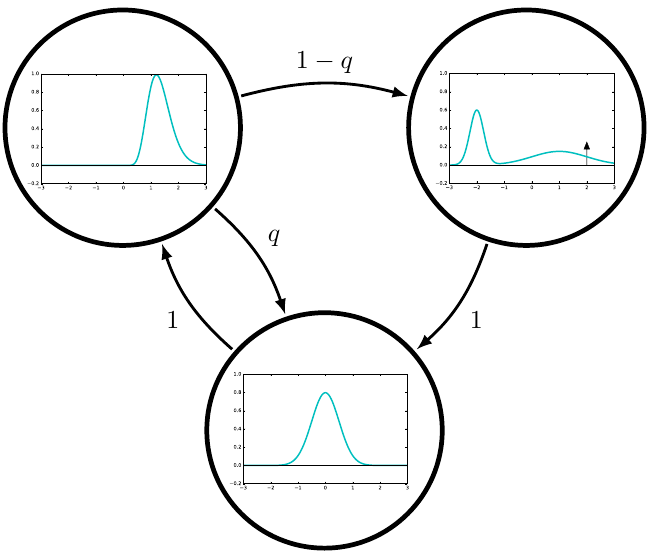}
\end{center}
\caption{Simple $3$-state HMM that generates a stochastic process according to
	the state-to-state transition dynamic $T$ and the probability density
	functions (PDFs) $\{ \pdf(X | s) \}_{s \in \SSet}$ associated with each
	state. Theorem \ref{thm:PSDequivalence} asserts that its power spectrum is
	the same (modulo constant offset) as the power spectrum generated from an
	alternative process where each state's PDF is solely concentrated at the
	average value $\langle X \rangle_{\pdf (X | s)}$ of the original PDF
	associated with the state.
	}
\label{fig:CC_HMM}
\end{figure}

Epitomizing the processes in the class considered, Fig.~\ref{fig:CC_HMM} presents a
rather simple HMM with continuous observable alphabet $\Abet = \mathbb{R}$,
whose samples are distributed according to the probability density function
shown within each hidden state. As seen in the HMM's top-right state, both
continuous probability density functions and discrete output probabilities can
be accommodated in this framework: Finite probability of a particular
observable is accomplished by an appropriately weighted Dirac $\delta$-function in
the probability density function. The memoryful structure in
Fig.~\ref{fig:CC_HMM} should be contrasted with the completely memoryless
process of sampled Gaussian white noise shown in Fig.~\ref{fig:WhiteNoises}.

\begin{figure}[t]
\begin{center}
\includegraphics[width=0.375\textwidth]{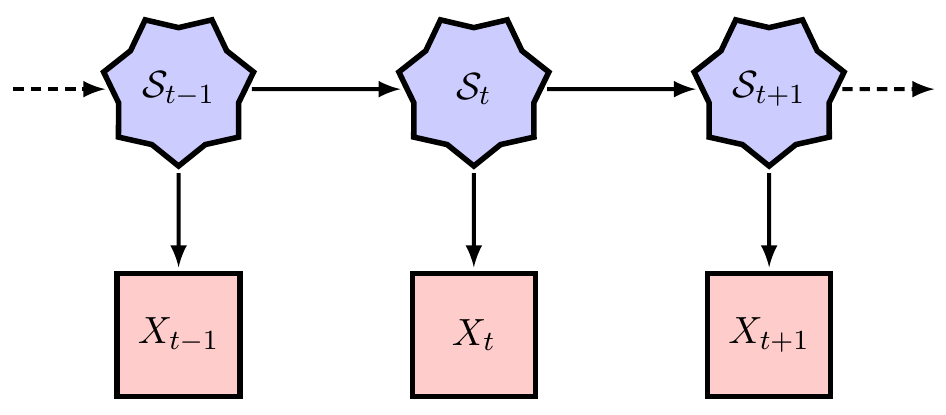}
\end{center}
\caption{Bayesian network for a state-emitting hidden Markov model graphically
	depicts the structure of conditional independence among random variables
	for the hidden state $\{ \St_n \}_{n \in \mathbb{Z}}$ at each time $n$ and
	the random variables $\{ X_n \}_{n \in \mathbb{Z}}$ for the observation at
	each time $n$.
  }
\label{fig:HMM_BayesNet}
\end{figure}

Figure~\ref{fig:HMM_BayesNet}'s Bayes network depicts the structure of conditional independence among the random variables for these memoryful signal generators. For example, for a generic HMM, 
$\pdf(X_t |  X_{t-N} \dots X_{t-2} X_{t-1} = x_{t-N}  \dots  x_{t-2} x_{t-1} )$ 
cannot be simplified since the condition on even arbitrarily 
distant past observables can influence the probability of the current observable.
However, when conditioning on hidden states, the situation can simplify
markedly. For example:
\begin{align*}
\pdf(X_t & | X_{t-N} \dots X_{t-2} X_{t-1}
  = x_{t-N} \dots x_{t-2} x_{t-1} , \\
  & \qquad \St_{t-N} \dots \St_{t-2} \St_{t-1} = s_{t-N} \dots s_{t-2} s_{t-1} ) \\
  & = \pdf(X_t | \St_{t-1} = s_{t-1} ) \\
  & = \sum_{s \in \SSet} T_{s_{t-1}, s} \, \pdf(X | s)
  ~.
\end{align*}

The general properties of HMMs allow one to calculate any statistic about the
generated process from the hidden-state-to-state transition matrix $T$ and set
$\mathcal{P}$ of conditional probability density functions. For simplicity in
the following, assume a finite set of hidden states and a single attracting
component. Then every transition matrix $T$ admits a unique stationary
distribution $\stationary$. This is determined as $T$'s left eigenvector
associated with the eigenvalue of unity: $\bra{\stationary} T =
\bra{\stationary}$. The eigenvector is normalized in probability:
$\braket{\stationary | \one} = 1$, where $\ket{\one}$ is the column vector of
all ones. Note also that $\ket{\one}$ is the right eigenvector of $T$
associated with the eigenvalue of unity, $T \ket{\one} = \ket{\one}$.
This property conserves state probability in hidden Markov chain evolution.

We can now provide the correlation functions and power spectral density in
general and in closed form for the entire class of stochastic process generated
by finite-state HMMs. Helpfully, for particular HMMs, the expressions become
analytic in the model parameters.

Appendix~\ref{sec:ACFforHMMs} shows that the autocorrelation function is given
by:
\begin{align}
\gamma(\tau) & = 
\begin{cases}
\bra{\stationary} \Omega \, T^{|\tau| } \, \overline{\Omega} \ket{\one} & \text{if } \tau \leq 1 \\
\bigl\langle \left| x \right|^2 \bigr\rangle & \text{if } \tau = 0 \\
\bra{\stationary} \overline{\Omega} \, T^{|\tau| } \, \Omega \ket{\one} & \text{if } \tau \geq 1
\end{cases}
~,
\label{eq:ExplicitCorrFnForm}
\end{align}
where 
$\Omega$ is the $|\SSet|$-by-$|\SSet|$ \emph{average-observation matrix} defined by:
\begin{align}
\Omega 
= \sum_{s \in \SSet} \braket{X}_{\pdf( X | s )} \ket{ s } \bra{ s } ~.
\label{eq:AvgObsMatrix}
\end{align}
We use the hidden-state basis in which $\ket{ s }$ is the column vector of all
$0$s except for a $1$ at the index corresponding to state $s$. $\bra{ s }$ is
simply its transpose. This yields a natural decomposition of the identity
operator: $I = \sum_{s \in \SSet} \ket{ s } \bra{ s }$. In the hidden-state
basis, then, the $\Omega$ matrix simply places state-conditioned average
outputs along its diagonal.

The power spectrum is calculated starting from Eq.~\eqref{eq:PSDfromACF}
together with Eq.~\eqref{eq:ExplicitCorrFnForm}, using the spectral
decomposition techniques developed for nonnormal and nondiagonalizable
operators in Ref.~\cite{Riec18c}. In the derivation it is important to treat
individual eigenspaces separately, as our generalized framework naturally
accommodates. Appendix~\ref{sec:PSDderivation} gives the derivation's full
details. Qualitatively, the power spectrum decomposes naturally into a discrete
part $P_\text{d}(\omega)$ (a weighted sum of Dirac $\delta$-functions) and a
continuous part $P_\text{c}(\omega)$ (a collection of diffuse peaks):
\begin{align*}
P(\omega) = P_\text{c}(\omega) + P_\text{d}(\omega)
  ~.
\end{align*}
For the power spectrum's continuous part the end result is:
\begin{align}
P_\text{c}(\omega) 
= \bigl\langle \left| x \right|^2 \bigr\rangle
 + 2 \, \text{Re} \bra{\stationary} \overline{\Omega} \, T \, \bigl( e^{i \omega} I - T \bigr)^{-1} \Omega \ket{\one} ~,
\label{eq:PcwFromResolvent}
\end{align}
where Re$(\cdot)$ denotes the real part of its argument.

Remarkably, \emph{all} of the $\omega$-dependence is in the apparently simple
term $\bigl( e^{i \omega} I - T \bigr)^{-1}$. This is the \emph{resolvent} of
$T$ along the unit circle in the complex plane. However, and central to our
main results, this frequency dependence is \emph{filtered} through
$\bra{\stationary} \overline{\Omega}$ and $\Omega \ket{\one}$. Notably, if the
average-observation matrix was proportional to the identity, then all
frequency dependence would be lost since $\text{Re} \bra{\stationary}  \bigl(
e^{i \omega} I - T \bigr)^{-1} \ket{\one} = -1/2$ is independent of 
$\omega$~\footnote{An easy 
explanation for this relies on the spectral decomposition of the resolvent, 
given in the next section.
}.
Frequency dependence of the power spectrum thus requires that there are
different averages associated with different states. Surprisingly though, none
of the structure of the conditional probability density functions $\{ \pdf(X|s)
\}_s$ matters for the power spectrum, except for the average value observed in
each state. Structure beyond averages is simply not captured.

\subsection{Apparent Structure}
\label{sec:ApparentStructure}

To fully appreciate the structure that \emph{is} captured by the power spectrum
requires a spectral decomposition of the transition matrix. The set $\Lambda_T$
of $T$'s eigenvalues is calculated as usual. However, since transition matrices
are generically nonnormal and often nondiagonalizable, the spectral projection
operators associated with $T$ deserve a brief review.

In particular, the spectral projection operator $T_{\lambda}$ associated with
eigenvalue $\lambda$ can be defined as the residue of $(z I - T)^{-1}$ as
$z \to \lambda$:
\begin{align}
T_{\lambda} = \tfrac{1}{2 \pi i} \oint_{C_\lambda} \bigl( zI - T \bigr)^{-1} \, dz
  ~,
\end{align}
where $z \in \mathbb{C}$ and $C_\lambda$ is a small counterclockwise contour
around the eigenvalue $\lambda$. Alternatively, the spectral projection
operators can be constructed from all left eigenvectors, generalized left
eigenvectors, right eigenvectors, and generalized right eigenvectors
associated with $\lambda$. The construction is given explicitly in
Ref.~\cite{Riec18c}. In the simple case where the eigenvalue is nondegenerate,
the eigenprojector takes on the simple form:
\begin{align*}
T_{\lambda}
  = \frac{ \ket{\lambda} \bra{\lambda} }{ \braket{ \lambda | \lambda } }
  ~.
\end{align*}

However, the left and right eigenvectors are \emph{not} simply
complex-conjugate transposes of each other, as they would be in the
normal-operator case familiar from closed quantum systems and undirected
networks. For example, the spectral projection operator associated with
stationarity---$T_{1} = \ket{\one} \bra{\stationary}$---can be interpreted as
the classical version of a density matrix but, typically, the stationary
distribution is not uniform and so $\bra{\stationary}$ is not proportional to
the transpose of $\ket{\one}$.

We will also use the broader class of spectral companion operators:
\begin{align}
T_{\lambda, m} = T_{\lambda} (T - \lambda I)^m 
  ~.
\end{align}
They have the useful property that $T_{\lambda, m} T_{\zeta, n} =
\delta_{\lambda, \zeta} T_{\lambda, m+n}$. Clearly, the spectral projection
operator is contained in this set, as $T_\lambda = T_{\lambda, 0}$. It should
be noted that $T_{\lambda, m} = \boldsymbol{0}$ for $m \geq \nu_\lambda$, where
$\nu_\lambda$ is the \emph{index} of the eigenvalue $\lambda$---i.e., the size
of the largest Jordan block associated with $\lambda$. One should keep in mind
that the transition matrix can be represented as:
\begin{align*} 
T = \sum_{\lambda} \bigl( \lambda T_{\lambda, 0} + T_{\lambda, 1}  \bigr)
  ~.
\end{align*} 
While the resolvent has the general spectral decomposition:
\begin{align} 
(z I - T)^{-1} & = 
  \sum_{\lambda \in \Lambda_T} \sum_{m = 0}^{\nu_\lambda - 1}
  \frac{1}{(z - \lambda)^{m+1}}  T_{\lambda,m} ~.
\label{eq:ResolventPartialFractExpansion}
\end{align}

\begin{figure*}[ht]
\begin{center}
\includegraphics[width=2.0\columnwidth]{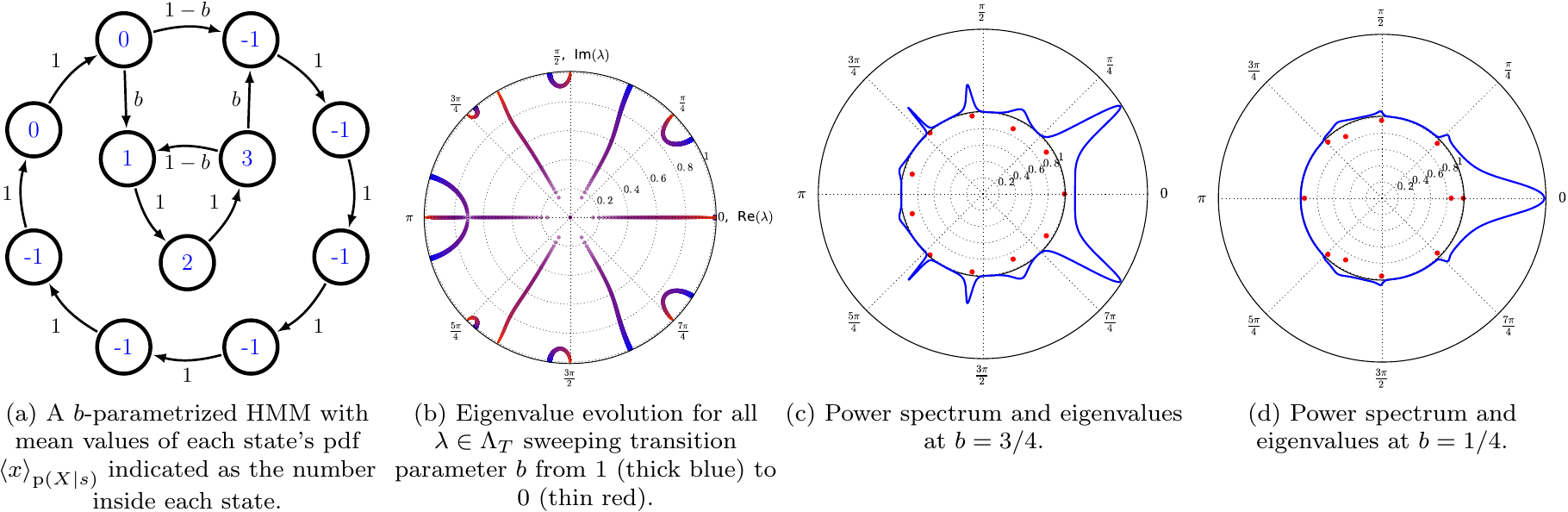}
\end{center}
\caption{Parametrized HMM of a stochastic process, its eigenvalue
	evolution, and two coronal spectrograms showing power spectra emanating
	from eigen-spectra.
	}
\label{fig:CC_DP}
\end{figure*}

The spectral expansion of the resolvent given by
Eq.~\eqref{eq:ResolventPartialFractExpansion} allows us to better interpret the
qualitative shape of the power spectrum Eq. (\ref{eq:PcwFromResolvent}):
\begin{align}
P_\text{c}(\omega) = \bigl\langle \left| x \right|^2 \bigr\rangle
  + \sum_{\lambda \in \Lambda_T} \sum_{m = 0}^{\nu_\lambda - 1}
  2 \, \text{Re}
  \frac{\bra{\stationary} \overline{\Omega} \, T \, T_{\lambda, m} \Omega \ket{\one}}{(e^{i \omega} - \lambda)^{m+1}}
  .
\label{eq:PcwFromDecomposedResolvent}
\end{align}
Note that $\bra{\stationary} \overline{\Omega} \, T \, T_{\lambda, m} \Omega
\ket{\one}$ is a complex-valued scalar and all of the frequency dependence now
handily resides in the denominator.
When $T$ is diagonalizable,
Eq.~\eqref{eq:PcwFromDecomposedResolvent} reduces to:
\begin{align*}
P_\text{c}(\omega) = \bigl\langle \left| x \right|^2 \bigr\rangle
  + \sum_{\lambda \in \Lambda_T}
  2 \, \text{Re} \biggl(
  \frac{\lambda \bra{\stationary} \overline{\Omega} \, T_{\lambda} \Omega \ket{\one}}{e^{i \omega} - \lambda} \biggr)  ~.
\end{align*}

The discrete ($\delta$-function) portion of the power spectrum is:
\begin{align}
P_\text{d}(\omega) \! & = \!\!\!\! \sum_{k = -\infty}^{\infty} 
  \!\! \sum_{\lambda \in \Lambda_T \atop |\lambda| = 1}
  \!\! 2 \pi  \, \delta( \omega \!-\! \omega_\lambda \!+\! 2 \pi k) 
 \text{Re} \! \bra{\stationary} \overline{\Omega} \, T_\lambda \Omega \ket{\one}
 \!,
\label{eq:PdwFromResolvent}
\end{align}
where $\omega_\lambda$ is related to $\lambda$ by $\lambda = e^{i
\omega_\lambda}$. Equation~\eqref{eq:PdwFromResolvent} is valid even when $T$
is nondiagonalizable: An extension of the Perron-Frobenius theorem guarantees
that $T$'s eigenvalues on the unit circle have index $\nu_\lambda = 1$. With
$T_1 = \ket{\one} \bra{\stationary}$, it is useful to note that
$\bra{\stationary} \overline{\Omega} \, T_1 \Omega \ket{\one} = \bigl|
\braket{x} \bigr|^2$, so that the $\delta$-function at zero frequency appears
whenever the average observation is nonzero.

When plotted as a function of the angular frequency $\omega$ around the unit
circle, the power spectrum suggestively appears to emanate from the eigenvalues
$\lambda \in \Lambda_T$ of the hidden linear dynamic. This is illustrated by
the \emph{coronal spectrograms} in Figs.~\ref{fig:CC_DP}(c) and (d); these are
discussed once the general phenomenon is explained.

$T$'s eigenvalues \emph{on} the unit circle yield Dirac $\delta$-functions in the
power spectrum. $T$'s eigenvalues \emph{within} the unit circle yield more
diffuse line profiles, increasingly diffuse as the magnitude of the eigenvalues
retreats toward the origin. Moreover, the integrated magnitude of each
contribution is determined from the amplitude $\bra{\stationary}
\overline{\Omega} \, T_{\lambda} \Omega \ket{\one}$. Finally, we note that
nondiagonalizable eigenmodes yield qualitatively different line profiles.

The spectral decomposition of the power spectrum offers several insights into
the minimal temporal structure required to generate the observed power
spectrum.  In particular, since (i) each local maxima in the power spectrum
emanates from an eigenvalue of the hidden state-to-state transition matrix and
(ii) since the number of unique eigenvalues is upper bounded by the number of
hidden states (i.e., $\left| \Lambda_T \right| \leq \left| \SSet \right|$), we
have the following result: Counting both diffuse peaks and $\delta$-functions, the
number of observed peaks in the power spectrum (from $\omega \in (-\pi, \pi]$
in the discrete-time setting) puts a lower bound on the number of hidden states
of \emph{any} model capable of generating the observed stochastic process.
Note further that all transition matrices must have an eigenvalue of unity
and that this eigenvalue can only produce a $\delta$-function at $\omega=0$ with
no other way to shape the power spectrum over other frequencies. This gives the
immediate consequence that all single-state HMMs (i.e., all IID processes) have
a flat power spectrum, as suggested earlier. In such cases, $\Lambda_T = \{ 1
\}$, and there are no other eigenvalues to shape the power spectrum.

Figure~\ref{fig:CC_DP} shows the power spectrum of a particular parametrized
family of stochastic processes. Figure~\ref{fig:CC_DP}(a) displays the HMM's
skeleton with state-to-state transition probabilities parametrized by $b$. The
mean values $\braket{x}_{\pdf(X|s)}$ observed from each state are indicated as
the blue number inside each state. The process generated depends on the actual
PDFs and the transition parameter $b$. Although, and this is one of our main
points, the power spectrum is ignorant to the PDFs' details.

The evolution of the eigenvalues $\Lambda_T$ of the hidden-state transition
dynamic is shown from thick blue to thin red markers in
Fig.~\ref{fig:CC_DP}(b), as we sweep the transition parameter $b$ from $1$ to
$0$. A subset of the eigenvalues pass continuously but very quickly through the
origin of the complex plane as $b$ passes through $1/2$. The continuity of this
is not immediately apparent numerically, but can be revealed with a finer
increment of $b$ near $b \approx 1/2$. Notice the persistent eigenvalue of
$\lambda_T = 1$, which is guaranteed by the Perron--Frobenius theorem.

Using coronal spectrograms, introduced in Refs.~\cite{Riec14b} and
\cite{Riec18b}, Figs.~\ref{fig:CC_DP}(c) and \ref{fig:CC_DP}(d) illustrate how
the observed power spectrum $P(\omega)$ emanates from the \emph{eigen}-spectrum
$\Lambda_T$ of the hidden linear state-dynamic. Specifically, in
Fig.~\ref{fig:CC_DP}(c) and again, at another parameter setting, in
Fig.~\ref{fig:CC_DP}(d), we show the power spectrum $P(\omega)$ (plotted around
the unit circle in solid blue) and the eigen-spectrum $\Lambda_T$ (plotted as
red dots on and within the unit circle) of the state-to-state transition matrix
for the $11$-state hidden Markov chain (Fig. \ref{fig:CC_DP}(a)) that generates
it. As anticipated from Eq.~\eqref{eq:PcwFromDecomposedResolvent}, the power
spectrum has sharper peaks when the eigenvalues are closer to the unit circle.
The integrated magnitude of each peak depends on $\bra{\stationary}
\overline{\Omega} \ket{\lambda} \bra{\lambda} \Omega \ket{\one}$.

{ \color{blue}
It is easy to verify for this example that the stationary distribution
$\bra{\stationary}$ is uniform for any $b \in (0, 1]$ and that there is no
$\delta$-function at zero frequency since the average observation is zero.
Nevertheless, as $b \to 1$, ten $\delta$-functions (with five different
integrated magnitudes) emerge (per 2$\pi$ band of angular frequency) as the
non-unity eigenvalues of the transition matrix approach the points $\{ e^{i 2 n
\pi / 11} \}_{n=1}^{10}$ on the unit circle. At $b=1$, the power spectrum is
(up to a constant offset) the same as its discrete part: $P(\omega) =
P_\text{d}(\omega)$ + \text{const}. Whereas for $b \in (0, 1)$, the power
spectrum is diffuse and is the same as its continuous part: $P(\omega) =
P_\text{c}(\omega)$. 
}

Interestingly, our continuous \emph{power} spectrum is the shadow of the
discrete \emph{eigen}-spectrum of nonunitary dynamics. (The former is closely
related to the continuous eigen-spectrum of unitary models of chaotic
dynamics.) This suggests that resonances in various physics domains concerned
with a continuous spectrum can be modeled as consequences of simpler nonunitary
dynamics. Indeed, hints of this already appear in
Refs.~\cite{Nare03,Soko06,Most09}.

{ \color{blue}
While we frame our main results in terms of HMMs, in fact, they apply broadly
to regularly-observed physical systems. Many physical systems have exact
representations as finite latent-state models, as in the examples of
Sec.~\ref{sec:FWN_in_PhysicalSystems}. However, even when the mapping is not
exact, most if not all dynamical systems encountered in physics can be
approximated to an arbitrary accuracy by either an autonomous or an input-dependent HMM~\cite{Blan02, Riec18a}. \emph{The eigen-decomposition then serves to re-express the physical system and its power spectrum in its
natural state space.}

The interpretation for discrete-state physical systems is obvious. While there
are additional mathematical nuances with a continuous state space, the overall
picture remains intact~\footnote{In general, the spectrum of a time-evolution
operator is highly sensitive to the function space it transforms. However, the point spectrum---i.e., the eigenvalues---reflects the operator's simplicity once the appropriate function space has been identified. Fortunately, convergence of eigenvalues and their spectral projections is guaranteed when using a stochastically-smoothed modification to Ulam's method, as in Ref.~\cite[Thm~4]{Blan02}. (Note that his can be extended to more general manifolds.) Measurement naturally induces an Ulam-type partitioning, while the impossibility of perfectly isolating a physical system inevitably yields the stochastic smoothing. Ulam-type partitioning of state space, with arbitrary observables on the partitions, turns dynamical systems into HMMs---effectively broadening the scope of our analysis. Extended dynamical mode decomposition (EDMD) can also be used to obtain the spectral projections of the time-evolution operator~\cite{Klus16}. The estimated eigenvalues and projectors from either method can then be used to apply our results to generic dynamical systems in physics.}. Specifically, Eq.~\eqref{eq:PcwFromDecomposedResolvent} (and our subsequent analysis) applies to most dynamical systems encountered in physics---including quantum systems represented in Liouville space~\cite{Petr96}---since these dynamical systems have a countable number of discrete eigenmodes.}

\subsection{Continuous-time processes}
\label{sec:ContTime}

For both simplicity and generality, we focused on discrete-time
dynamics~\footnote{Indeed, discrete-time dynamics are, in a sense, more
general than continuous-time dynamics. That is, continuous-time dynamics are
obtained as the limiting behavior of discrete-time dynamics.}. However,
correlation and power spectra are often applied to continuous-time processes.
This section makes a more explicit connection to continuous-time processes and
points out important features.

First, continuous-time processes are typically observed not continuously, but
periodically at some sampling frequency $f_0$. The duration $\tau_0 = 1/ f_0$
between observations thus induces a discrete-time transition operator
$T_{\tau_0}$ between states in that time interval. In such cases, the
discrete-time transition matrix is related to the continuous-time generator $G$
of time evolution by $T_{\tau_0} = e^{\tau_0 G}$. Accordingly, the
continuous-time generator can be obtained from the discrete-time dynamic via $G
= f_0 \ln T_{\tau_0}$~\footnote{One selects the simplest branch of the
complex logarithm since the apparent freedom offered through its different
branches corresponds to unobserved cyclic behaviors at inaccessibly small
timescales that, if they exist, are aliased by the sampling frequency $f_0$. In
other words, the different branches are a gauge freedom supporting all dynamics
that could exist but are suppressed by the mapping $G \mapsto e^{\tau_0 G}$.}
And, the eigenvalues of $T_{\tau_0}$ and $G$ are simply related by
$\Lambda_{T_{\tau_0}} = \bigcup_{\zeta \in \Lambda_G} \{ e^{\tau_0 \zeta } \}$
\footnote{The relationship between discrete and continuous time is the same as
that yielding the well-known conformal mapping of the interior of the unit
circle in the complex plane to the left-half of the complex plane, which also
relates $z$-transforms and Laplace transforms.}.

\subsubsection{Autocorrelation and power spectra}
\label{sec:ContTimePS}

Continuous-time representations can be analyzed directly, though. Consider the
generic case of a continuous-time dynamic over a hidden state-space, with two
types of example in mind:
\begin{enumerate}
\setlength{\topsep}{-5pt}
\setlength{\itemsep}{-5pt}
\setlength{\parsep}{-5pt}
\item The system evolves through a continuous state-space. This describes both
	typical linear and nonlinear systems, including chaotic dynamical systems
	and Fokker--Planck dynamics. Then $G$ is the generator that induces the
	finite-time Ruelle--Perron--Frobenius operator. Or,
\item Observations are functions of a finite-state space with continuous-time
	transition rates. An example is current flowing or not, depending on the
	conformation of a biomolecular ion channel. Then $G$ is the rate matrix of
	the master equation.
\end{enumerate}
These different settings have the same expression for the autocorrelation
and power spectrum. We now give these in closed-form.

For real-valued $\tau > 0$, the autocorrelation is:
\begin{align}
\gamma(\tau) 
  & = \Braket{ \, \overline{X}_{\realtime}  X_{\realtime + \tau}}
  = \bra{\stationary}  \overline{\Omega}  \, e^{\tau G } \, \Omega \ket{\one} 
~.
\end{align}
From this, we derive the continuous part of the power spectrum $P_\text{c}(f)$
with respect to frequency $f \in \mathbb{R}$, with the result that:
\begin{align}
P_\text{c}(f) = 2 \, \text{Re} \bra{\stationary}
 \overline{\Omega} \, \bigl( 2 \pi i f I - G \bigr)^{-1} \Omega \ket{\one}
 ~.
\end{align}
Appealing to the resolvent's spectral expansion again allows us to better
understand the possible shapes of the power spectrum:
\begin{align}
P_\text{c}(f) = \sum_{\lambda \in \Lambda_G} \sum_{m = 0}^{\nu_\lambda - 1}
  2 \, \text{Re} \frac{
  \bra{\stationary} \overline{\Omega} \, G_{\lambda, m} \Omega \ket{\one}}{(2
  \pi i f - \lambda)^{m+1}}
\label{eq:ContTimePSD}
  ~.
\end{align}
Since all of the frequency-dependence is isolated in the denominator and since
$\bra{\stationary} \overline{\Omega} \, G_{\lambda, m} \Omega \ket{\one}$ is a
frequency-independent complex-valued constant, peaks in $P_\text{c}(f)$
arise only via contributions of the form Re$\frac{c}{(2 \pi i f -
\lambda)^n}$ for $c \in \mathbb{C}$, $f \in \mathbb{R}$, $\lambda \in
\Lambda_G$, and $n \in \mathbb{Z}_+$.

\subsubsection{Applications}
\label{sec:ContTimeApps}

{ \color{blue} Equation~\eqref{eq:ContTimePSD} helps explain the shapes of
power spectra of chaotic dynamical systems, as appeared some time ago, e.g.,
in Ref.~\cite{Farm80}. In that case, the eigenvalues of the time-evolution
operator---whether the Ruelle--Perron--Frobenious transfer operator or the
Koopman operator~\cite{Klus16}---are known as Ruelle--Pollicott
resonances~\cite{Poll85, Ruel86, Gasp05}, and $\bra{\stationary}$ is the
stationary distribution on the attractor. Stochastic differential equations
leading to Fokker--Planck dynamics, ubiquitous in statistical physics, also
obey Eq.~\eqref{eq:ContTimePSD}. In these cases, the spectral projection
operators describe the decay modes of probability densities on the continuous
state space.

Even when the exact operator for time evolution is unknown,
Eq.~\eqref{eq:ContTimePSD} can be used for the inverse problem of inferring the
hidden linear dynamic from data---since the empirical power spectrum constrains
the system's eigen-spectrum.

It should be noted however that power spectra obtained either experimentally or
numerically at finite sampling rate can deviate significantly from
Eq.~\eqref{eq:ContTimePSD} as $f \to f_0/2$. Equation~\eqref{eq:ContTimePSD}
only describes the empirical power spectrum of continuous-time processes for
frequencies much less than the sampling frequency such that $f/f_0 \ll 1$.
Whereas Eq.~\eqref{eq:PcwFromDecomposedResolvent} describes the empirical power
spectrum exactly over all frequencies. The empirical power spectrum will
approach Eq.~\eqref{eq:ContTimePSD} over any finite frequency band as the
sampling frequency is increased, coinciding in the limit that $f_0 / f \to
\infty$.
}

{ \color{blue} 
\subsubsection{Lorentzians and $1/f$ noise}
\label{sec:one_over_f_analytics}

When $c_{\lambda} \equiv \bra{\stationary} \overline{\Omega} \, G_{\lambda, 0}
\Omega \ket{\one}$ is real-valued, then the eigenmode's contribution to the
power spectrum is $c_{\lambda}$ times a Cauchy--Lorentz distribution over
frequencies, centered at $f= \text{Im}(\lambda)/ 2 \pi $ with full width at
half maximum (FWHM) of Re$(\lambda) / \pi$. This becomes a delta function in
the limit Re$(\lambda) \to 0$. It is notable that nondiagonalizable eigenmodes
contribute qualitatively distinct line profiles to the power spectrum.

Still one may wonder---since Eq.~\eqref{eq:ContTimePSD} is fully general for
continuous-time dynamics---where the commonly encountered feature of
$1/f$-noise could possibly originate. Inspired by Bernamont's 1937 insight that
superposed Lorentzians can lead to $1/f$ noise~\cite{Bern37b}, we can identify
a source of $1/f$ noise in our more general setting.

\begin{Def}
\label{def:UniformDim}
An observable continuous-time process has \emph{\DHD} if its:
\begin{enumerate}
\setlength{\topsep}{-5pt}
\setlength{\itemsep}{-5pt}
\setlength{\parsep}{-5pt}
\item Time-evolution generator $G$ is diagonalizable and has $N+1$ evenly
	spaced eigenvalues along the real line $\Lambda_G = \{ -n a \}_{n=0}^N$
	for some $a>0$, and 
\item Spectral intensity fades with increasing frequency according to
	$c_{-na} = c / n$ for $n \geq 1$ and some $c \in \mathbb{R}$.
\end{enumerate}
\end{Def}

Appendix~\ref{sec:OneOverfApp} shows that any process with \DHD\ produces $1/f$
noise over a frequency bandwidth proportional to $N$, such that:
\begin{align*}
P(f) \sim
\begin{cases}
\text{constant} & \text{if } f < 3 a / 2 \pi^2 \\
1/f  & \text{if } 3 a / 2 \pi^2 < f \lesssim a N / 4 \pi^2 \\
1/{f^2} & \text{if } f \gtrsim  a N / 4 \pi^2
\end{cases} ~.
\end{align*}
Note that the power spectrum's $1/f$ portion can start at very low frequencies,
if $a$ is small.

The surprising prevalence of $1/f$ noise in nature can now be reframed, in
light of our spectral results: Why would \DHD\ be so common in nature? We
suggest that \DHD\ is a consequence of common motifs of causal dependence
in processes. These dependencies impose structural constraints on transition
rate matrices that could characteristically shape their spectral properties.
Hopefully, this spectral reframing of $1/f$ noise will stimulate further
attempts to explain its ubiquity.
}

{ \color{blue} 

\subsection{Transducing structured noise}
\label{sec:DeterministicTransduction}

For certain dynamics, it is profitable to split the generator into
deterministic and random components. This is especially useful when a linear
time-invariant (LTI) system takes the structured noise as input. Random thermal
motion in a harmonic trap is a simple example.

When a LTI system transduces structured noise---taking process $X$ to
process $Y$---the output is generically a simple transformation of the noise,
modulated by the square magnitude of the LTI system's transfer function, $H_{X
\to Y}(\omega)$ or $H_{X \to Y}(f)$~\cite{Enge18}. In discrete-time the power
spectrum is:
\begin{align}
P_{YY}(\omega) = | H_{X \to Y} (\omega) |^2  P_{XX}(\omega) ~.
\label{eq:TransducedPSD}
\end{align}
This requires modification, however, when the eigenvalues of the noise 
coincide with the poles and zeros of the LTI system's transfer function.

Consider a LTI system described by polynomials $\mathscr{P}(\mathcal{D})$ and
$\mathscr{Q}(\mathcal{D})$ of either the discrete-time delay operator (i.e.,
$\mathcal{D} Y_t = Y_{t-1}$) or the continuous-time differential operator
(i.e., $\mathcal{D} Y_t = \tfrac{d}{dt} Y_t $) such that:
\begin{align*}
\mathscr{P}(\mathcal{D}) Y_t = \mathscr{Q}(\mathcal{D}) X_t ~.  
\end{align*}
Then the square-magnitude of the transfer function is given by:
\begin{align*}
 | H_{X \to Y} (\omega) |^2 = \frac{|\mathscr{Q}(e^{i \omega})|^2}{|\mathscr{P}(e^{i \omega})|^2} \, \,
\end{align*}
or:
\begin{align*}
 | H_{X \to Y} (f) |^2 = \frac{|\mathscr{Q}(i 2 \pi f)|^2}{|\mathscr{P}(i 2 \pi f)|^2} 
\end{align*}
for discrete-time or continuous-time models, respectively. In particular, $X_t$
can be generated from a noise model that can be any HMM type discussed here.

For example, each spatial dimension of a Brownian trajectory simply integrates
a white noise $X_t$ according to the finite-difference equation: $Y_t - Y_{t-1}
= X_t$. Appendix \ref{sec:AppendixOnDeterministicTransduction} shows this leads
to the well-known power spectrum of Brownian noise $\sim 1/f^2$ in the limit of
$f_0/f \to \infty$ and gives the correction for finite sampling rates. More
generally, Eq.~\eqref{eq:TransducedPSD} can be used to evaluate the power
spectrum from Langevin-type differential equations that transduce arbitrarily
sophisticated noise processes.

Notably, any noise structure not revealed by $X_t$'s power spectrum
$P_{XX}(\omega)$ remains veiled by $P_{YY}(\omega)$ after passing through any
LTI system. This begs the question of what has been hidden.
}

\section{Hidden Structure}
\label{sec:HiddenStructure}

In fact, quite a lot is hidden. Remarkably, the power spectrum generated by any
hidden-Markov process with continuous random variables for the
state-observables is the same as that generated by a potentially much simpler
process---one that is a function of the same underlying Markov chain that
instead emits the \emph{expectation value} of the state observable.

{\The \label{thm:PSDequivalence} 
Let $\mathcal{P} = \bigl\{ \pdf(X | s) \bigr\}_{s \in \SSet}$ specify any
state-paired collection of probability density functions over the domain $\Abet
\subseteq \mathbb{C}$. Let $\mathcal{B} = \bigl\{ \braket{X}_{\pdf(X | s)}
\bigr\}_{s \in \SSet}$ and let $\mathcal{Q} = \bigl\{ \delta(X -
\braket{X'}_{\pdf(X' | s)} ) \bigr\}_{s \in \SSet}$. Then, the power spectrum
generated by any hidden Markov model $\mathcal{M} = \bigl( \SSet , \Abet ,
\mathcal{P}, T  \bigr)$ differs at most by a constant offset from the power
spectrum generated by the hidden Markov model $\mathcal{M}' = \bigl( \SSet ,
\mathcal{B} , \mathcal{Q}, T \bigr)$ that has the same hidden Markov chain but
in any state $s \in \SSet$ emits, with probability one, the state-conditioned
expected value $\braket{X}_{\pdf(X | s)}$.
}

{\ProThe 
From Eqs.~\eqref{eq:PcwFromResolvent} and \eqref{eq:PdwFromResolvent}, we see
that $P_\text{c}(\omega) + P_\text{d}(\omega) - \bigl\langle \left| x \right|^2
\bigr\rangle$ depends only on $T$ and $\bigl\{ \braket{X}_{\pdf(X|s)} \}_{s \in
\SSet}$. Thus, all HMMs sharing the same $T$ and $\bigl\{
\braket{X}_{\pdf(X|s)} \}_{s \in \SSet}$ have the same power spectrum $P(\omega)
= P_\text{c}(\omega) + P_\text{d}(\omega)$, modulo a constant offset determined
by differences in $\bigl\langle \left| x \right|^2 \bigr\rangle$.
}

\begin{figure}[t]
\begin{center}
\includegraphics[width=0.99\columnwidth]{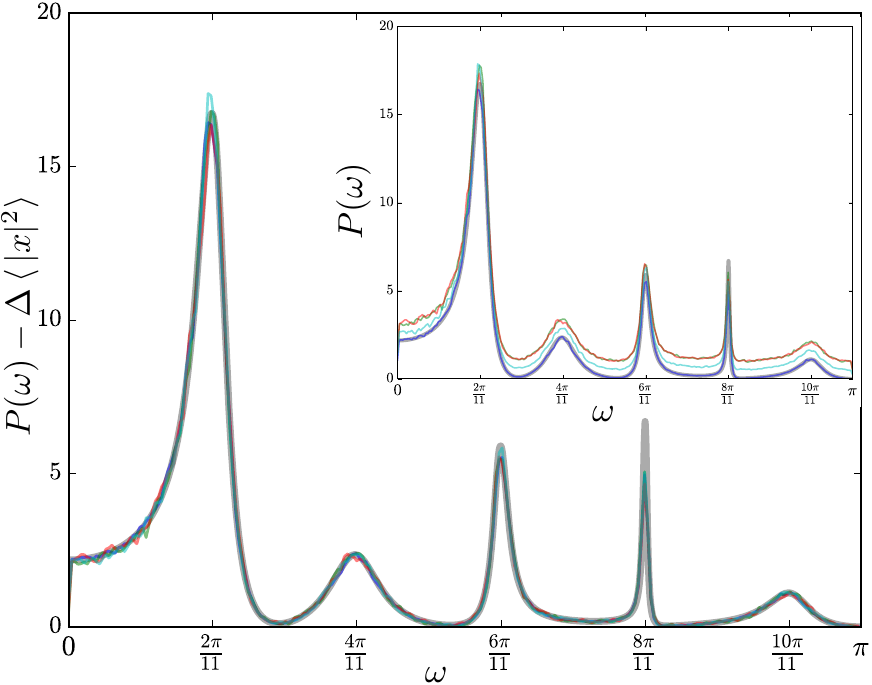}
\end{center}
\caption{Demonstrating Thm.~\ref{thm:PSDequivalence} for the processes
	generated by the HMM skeleton of Fig.~\ref{fig:CC_DP}(a), using transition
	parameter $b = 3/4$ as in Fig.~\ref{fig:CC_DP}(c). Besides an
	overall constant offset of $\braket{ \left| x \right|^2 }$, the power
	spectrum is insensitive to all details of the state-conditioned PDFs except
	for their averages. On top of the theoretical curve (thick gray) given by
	Eq.~\eqref{eq:PcwFromResolvent} we overlay horizontal offsets of the power
	spectra calculated numerically for stochastically generated time series.
	The state-conditioned PDFs used to define the different stochastic processes are: 
	(i) single $\delta$-functions, (ii) single Gaussians, 
	(iii) two symmetrically spaced $\delta$-functions (with no support at the mean), 
	and (iv) weighted $\delta$-functions with asymmetric spacing. 
	For each, a time series of length $2^{18}$ was generated. The Welch method 
	was used to calculate the average power spectrum for each process using FFTs of
	segments of length $2^9$. The inset shows the raw power spectrum 
	for each process without the offset.
	}
\label{fig:Thm1demo}
\end{figure}

Figure~\ref{fig:Thm1demo} demonstrates Thm.~\ref{thm:PSDequivalence} for the
power spectrum in Fig.~\ref{fig:CC_DP}(c).

One immediate consequence is the following.

{\Cor \label{cor:PSD_from_ZeroMean_pdfs} 
Any hidden Markov chain with any arbitrary state-paired collection of zero-mean
distributions, i.e.: 
\begin{align*}
\mathcal{P} \in \bigl\{ \{ \pdf( X | s) \}_{s \in \SSet} : \braket{X}_{\pdf( X |s)} = 0 \text{ for all } s \in \SSet \bigr\}
  ~,
\end{align*}
generates a flat power spectrum indistinguishable from white noise.
}

{\ProCor 
This follows immediately from Thm.~\ref{thm:PSDequivalence} and the fact that
the all-zero sequence has a power spectrum that is zero everywhere. Thus, the
corresponding power spectrum of the actual process is a flat (nonzero) 
power spectrum of uniform height $\braket{ \left| x \right|^2 }$.
}

We can relax the corollary to include cases where the state-conditioned PDFs
are all equal to a potentially-nonzero constant. Although, a $\delta$-function at
zero frequency (of integrated magnitude equal to the square magnitude of the
constant) will then be observed in addition to the flat power spectrum.

\begin{figure}[t]
\begin{center}
\includegraphics[width=0.99\columnwidth]{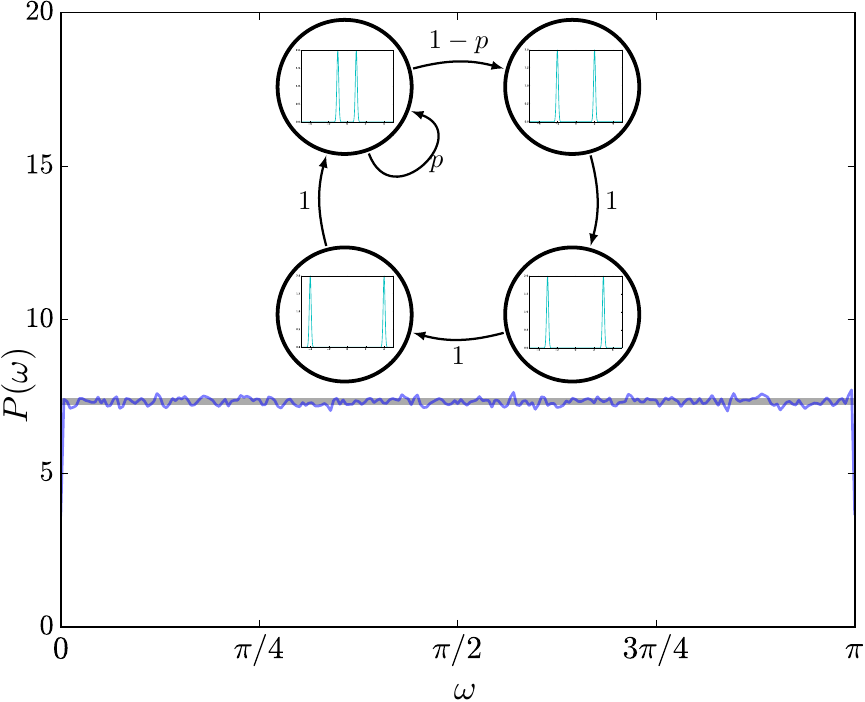}
\end{center}
\caption{Demonstrating Cor.~\ref{cor:PSD_from_ZeroMean_pdfs} on the
	\emph{Noisy Phase-Slip Process}: The overtly-structured stochastic process
	generated by the HMM (inset) has a flat power spectrum for all values of
	the phase-slip transition parameter $p \in [0, 1]$. The flat power spectrum
	is shown analytically (thick gray) and numerically (thin blue) for
	$p=1/10$.  The numerical power spectrum was calculated from a simulated
	time series of length $2^{20}$ using the Welch method, performing FFTs on
	segments of length $2^9$.
	}
\label{fig:Cor1demo}
\end{figure}

The corollary's implications are striking. It is quite surprising, to consider
one broad class of examples, that a power spectrum can be completely flat even
when a ring of sequential states are visited that emit observables with
probability density functions having no overlapping support.
Figure~\ref{fig:Cor1demo} gives an example. In such a case, any cogent observer
immediately detects the obvious structure in the mismatched supports---observed
values are distinct---and forbidden realizations. Yet the power spectrum
remains silent about this structure, reporting only the featureless signature
of white noise.

In other more challenging settings, structure is not always so obvious without
a reliable aid. Indeed, structure becomes increasingly difficult to detect
(by any means) when the state-conditioned probability density functions have
overlapping support. This is the generic case of non-Markovian processes. 
The hidden states cannot be detected via casual inspection.

While they give a concrete sense of missing structure, these cases fall far
short of telling the full story of how power spectra mask structure. The
following sections, culminating in Thm.~\ref{thm:GenFlatPS}, address additional
ways white noise appears without needing to meet the requirements of
Cor.~\ref{cor:PSD_from_ZeroMean_pdfs}. 

\subsection{Nonlinear Pairwise Correlation}

In a sense, the structure of the stochastic process in Fig.~\ref{fig:Cor1demo}
was hidden as shallowly as possible to evade appearing in the power spectrum.
As mentioned, the structure should be trivial to detect by other means. Indeed,
while the \emph{linear} pairwise correlation $\gamma(\tau)$ vanished for all
$\tau > 0$, there is still pairwise dependence between the generated random
variables, which is nonlinear. This pairwise dependence can be teased out using
the pairwise mutual information $\I(X_0 ; X_\tau)$ between observables at
different times \cite{Cove06a}. For the process of Fig.~\ref{fig:Cor1demo}, if
we take the limit of the narrow Gaussians in the state-conditioned PDFs to be
pairs of $\delta$-functions, then the pairwise mutual information can be
calculated exactly as shown in App.~\ref{sec:PairwiseMI}. In fact, $\I(X_0 ;
X_\tau)$ will be unchanged for any set of four PDFs we could have chosen for
the states of the example HMM, as long as the PDFs all have mutually exclusive
support for the observable output. (This then makes the hidden state a function
of the instantaneous observable.)

A concise summary of the pairwise mutual information is provided via
Ref.~\cite{Riec18b}'s \emph{power-of-pairwise-information} (POPI) spectrum: 
\begin{align*}
\mathcal{I}( \omega ) = - \H(X_0) + \lim_{N \to \infty} \sum_{\tau = -N}^N 
  e^{- i \omega \tau } \I(X_0 ; X_\tau)
  ~,
\end{align*}
where $\H(\cdot)$ is the Shannon entropy of its argument \cite{Cove06a}.
Examining the pairwise mutual informations and the POPI spectrum for this
example (see Figs. \ref{fig:PMI} and \ref{fig:POPI} in
App.~\ref{sec:PairwiseMI}), we find the decay of pairwise information to scale
intuitively with the phase-slip-parameter $p$. While Fig.~\ref{fig:Cor1demo}'s
example has no \emph{linear} correlation, nevertheless it \emph{does} have
pairwise structure. Thus, the structure of the example process was hidden from
power spectra, but not hidden from the POPI spectrum.

The following sections continue investigating temporally-structured processes,
but focus on those with no linear pairwise correlation (and so a flat power
spectrum) and no pairwise mutual information (and so a flat POPI spectrum).
These will lead us to introduce a general condition for flat power spectra.
And, since power spectra fail so often to detect structure, we turn from
criticizing them to being constructive: introducing ways to detect hidden
structure.

\subsection{Sophisticated Fraudulent White Noise}

Theorem~\ref{thm:PSDequivalence} established that the power spectrum from
processes with \emph{continuous} observable random variables is the same as the
power spectrum from much simpler corresponding processes with \emph{discrete}
observable random variables. Accordingly, Thm.~\ref{thm:PSDequivalence}
motivates studying the power spectra of processes with discrete observable
random variables to determine if there are further ways to achieve a flat power
spectrum, beyond Cor.~\ref{cor:PSD_from_ZeroMean_pdfs}'s possibilities. For
observables that are discrete random variables, it is sufficient to consider
their probability \emph{distributions} rather than their probability
\emph{density} functions. 

We begin this next step of the development by establishing the following
simple lemma:

{\Lem \label{lem:FlatProcess} 
Any stochastic process (not necessarily stationary) with the
\emph{Single-Condition-Independent Property} (SCIP):
\begin{align*}
\Pr(X_t | X_{t'} = x) & = \Pr(X_t) \\
  & = \Pr(X_{t'})
  ~,
\end{align*}
for all $x \in \Abet$ and all $t \neq t'$, generates a flat power spectrum,
mimicking white noise.
}

{\ProLem
See App.~\ref{sec:Lemma1Proof}
}

SCIP processes not only have a flat power spectrum but also a flat POPI
spectrum. SCIP implies $\I(X_0; X_\tau) = 0$ for all $\tau \neq 0$ which, in
turn, implies $\mathcal{I}( \omega ) = 0$. These processes completely lack any
pairwise correlation, whether linear or nonlinear.

Notably, Lem.~\ref{lem:FlatProcess} is not covered by
Cor.~\ref{cor:PSD_from_ZeroMean_pdfs}; nor is
Cor.~\ref{cor:PSD_from_ZeroMean_pdfs} subsumed by Lem.~\ref{lem:FlatProcess}.
Accordingly, the following develops a single simple condition (culminating in
Thm.~\ref{thm:GenFlatPS}) that covers all of these cases of fraudulent white
noise.

Crucially, the class of potentially-fraudulent-white-noise processes suggested
by Lem.~\ref{lem:FlatProcess} is nontrivial. In addition to IID processes, this
class includes non-Markovian processes that hide all of their structure beyond
pairwise correlations.

The Random--Random--XOR process (RRXOR), discussed at length in
Ref.~\cite{Riec18b}, is an example. Over blocks of length $3$, the first two
bits are generated randomly from a uniform distribution and the third bit is
then the logical XOR operation of the last two. Explicitly:  
\begin{align*}
X_{3n + \phi} & = \text{XOR}(X_{3n-2 + \phi}, X_{3n-1 + \phi})~,~\text{whereas} \\
X_{3n-2 + \phi} & \sim \left(\tfrac{1}{2}, \tfrac{1}{2}\right) ~\text{and} \\
X_{3n-1 + \phi} & \sim \left(\tfrac{1}{2}, \tfrac{1}{2}\right)
  ~,
\end{align*}
for all $n \in \{ 1, 2, \dots \} $. As a SCIP process, the RRXOR process has a
flat power spectrum although it does not fall under the purview of
Cor.~\ref{cor:PSD_from_ZeroMean_pdfs}. Indeed, the RRXOR process has no
pairwise correlation at all since $\I(X_0; X_\tau) = 0$ for all
$\tau > 0$.  Accordingly, the POPI spectrum is zero over all frequencies. The
structure in this process is strictly \emph{three-way} correlation. In
Ref.~\cite{Riec18b}, the phase $\phi$ itself is a random variable, and
synchronizing to the phase is a surprisingly difficult task~\footnote{Indeed,
the burden of resolving this phase ambiguity renders the RRXOR process infinite
Markov order.  
}.
No matter, whether or not the phase $\phi$ is given, the process has no
pairwise correlation---resulting in a flat power spectrum and flat POPI
spectrum---and only reveals correlation in its three-way structure.

It is interesting to note that the related RRX\emph{NOR} process, where $X_{3n}
= \text{XNOR}(X_{3n-2}, X_{3n-1})$, also has a flat power spectrum. In fact,
this suggests a new method to hide structure: embed a correlated message into a
sequence of RRXOR and RRXNOR 3-bit sequences that lifts all correlation beyond
pairwise. Specifically, the original message is transformed into a sequence of
choices about whether to use XOR or XNOR on the previous two random bits. As
long as the read frame and the embedding mechanism is known, the message can be
easily extracted. But, if it is not known that a message is embedded, it cannot
be detected simply by looking for pairwise correlations.

Through similar construction, structure can be shifted up to arbitrarily-high
orders of correlation. 
Stochastic processes can be constructed
with $N$-way correlation but no $n$-way correlation for all $n < N$. Moreover,
an arbitrarily correlated message can be embedded within such a process, such
that its structure is lifted beyond any desired order of correlation.

\subsection{Content-preserving Whitening}

Corollary~\ref{cor:PSD_from_ZeroMean_pdfs} gave a method to construct an
arbitrarily complex process with a truly flat power spectrum, so long as all
hidden states have the same average output. Here, we introduce an alternate
method to construct arbitrarily complex processes with truly flat power
spectra. These processes, in addition, are devoid of $n$-way correlation for
all $n < N$.

\begin{enumerate}
\setlength{\topsep}{-5pt}
\setlength{\itemsep}{-5pt}
\setlength{\parsep}{-5pt}
\item Choose an embedding block length $N \geq 3$.
\item Choose any stochastic process (``Process A'') with a binary output
	alphabet.
\item Construct ``Process B'' as follows:\\[-20pt]
    \begin{itemize}
	\setlength{\topsep}{-5pt}
	\setlength{\itemsep}{-5pt}
	\setlength{\parsep}{-5pt}
    \item Whenever Process A would produce a $0$, Process B will sample a word
		uniformly from the set of all words of length $N$ with an even number
		of $1$s.
    \item Whenever Process A would produce a $1$, Process B will sample a word
		uniformly from the set of all words of length $N$ with an odd number
		of $1$s.
    \end{itemize}
\end{enumerate}    

Any Process B constructed in this manner has a truly flat power spectrum.
Process B will also be devoid of $n$-way correlation for all $n < N$.
Moreover, if A is a stationary process such that its statistical complexity
$\Cmu$(A) is well defined~\cite{Crut88a, Crut12a}, then Process B is
also a stationary process with $\Cmu$(B) $\geq \Cmu$(A).

This also works for ``infinitely structured'' processes, those with divergent
statistical complexity. Choose \emph{any} binary Process-A family with $\Cmu
\to \infty$. This can be, for example, Ref.~\cite{Trav11b}'s \emph{Heavy-Tailed
Periodic Mixture Process} that has infinite past--future mutual information:
$\EE \to \infty$. Then add some structure, via content-preserving
whitening, to obtain a binary Process-B family with $\Cmu \to \infty$ and a
truly flat power spectrum.

Similar constructions can also be developed for processes with larger alphabets.

Through the lens of pairwise correlation, such structure is simply missed.
However, before moving on to consider more advanced methods to detect such
structure, we finish our investigation of flat power spectra from structured
processes. The next section addresses a broad class of possibly-input-dependent
process generators and we give a very general condition for when a flat power
spectrum results.

\subsection{Input-dependent Generators and Fraudulent White Noise}

Probing fraudulent white noise more broadly, consider an arbitrarily correlated
message $\vec{m}$ and an input-dependent generator $\mathcal{M}(\vec{m})$ of an
observable output process $\{ X_t \}_{t \in \mathcal{T}}$. The lengths of the
inputs and outputs need not be commensurate, and the input and output alphabets
may also be distinct. The generator is fully specified by the tuple
$\mathcal{M}(\vec{m}) = \bigl( \SSet, \Abet, \mathcal{P}, \{ T_t(\vec{m}) \}_t
, \boldsymbol{\mu}_1 \bigr)$. That is, the internal states $\SSet$, output
alphabet $\Abet$, and state-dependent PDFs $\mathcal{P}$ are static. However,
the hidden-state-to-state transition matrix $ T_t(\vec{m})$ at time $t$ is
potentially a function of the full input $\vec{m}$. Since stationarity is no
longer assumed, the initial distribution $\boldsymbol{\mu}_1$ over hidden
states must be specified for the statistics of the output process to be well
defined.

\begin{figure}[h]
\begin{center}
\includegraphics[width=0.4\textwidth]{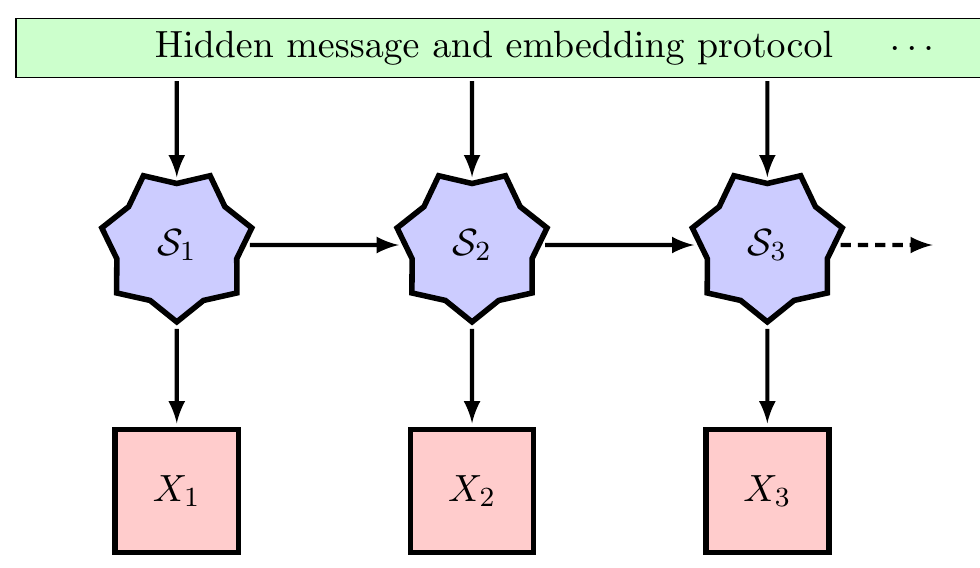}
\end{center}
\caption{Bayesian network for memoryful input-dependent generators.
  }
\label{fig:InputDepGen_BayesNet}
\end{figure}

Figure~\ref{fig:InputDepGen_BayesNet} shows the relevant Bayes network for this
general type of input-dependent generator. Contrast this with
Fig.~\ref{fig:HMM_BayesNet}, which showed the Bayes network of autonomous HMM
generators.  Autonomous HMMs can be seen as a special case of these
possibly-input-dependent generators when the process $\mathcal{M}(\vec{m}) =
\mathcal{M}$ is input-independent and the initial distribution
$\boldsymbol{\mu}_1 = \boldsymbol{\pi}$ is taken to be the stationary
distribution $\bra{ \boldsymbol{\pi}} = \bra{ \boldsymbol{\pi} } T$ of the
time-independent transition matrix $T_t(\vec{m}) = T$.

The memoryful input-dependent generators we now consider also generalize the
memoryful transducers introduced in Ref.~\cite{Barn15} to use
continuous-variable outputs and allow the lengths of input and output to be
incommensurate. Via any of the above models, very general message-embedding
schemes can be developed that produce sophisticated fraudulent white noise.

Even with all the generalizations, we can determine autocorrelation and power
spectra. Similar to the derivation for HMMs, we find that \emph{if} the process
is wide-sense stationary then (for $\tau \geq 1$):
\begin{align}
\gamma(\tau) &= \braket{ \boldsymbol{\mu}_t | \overline{\Omega} \, T_{t:t+\tau}(\vec{m}) \, \Omega | \one }
  ~,
\label{eq:AutocorrForInputDepModels}
\end{align}
which must overall be $t$-independent (so long as $t \geq 1$). Here,
$\bra{\boldsymbol{\mu}_t} = \bra{ \boldsymbol{\mu}_1 } T_{1:t} (\vec{m})$ and
$T_{a:b}(\vec{m}) = \prod_{t=a}^{b-1} T_{t} (\vec{m})$, and $\Omega$ is again
given by Eq.~\eqref{eq:AvgObsMatrix}. (Notice that $T_{a:a+\tau}(\vec{m}) =
T^\tau$ for the special case of autonomous HMMs.)

Thus, autocorrelation for $\tau \geq 1$ can be calculated as $\braket{ \boldsymbol{\mu}_1 |
\overline{\Omega} \, T_{1:1+\tau}(\vec{m}) \, \Omega | \one }$, assuming that
the pairwise statistics are stationary. This can also be written as:
\begin{align}
\gamma(\tau) = \big\langle  \braket{\overline{x}}_{\pdf(X | \St_t)} \braket{x}_{\pdf(X | \St_{t+\tau})} \big\rangle_{\Pr(\St_t, \St_{t+\tau}) } ~,
\label{eq:AutocorrAsAvgAvgs}
\end{align}
where we treat $\braket{x}_{\pdf(X | \St_t)}$ as a random variable that depends
on $\St_t$ and the whole expression becomes $t$-independent assuming stationary
pairwise statistics. Accordingly, the autocorrelation function is constant and
the power spectrum is flat whenever:
\begin{align*}
\Pr \bigl( \braket{x}_{\pdf(X | \St_{t+\tau})} |  \St_t = s \bigr)
  & = \Pr \bigl( \braket{x}_{\pdf(X | \St_{t+\tau})} \bigr) \\
  & = \Pr \bigl( \braket{x}_{\pdf(X | \St_t)} \bigr)
  ~,
\end{align*}
for all $\tau$, for all $t \in \mathcal{T}$, and for all $s \in \SSet$.

However, this requirement is too strict to cover all cases of interest.  For
example, it does \emph{not} yet imply the flat power spectrum of the RRXOR
process. More generally, constant autocorrelation and flat power spectra can be
guaranteed by an even weaker condition.

To appreciate this, define the set $\Xi$ of average outputs emitted by the
states: $\Xi \equiv \bigcup_{s \in \SSet} \bigl\{ \braket{x}_{\pdf(X | s)}
\bigr\}$. Furthermore, we define $\SSet_\xi \subset \SSet$ as the set of states
that all emit the same average output $\xi \in \Xi$. Explicitly, $\SSet_\xi
\equiv \{ s \in \SSet :  \braket{x}_{\pdf(X | s)} = \xi \}$. Using these
quantities, we can state our result more precisely as the following theorem.

{\The \label{thm:GenFlatPS} 
Let $\{ X_t \}_t$ be a stochastic process generated by any of the hidden-state
models $\mathcal{M}(\vec{m})$ discussed above, including autonomous HMMs and
input-dependent generators, $X_t$ the random variable for the observable at
time $t$, and $\St_t$ the random variables for the hidden state at time $t$.
Such processes have \emph{constant autocorrelation and a flat power spectrum} 
if:
\begin{align}
\Pr(\St_{t+\tau} \in \SSet_{\xi'} | \St_t \in \SSet_\xi )
  & = \Pr(\St_{t+\tau} \in \SSet_{\xi'} ) \nonumber \\
  & = \Pr(\St_{t} \in \SSet_{\xi'} ) 
  ~,
\nonumber  
\end{align}
for all separations $\tau > 0$, for all $t \in \mathcal{T}$, and for all $\xi, \xi' \in \Xi$.
}

{\ProThe
See App.~\ref{sec:Thm2Proof}
}

Theorem~\ref{thm:GenFlatPS} says that a flat power spectrum results whenever
the average output of the future hidden state is independent of the average
output of the current latent state.

This generalized condition for flat power spectra covers the special case for
HMMs as well as fraudulent white noise from message-embedding schemes with
stationary pairwise statistics, but nonstationary high-order statistics.
Appendix~\ref{sec:MealyForQuantum} shows that a modified version of
Thm.~\ref{thm:GenFlatPS} also applies to another class of generators that can
be more natural for measured quantum systems and systems with computational
dependencies. Theorem~\ref{thm:GenFlatPS} subsumes
Cor.~\ref{cor:PSD_from_ZeroMean_pdfs} as well as Lem.~\ref{lem:FlatProcess}.
And, it offers the most general guarantee yet for constant autocorrelation and
flat power spectrum.

By way of contrast consider the following. While zero pairwise mutual
information is always a sufficient condition for flat power spectrum, it is not
a necessary condition. Here, in Thm.~\ref{thm:GenFlatPS}, we find a very
general condition for a flat power spectrum.
Appendix~\ref{sec:GenThmForTimeDepPDFs} established a related theorem
(Thm.~\ref{thm:GenFlatPSforModelswTDepPDFs}) that further generalizes the
condition for flat power spectra, allowing for time-dependent PDFs associated
with each state.  Moreover, Thm.~\ref{thm:GenFlatPS} and
Thm.~\ref{thm:GenFlatPSforModelswTDepPDFs} constructively suggest how to design
such processes. Notably, these generalized conditions do not require a
stationary dynamic over the hidden states of the observation-generating
mechanism, which furthermore allows messages to hitchhike undetected aboard
fraudulent white noise.

{\color{blue}
More broadly, we may ask when two processes generate the same power spectrum,
whether or not it is flat.

{\The \label{thm:SamePS} 
Let $\{ X_t \}_t$ and $\{ Y_t \}_t$ be two stochastic processes generated by any of the hidden-state
models $\mathcal{M}(\vec{m})$ discussed above, including autonomous HMMs and
input-dependent generators, $X_t$ and $Y_t$ the random variables for the observables at
time $t$, and $\St_t \in \SSet$ and $\mathcal{R}_t \in \boldsymbol{\mathcal{R}}$ 
the random variables for the respective hidden states at time $t$.
These processes have \emph{identical power spectra}, up to a constant offset,
if:
\begin{align}
\Pr(\St_t \in \SSet_\xi , \St_{t+\tau} \in \SSet_{\xi'} )
  & = \Pr( \mathcal{R}_t \in \boldsymbol{\mathcal{R}}_\xi , \mathcal{R}_{t+\tau} \in \boldsymbol{\mathcal{R}}_{\xi'} )
  ~,
\nonumber  
\end{align}
for all separations $\tau > 0$, for all $t \in \mathcal{T}$, and for all $\xi, \xi' \in \Xi$.
}

{\ProThe
See App.~\ref{sec:ProofOfSamePSD}
}

Section~\ref{sec:SilentCrystals} below leverages Thm.~\ref{thm:SamePS} to
determine the degeneracy of diffraction patterns from distinct physical
structures.
}

This suite of results emphasizes our main argument's generality: Power spectra
are mute when detecting a broad range of observable structure. Whether
observing physical, biological, or social systems, we seek structure that
reveals mechanism and begets predictability. Through the lens of power spectra,
or pairwise correlation more generally, much structure is simply missed. The
challenge then is to look for structure beyond pairwise.
{\color{blue}
Section~\ref{sec:SeekingStructureInNoise} addresses this challenge shortly.
First, though, to motivate the extra effort, we show that fraudulent white
noise is indeed a feature of real physical systems.
}

{ \color{blue} 
\section{Hidden Physical Structure}
\label{sec:FWN_in_PhysicalSystems}

To ground the theoretical consequences in natural, even familiar phenomena,
this section takes on three, rather disparate physical systems. It draws out
important physical implications of fraudulent white noise and power spectral
degeneracy in quantum entanglement, chaotic crystallography, and
neural-membrane ion channels.

\subsection{Fraudulent White Noise From Quantum Entanglement}
\label{sec:FWN_fromEntanglement}

Correlated measurements of entangled quantum systems indelibly confirmed the
reality of nonlocal physical states. In particular, Bell tests conclusively
showed that no local hidden variable theory is consistent with certain
strongly-correlated observations~\cite{Hens15, Gius15, Shalm15}. Detecting
correlation in more general quantum states should similarly yield a deeper
appreciation of quantum correlation's important role in everything from
thermodynamics~\cite{Espo10} to gravity~\cite{Van10, Suss16}. But what if our
tools mask correlations?

Entangled many-body systems, as it turns out, easily generate fraudulent white
noise when they are measured. The following demonstrates that repeated
measurements of even quite simple entangled states leads to fraudulent white
noise. As a consequence, one is at risk of inadvertently inferring randomness
where there is essential correlation. One tends to assume, of course, that
measurements would necessarily confirm the ubiquity of entanglement and reveal
the high-order correlations naturally induced in the time evolution of physical
systems.

As a particular example, consider the entangled three-body quantum state:
\begin{align*}
\ket{\Xi}
\equiv 
\frac{1}{2} \Bigl( \ket{000} + \ket{011} + \ket{101} + \ket{110}  \Bigr) 
  ~,
\end{align*}
where, for example,
$\ket{011} = \ket{0}_\text{A} \otimes \ket{1}_\text{B} \otimes \ket{1}_\text{C} $.
The quantum circuit diagram:
\begin{align*}
\Qcircuit @C=1em @R=.7em {
	\lstick{\ket{0}} & \gate{H} & \qw & \ctrl{2} & \qw \\ 
	\lstick{\ket{0}} & \gate{H} & \ctrl{1} & \qw & \rstick{ \, \Bigg\} \ket{\Xi}} \qw \\ 
	\lstick{\ket{0}} & \qw 		& \targ & \targ & \qw \\ 
} 
\end{align*}
shows that $\ket{\Xi}$ is directly generated by a sequence of two Hadamard
gates and two controlled-NOT (CNOT) gates applied to the unentangled state
$\ket{ 0 0 0 }$. Recall that the Hadamard gate $H_\text{A}$ maps
$\ket{0}_\text{A}$ to $\ket{+}_\text{A} \equiv \bigl(
\ket{0}_\text{A} + \ket{1}_\text{A} \bigr) / \sqrt{2}$.

When measured in the \emph{computational basis} of $0$s and $1$s, repeated
preparation and measurement of $\ket{\Xi}$ states leads exactly to the RRXOR
process discussed above. This quantum preparation and measurement setup is
shown explicitly in Fig.~\ref{fig:HMMsFromEntanglement}(d). Certainly,
observations contain predictable correlations. A pairwise analysis of the
observation sequence, however, gives the statistics white noise. This holds
whether the analysis used either power spectra or POPI spectra or, indeed, any
analysis that can be performed in one-on-one meetings among Alice, Bob, and
Charlie who each hold one of the component qubits.

\begin{figure}[h]
\begin{center}
\includegraphics[width=0.47\textwidth]{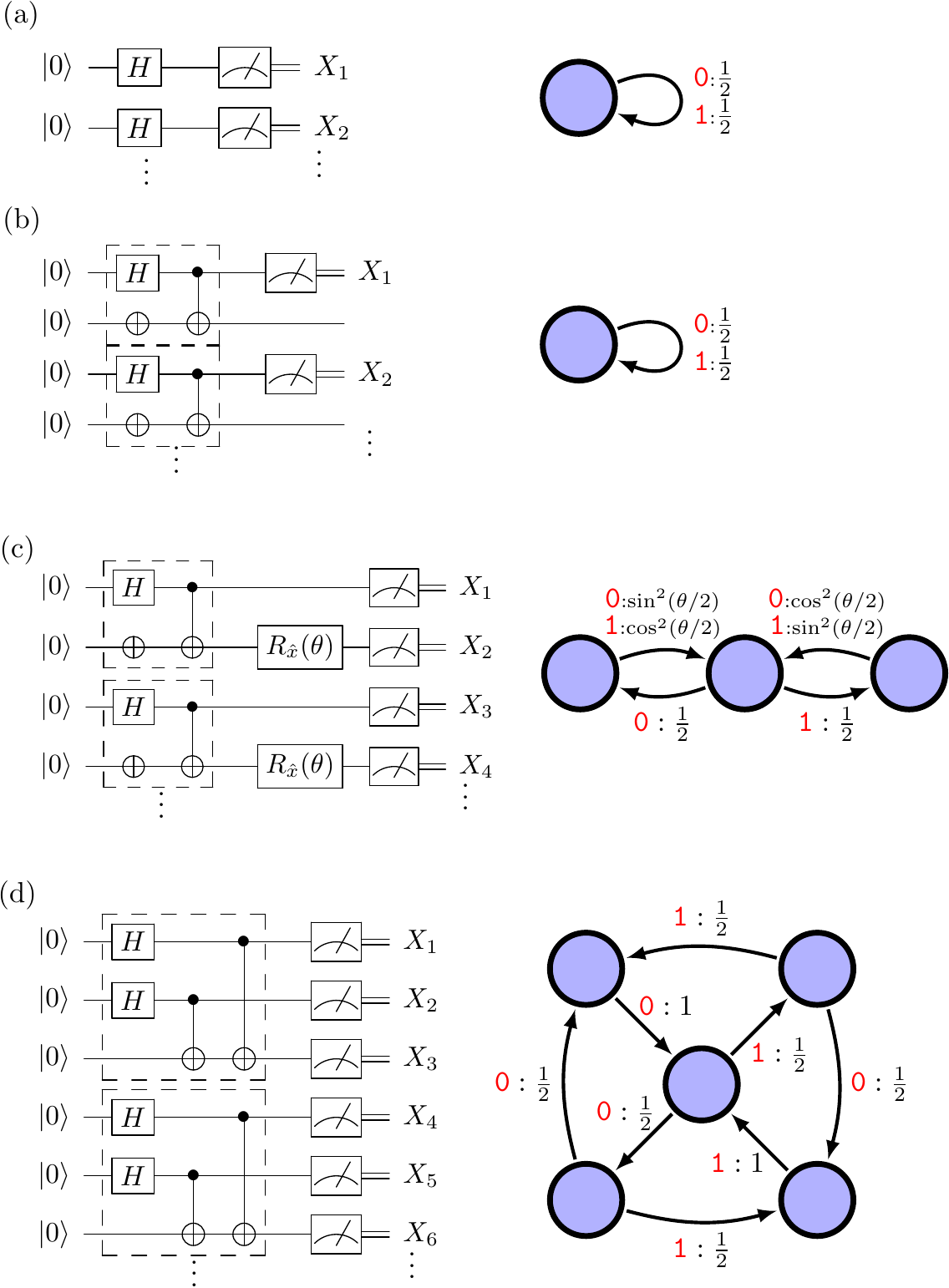}
\end{center}
\caption{Stochastic processes generated by fixed measurements of
	unitarily-transformed blank quantum inputs. These include: (a)
	measurement-basis-dependent genuine white noise; (b)
	measurement-basis-independent uniform white noise; (c) a correlated Bell
	process; and (d)  entanglement-enabled fraudulent white noise. Dashed boxes
	are drawn around the entangling unitary  modules in each case; except (a),
	where there is no entanglement. The induced Mealy-type HMMs shown on the
	right are the minimal descriptors of the output process. The edge label
	``${\color{red} \texttt{x}} : q $'' on the transition from state $s$ to
	$s'$ indicates the joint probability $\Pr(X_t = {\color{red} \texttt{x}},
	\St_{t+1} = s' | \St_t = s ) = q$ of observing ${\color{red} \texttt{x}}
	\in \Abet$ and transitioning to $s'$, given the current state $s$.
	Mealy-type HMMs are a simple case of the more general Measurement Feedback
	Models discussed in App.~\ref{sec:MealyForQuantum}.
  }
\label{fig:HMMsFromEntanglement}
\end{figure}

Figure~\ref{fig:HMMsFromEntanglement} compares additional examples of
stochastic processes generated by fixed measurement of unitarily-transformed
blank quantum inputs~\footnote{Time-varying measurement protocols
further enrich the observation sequence. See, for example, the recent
examples in Ref.~\cite{Vene19}---states that are exquisitely detailed even
without entanglement. Generators of classical stochastic processes may also use
persistent quantum memory, in which case they require less memory than
otherwise classically necessary~\cite{Gu12, Maho16, Riec15b, Bind18, Liu19}.
It is interesting that the quantum generators (of classical stochastic
processes) considered in this section appear to require no supplementary memory
at all.}.
Panel \ref{fig:HMMsFromEntanglement}(a) reminds us that almost any measurement
of a quantum system yields some randomness. The amount of uncertainty, though,
depends on how well the measurement basis aligns with the system's quantum
state. However, Panel \ref{fig:HMMsFromEntanglement}(b) reminds us that local
properties of a maximally-entangled state are maximally unpredictable,
regardless of the local measurement basis. The entire structure of a
maximally-entangled state exists only nonlocally among constituents, yielding
correlations when measurements on different parts of the system are
compared---as in the Bell process of Panel \ref{fig:HMMsFromEntanglement}(c).

When the number of entangled parties is larger than two, correlation becomes
much harder to detect. Nevertheless, in each case the physical input, unitary
transformation, and measurement protocol together determine the HMM that
exactly describes the correlated output process.
Figure~\ref{fig:HMMsFromEntanglement}(d) shows how fraudulent white noise in
the form of the RRXOR process can arise from measurements of entangled
three-bodied quantum systems. Moreover, adding two swap gates:
\begin{align*}
\Qcircuit @C=1em @R=.7em {
	\lstick{\ket{0}} & \gate{H} & \qw & \ctrl{2} &  \qw &  \qw & \qw & \qw &  \meter &   \rstick{X_1} \cw \\
	\lstick{\ket{0}} & \gate{H} & \ctrl{1} & \qw  &  \qw & \qswap & \qw & \qw &  \meter &  \rstick{X_2} \cw \\
	\lstick{\ket{0}} & \qw 		& \targ & \targ & \qswap & \qswap \qwx & \qw & \qw & \meter &  \rstick{X_3} \cw 
	\\
	 \lstick{\ket{0}} & \gate{H} & \qw & \ctrl{2}  &  \qw &  \qw & \qw & \qw & \meter & \rstick{X_4} \cw \\
	\lstick{\ket{0}} & \gate{H} & \ctrl{1} & \qw & \qswap \qwx[-2] &  \qw & \qw &  \qw & \meter  & \rstick{X_5}  \cw \\
	\lstick{\ket{0}} & \qw 		& \targ & \targ & \qw & \qw &  \qw & \qw & \meter & \rstick{X_6} \cw
	 \gategroup{1}{2}{6}{7}{.7em}{--}
}
\end{align*}
transforms the output into the even more cryptic Interlaced RRXOR process,
discussed shortly in Sec.~\ref{sec:BecomingInformed}.

These constructions demonstrate that \emph{simple sequences of two-body
interactions can generate high-order correlations while revealing no low-order
correlation whatsoever}.
}

{ \color{blue} 
\subsection{Silent Crystals}
\label{sec:SilentCrystals}

One icon of natural structure is found in the atomic placements encoded in
crystals. Semiconductor crystal structures tell electrons and light how to move
within them, while the aperiodic crystals of our genetic script instruct our
cells how to behave. The typical way to probe crystal structure is X-ray
diffraction---the power spectrum of a crystal's electron density. Thus, the
preceding results on the degeneracy of power spectra highlight \emph{which
features of crystal structure can be inferred from diffraction patterns}.

\emph{Close-packed structures}, which mimic the dense packing of hard spheres,
offer an interesting case study due to their multiplicity and natural
abundance~\cite{Edwa42, Riec14a}. All close-packed structures are composed of
\emph{modular layers} $\{ A, B, C \}$, with a material-dependent basis attached
to a 2-dimensional hexagonal crystal lattice. Assembling these modular layers,
there are two choices for how to nestle the next layer to fill the holes as
tightly as possible. For a particular material, differences in diffraction
patterns arise from this sequence of stacking choices~\cite{Hend42, Trea91,
Varn12a, Riec14b}.

Besides the ABABAB$\dots$ period-two stacking of close-packed two-dimensional
monolayers that leads to \emph{hexagonal close packing} (hcp) and the
ABCABC$\dots$ period-three stacking of these monolayers that leads to
\emph{cubic close packing} (ccp), there is an infinite number of ways to stack
the monolayers as tightly as possible. The only constraint is the stacking rule
that no layer (whether A, B, or C) can appear twice in succession. Nature, it
turns out, is fully aware of all the possibilities.

For close-packed materials, the net energy from nearest-neighbor interactions
is indifferent to which of the infinitely many close-packed structures is
realized. This facilitates great diversity, in both natural and fabricated
materials, via \emph{polytypism} and \emph{random stacking}~\cite{Kitt76}.
Prominent examples of polytypic layered structures include SiC, ZnS, stacked
graphene, and ice~\cite{Varn15}. Different polytypes of the same material can
have very different electronic, optical, and mechanical properties~\cite{Bao11}.

\subsubsection{Diffraction theory of layered structures}

Appendix \ref{sec:DPsAsPSs} reviews the basics of diffraction theory and shows
that the diffracted intensity (as a function of the scattering vector
$\vec{q}$) can be written as a power spectrum of layer \emph{form factors} $X_n
= F^{(n)}(\vec{q}) \in \mathbb{C}$. Each layer form factor is the Fourier
transform of the spatially-extended scatterer density (e.g., the electron
density) associated with the layer. In particular, the expected diffracted
intensity can be written as:
\begin{align}
\braket{ I_\text{diff}(\vec{q}) } & = c N P(\omega) \nonumber \\
  & = c \Braket{ \left|  
  \sum_{n = 1}^{N} F^{(n)}(\vec{q}) e^{-i \omega  n }
  \right|^2 }
  ~,
\label{eq:DPasPSD}
\end{align}
where $\omega = \tau_0 \vec{q} \cdot \hat{\ell}$ quantifies the change in
wavenumber along the stacking direction $\hat{\ell}$ of $N$ sequential layers
of thickness $\tau_0$. 

For typical layered structures, there is only a small number of layer types.
For close-packed structures, to take one example, each layer realizes one of
only three allowed relative offsets in its plane. Yet, in detail, we know that
each layer type is subject to both thermal fluctuations and quantum uncertainty
of atomic positions. What are the consequences for the diffraction pattern?

Suppose there is a hidden-state model $\mathcal{M}(\vec{m}) = \bigl( \SSet,
\Abet, \mathcal{P}, \{ T_t(\vec{m}) \}_t , \boldsymbol{\mu}_1 \bigr)$ that
generates the correct statistics of the layer form factors in the
material---taking the stochastic stacking process, thermal motion, and quantum
uncertainty into account. Theorems \ref{thm:PSDequivalence} and
\ref{thm:SamePS} imply that---up to a constant offset---\emph{the diffraction
pattern will be the same if we instead consider the much simpler hidden-state
model $\mathcal{M}'(\vec{m}) = \bigl( \SSet, \mathcal{B}, \mathcal{Q}, \{
T_t(\vec{m}) \}_t , \boldsymbol{\mu}_1 \bigr)$ that outputs only the
\emph{expected} layer form factor from each hidden state}.

Appendix~\ref{sec:FWN2DWT} shows that this allows us to easily and rigorously
produce the surprising results of Debye--Waller theory for exactly periodic
lattices in the general setting of randomly stacked structures. Specifically,
the state-conditioned expectation value of the form factor directly leads to
the Debye-Waller factor that exponentially suppresses the intensity of the
diffraction pattern at large magnitude of the scattering vector. Surprisingly,
however, thermal and quantum fluctuations do not lead to broadening of the
diffraction pattern.

For close-packed structures, there are only three types of layers, differing
only via relative displacements of $1/3$ of a lattice translation vector in the
plane of the layer. As a result, if type-$A$ layers have an expected layer form
factor of $A=\psi$, then the other layer types are simply related by the third
roots of unity:
\begin{align*}
A = \psi , \; \; \;
B = \psi e^{i 2 \pi / 3} , ~\text{and}~ \; \; \;
C = \psi e^{-i 2 \pi / 3} ~.
\end{align*}
(See App.~\ref{sec:ClosePackedFormFactors}). The three possible
state-conditioned expectation values for the layer form factors serve as the
alphabet $\mathcal{B} = \{ A, B, C \}$ for the stacking process $\{ X_n \}_n$,
where $n$ indexes the layer and adjacent layers are separated by a distance of
$\tau_0$. The diffracted intensity from any close-packed structure is then
given by the power spectrum of the stacking process.

Information about the stacking process is most directly revealed via $P(\omega)
/ |\psi|^2$ wherever $|\psi|^2$ is nonzero, which discounts the expected
diffraction pattern of a single layer~\footnote{Dividing $P(\omega)$ by
$|\psi|^2$ is the same as analyzing the `corrected' diffraction pattern (which
discounts any $\omega$-dependence of $\psi$) along a row in reciprocal space
satisfying the ($h-k=1 \; \text{mod } 3$)-restriction of the Miller indices
associated with the crystalline modular layers.}.

Traditional crystals are described by periodic patterns. Much more generally,
crystal structure can be defined by the stochastic process that generates it.
(Traditional crystals, then, are the special case in which the stochastic
stacking process is deterministic and periodic or aperiodic.) For close-packed
structures layered according to a stochastic process that can be expressed by a
hidden Markov model, our results imply that the diffraction spectrum is
intimately related to the HMM's eigenspectrum.

\subsubsection{Random stacking example}

Both hcp and ccp crystals are described by very simple deterministic Markov
models. More generally, crystal structure can integrate both features of
randomness and features of determinism. Moreover, the randomness need not be
simply statistically-independent errors (possibly, \emph{faults}) in an
otherwise periodic parent crystal. Rather, the randomness itself can have a
rich causal architecture.

As a first example, consider the $p$-parametrized family of stochastic stacking
processes depicted in Fig.~\ref{fig:Crystal1Example}. For $p=1$, we recover the
deterministic period-two hcp structure. The period-two nature is reflected in
the Bragg reflection at $\omega=\pi$. For $p=0$, we recover the deterministic
period-three ccp structure. The period-three nature is reflected in the Bragg
reflection at $\omega=2\pi/3$. For other values of $p$, the structure is
described by a stochastic stacking process. 

\begin{figure}[h!]
\begin{center}
\includegraphics[width=0.98\columnwidth]{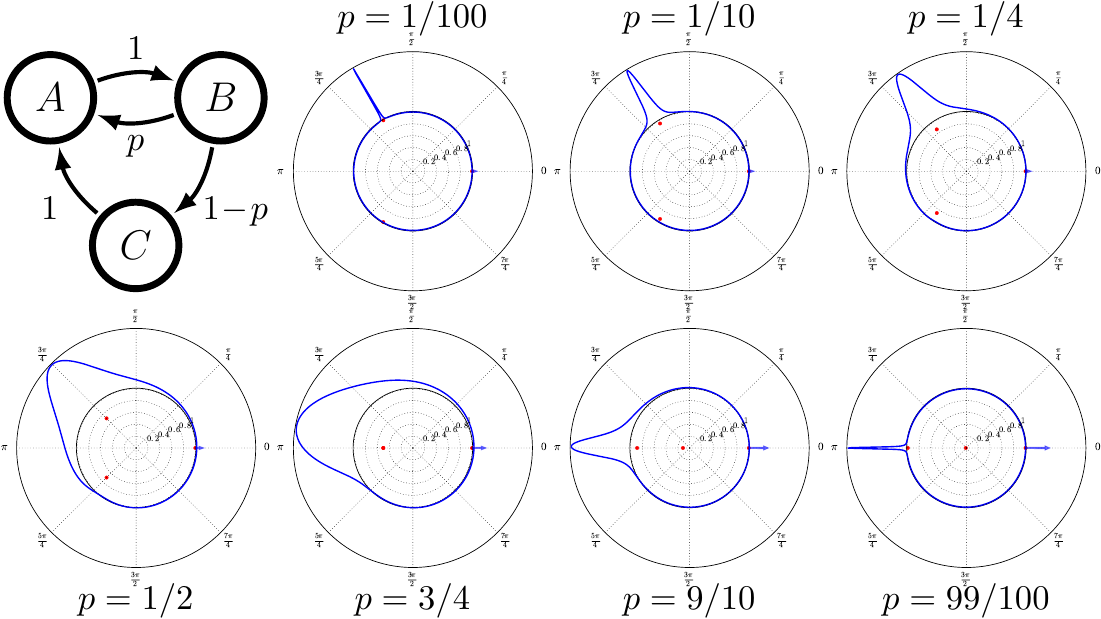}
\end{center}
\caption{Parametrized HMM that generates a family of stochastic stacking
	processes (top left) and the diffracted intensity $P(\omega) / |\psi|^2$ at
	different values of the faulting parameter $p$. Plotted as coronal
	spectrograms, it is clear that the diffraction spectrum emanates from the
	eigenspectrum of the HMM that generates the crystal.
	}
\label{fig:Crystal1Example}
\end{figure}

For any $p$, the transition matrix and average-observation matrix are:
\begin{align*}
T = 
\begin{bmatrix}
0 & 1 & 0 \\
p & 0 & 1-p \\
1 & 0 & 0
\end{bmatrix}
\quad \text{and } \quad
\Omega =  
\begin{bmatrix}
A & 0 & 0 \\
0 & B & 0 \\
0 & 0 & C
\end{bmatrix}
  ~,
\end{align*}
respectively. The transition matrix eigenvalues are $\Lambda_T = \bigl\{ 1, -
\tfrac{1}{2} \pm \sqrt{p - \tfrac{3}{4}} \bigr\}$. The transition matrix is
diagonalizable unless $p=3/4$, where it becomes nondiagonalizable.

For $p \neq 3/4$, each spectral projection operator is given by $T_\lambda =
\ket{\lambda} \bra{\lambda}$, with $\bra{\lambda} = \tfrac{1}{3 \lambda^2 - p}
\begin{bmatrix} \lambda & 1 & \lambda^2 - p \end{bmatrix}$ and $\ket{\lambda} =
\begin{bmatrix} \lambda & \lambda^2 & 1 \end{bmatrix}^\top$, where $\top$
denotes transposition. Recall that the stationary distribution is the left
eigenvector $\bra{\stationary} = \bra{1} = \tfrac{1}{3-p} \begin{bmatrix} 1 & 1
& 1-p\end{bmatrix}$. From these elements, we can calculate $\bra{\stationary}
\overline{\Omega}  \, T_{\lambda} \Omega \ket{\one} $ and the diffracted
intensity analytically as a function of the transition parameter $p$.
Appendix~\ref{sec:DPsFromHMMs} gives the calculation details.

\paragraph*{Bragg reflections without periodicity}
For $p \in (0, 1)$, the transition matrix $T$ only has a single eigenvalue on
the unit circle, so the discrete (Bragg) spectrum has a single contribution
from the eigenvalue of unity:
\begin{align*}
P_\text{d}(\omega) &= \frac{ 2 \pi p^2 |\psi|^2 }{(3-p)^2}  \! \sum_{\ell = -\infty}^{\infty} 
  \!  \delta( \omega \!+\! 2 \pi \ell) ~
 .
\end{align*}
It is interesting that this Bragg reflection persists despite the lack of any
long-range deterministic periodicities for $p \in (0,1)$. This rather reflects
a different type of long-range order: the persistent imbalance of layer types
within each realization of the stochastic stacking process. More generally,
Bragg reflections can be attributed to statistical symmetry breaking.
Deterministic periodicities are but one special case.

\paragraph*{Diffuse spectrum}
There is a diffuse contribution to the power spectrum for all $p \in (0, 1)$.
For $p \in (0, 3/4) \cup (3/4, 1)$, this contribution is:
\begin{align*}
P_\text{c}(\omega) = |\psi|^2 \Bigl(  1 - \tfrac{ p^2 }{(3-p)^2}  \Bigr) +  \!\!
  \sum_{\lambda \in \Lambda_T \setminus \{ 1 \} } \!\!\!\!
  2 \, \text{Re}
  \frac{ \bra{\stationary} \overline{\Omega}  \, T_{\lambda} \Omega \ket{\one}}{e^{i \omega} / \lambda - 1} 
  ~.
\end{align*}
However, the expanded expressions are significantly different for $p > 3/4$,
where all eigenvalues are real valued and distinct, and for $p < 3/4$, where
two of the eigenvalues are complex conjugate pairs.

\paragraph*{Nondiagonalizable diffraction profiles}

At $p=3/4$, the transition matrix of the stochastic stacking process becomes
nondiagonalizable. Curiously, this nondiagonalizability is a generic feature of
parametrized transition matrices at the point where real eigenvalues collide
and interact to gain complementary imaginary components. That is,
nondiagonalizability marks the onset of new behavior. In this case,
nondiagonalizability marks the transition from primarily period-$2$ to
primarily period-$3$ behavior. This critical point of nondiagonalizability is
accompanied by a qualitatively distinct diffraction profile---no longer
exhibiting the typical Lorentzian line profile. Observing such a line profile
experimentally indicates a material at the crossroads of structural
transformation.

\subsubsection{Degenerate diffraction patterns}

Our general results on the degeneracy of power spectra directly bear on the
degeneracy of diffraction patterns from different crystals. The enhanced
understanding of this degeneracy, in turn, sheds new light on the well-known
difficulty of the inverse problem of discovering crystal structure from
diffraction patterns~\cite{Tayl10}.

Consider a chaotic crystal with a stochastic stacking process described by the
simple HMM shown in Fig.~\ref{fig:Crystal2Example}a. The transition matrix
eigenvalues are $\Lambda_T = \bigl\{ 0, \pm 1 \bigr\}$.
Appendix~\ref{sec:ChaoticCrystalEx2App} shows that $\bra{\stationary}
\overline{\Omega}  \, T_{1} \Omega \ket{\one}  = \tfrac{1}{16} |\psi|^2$ and
$\bra{\stationary} \overline{\Omega}  \, T_{-1} \Omega \ket{\one}  =
\tfrac{9}{16} |\psi|^2$.  The resulting diffraction pattern consists of a flat
``white noise'' background:
\begin{align*}
P_\text{c}(\omega) &= \tfrac{ 3 }{8} |\psi|^2 ~
\end{align*}
together with two Bragg reflections per $2 \pi $ of angular frequency bandwidth:
\begin{align*}
P_\text{d}(\omega) &= \frac{ \pi |\psi|^2 }{8}  \! \sum_{\ell = -\infty}^{\infty} 
  \!  \bigl[ \delta( \omega \!+\! 2 \pi \ell) + 9 \delta( \omega \!-\! \pi \!+\! 2 \pi \ell) \bigr]
  ~.
\end{align*}

\begin{figure}[h!]
\begin{center}
\includegraphics[width=0.99\columnwidth]{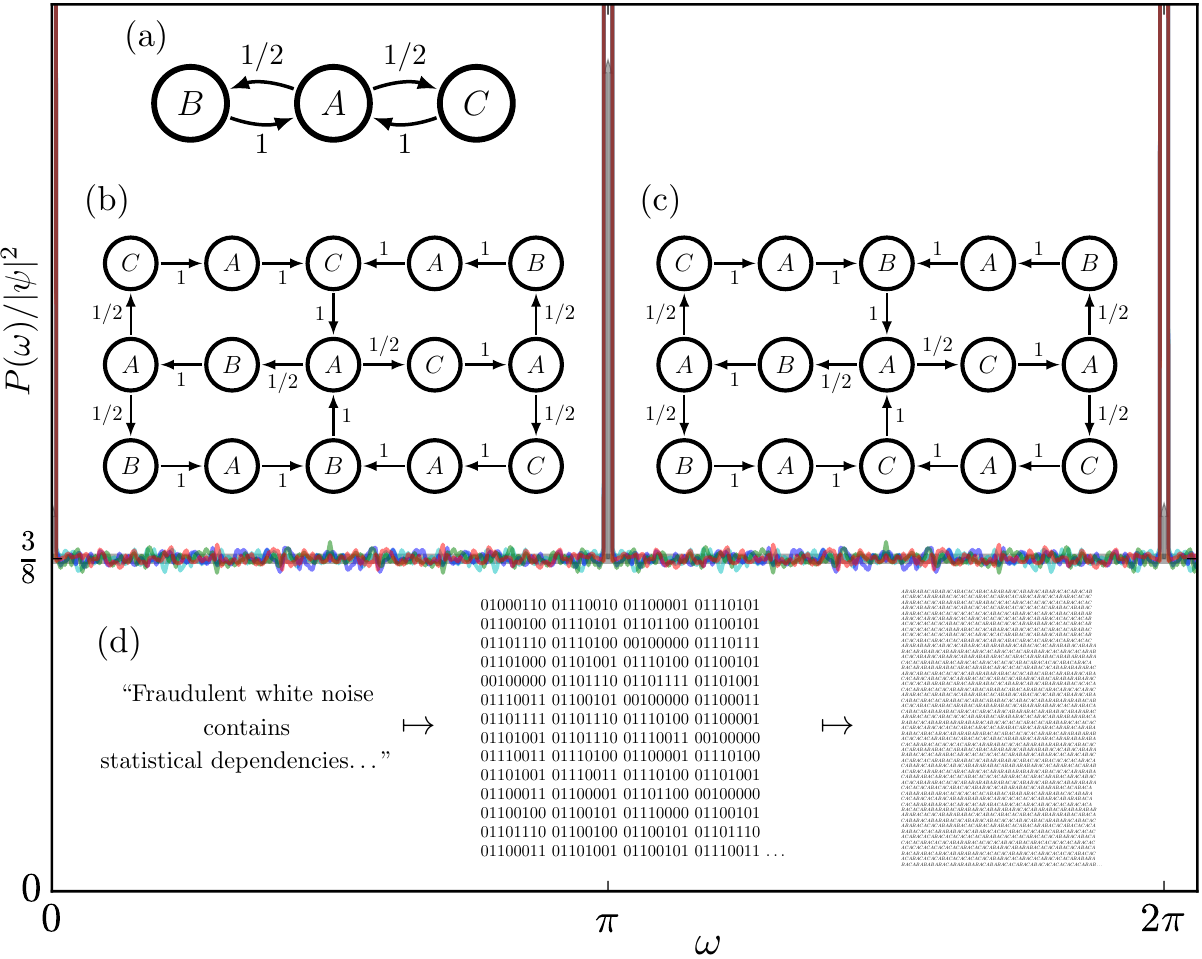}
\end{center}
\caption{Diffraction pattern (overall figure) consisting of a white noise
	background with two Bragg reflections per $2 \pi$ change of angular
	frequency along the stacking direction. This pattern will be observed from
	infinitely-many distinct stochastic processes that generate close-packed
	structures. The flat diffraction pattern is given analytically (thick gray)
	by Eq.~\eqref{eq:CommonDP}. We verify numerically that this diffraction
	pattern is observed from a crystal stacked according to the simple
	stochastic process of panel (a) (thin blue).  The same diffraction pattern
	results from the stochastic processes of panels (b) (thin cyan) and (c)
	(thin green) that have distinct nontrivial higher-order correlations. And,
	the same diffraction pattern results also from a crystal that contains the
	information needed to faithfully reconstruct the entire contents of the
	present manuscript. (d) To demonstrate this, we extracted an extended
	excerpt from the manuscript, converted the text to binary ASCII, and then
	converted each binary character to six layers of the crystal---sampled from
	process-(b) if $0$ and sampled from process-(c) if $1$, starting in the
	central $A$ state each time. The corresponding diffracted intensity is
	shown in thin red, coinciding with the others.
	}
\label{fig:Crystal2Example}
\end{figure}

The exact same diffraction pattern, however, results from an infinite number of
distinct and arbitrarily-complex stochastic stacking processes. In these cases,
the flat diffraction background is a fraudulent white noise, belies the
material's sophisticated correlated structure.

For example, the HMMs shown in Figs.~\ref{fig:Crystal2Example}b and
\ref{fig:Crystal2Example}c each contain nontrivial high-order correlation
between layer types. However, each produces the same diffracted intensity as
before:
\begin{align}
P(\omega) & = \frac{3}{8} |\psi|^2 \nonumber \\
  & \quad
  + \frac{ \pi |\psi|^2 }{8} \!\! \sum_{\ell = -\infty}^{\infty} 
  \!\!  \bigl[ \delta( \omega \!+\! 2 \pi \ell)
  \! + \! 9 \delta( \omega \!-\! \pi \!+\! 2 \pi \ell) \bigr]
   .
\label{eq:CommonDP}  
\end{align}

As another example that helps to drive home the point, a binary encoding of the
entire contents of this manuscript can be stored in the stacking sequence of a
close-packed crystal with exactly the same diffraction pattern as
Eq.~\eqref{eq:CommonDP}. In fact, \emph{any} sufficiently long binary sequence
can be encoded in a crystal with this diffraction pattern.

To construct this crystal, each $0$ is mapped to one of the layer sequences in
$\mathcal{L}_0 = \{ ABABAB, ABACAC, ACABAC, ACACAB \}$ with equal probability,
while each $1$ is mapped to one of the layer sequences in $\mathcal{L}_1 = \{
ABABAC, ABACAB, ACABAB, ACACAC \}$ with equal probability.  This is equivalent
to applying six iterations of the transition dynamic of
Fig.~\ref{fig:Crystal2Example}b for each $0$ and then applying six iterations
of the transition dynamic of Fig.~\ref{fig:Crystal2Example}c for each $1$,
starting in the central $A$ state each time.  \emph{Theorem \ref{thm:SamePS}
guarantees that the diffraction pattern of the resulting crystal is always
given by Eq.~\eqref{eq:CommonDP}}.  This is demonstrated in
Fig.~\ref{fig:Crystal2Example}.  Diffracted intensity is completely blind to
these correlated binary messages, but the original binary message can
nevertheless be recovered by other means (e.g., via scanning tunneling
microscopy (STM)).

A similar story can be told for our human genome---our DNA is the prototypical
``aperiodic crystal''~\cite{Schr44a}. Its diffraction pattern allowed scientists
to uncover its general double-helix
structure~\cite{Wats53a,Wilk53,Fran53,Wats53b}. However, the particular content
encoded by the DNA can only be extracted by more refined structure-detection
methods---carried out by a team of cooperative enzymes in vivo.

To summarize our view of diffraction spectra for chaotic crystals, we showed
that: (i) State-conditioned expectations of layer form factors simplify
diffraction analyses. (ii) Bragg reflections persist in close-packed structures
without periodic order. (iii) Nondiagonalizability heralds structural
transformation and yields qualitatively distinct line profiles. And, (iv) an
infinite number of arbitrarily complex crystal structures all produce the same
flat diffraction pattern (plus two Bragg reflections). These lessons supplement
a growing awareness of the diversity of ``order'' in solid-state physical
systems---order beyond what can be described by Patterson autocorrelation
functions and diffraction patterns~\cite{Cart12, Grim15, Varn15}.
} 

{ \color{blue} 
\subsection{Which ion channels features do power spectra capture?} 

Voltage-gated ion channels embedded in cellular membranes are the engines that
propagate signals among cells---coordinating electrical communication in our
brains, hearts, and throughout our bodies. Better understanding the dynamics
among the macromolecular conformations of these ion channels allows a better
understanding of biological function, malfunction, and possible intervention.
However, the ion channel conformations cannot be observed directly. Rather, it
is only possible to observe a \emph{function} of the hidden conformational
state---whether the instantaneous conformation allows current to flow or not.
This non-Markovian observable makes the inverse problem (of inferring the
dynamic over hidden conformational states) a difficult task~\cite{Blac57a,
Siek12}.

Fortunately, a large body of investigation over many decades elucidated the
biology of ion channels~\cite{Hodg52, Jian03, Kasa16, Daya05}. Nevertheless,
questions remain about how the measured power spectral features, like $1/f$
noise, arise in electrical measurements of ion channels. Does it derive
from the conformational switching dynamics? Is it from current fluctuations in
a particular conformation? If only power spectra are available, what can be
inferred?

Our results offer insight into which features of the power spectrum can be
attributed to the channel's conformational switching dynamics. Most notably,
our Thm.~\ref{thm:PSDequivalence} says that the conditionally-IID distributions
associated with each conformational state cannot possibly change the observed
power spectrum, so long as the average output from each state is left
unchanged. So, for example, state-dependent (conditionally-IID) noise cannot be
the source of $1/f$ noise since it cannot modulate the power spectrum.
Previously, this and related questions could only be explored experimentally
and numerically~\cite[Fig.~3]{Siwy02}.

\begin{figure}[h!]
\begin{center}
\includegraphics[width=0.9\columnwidth]{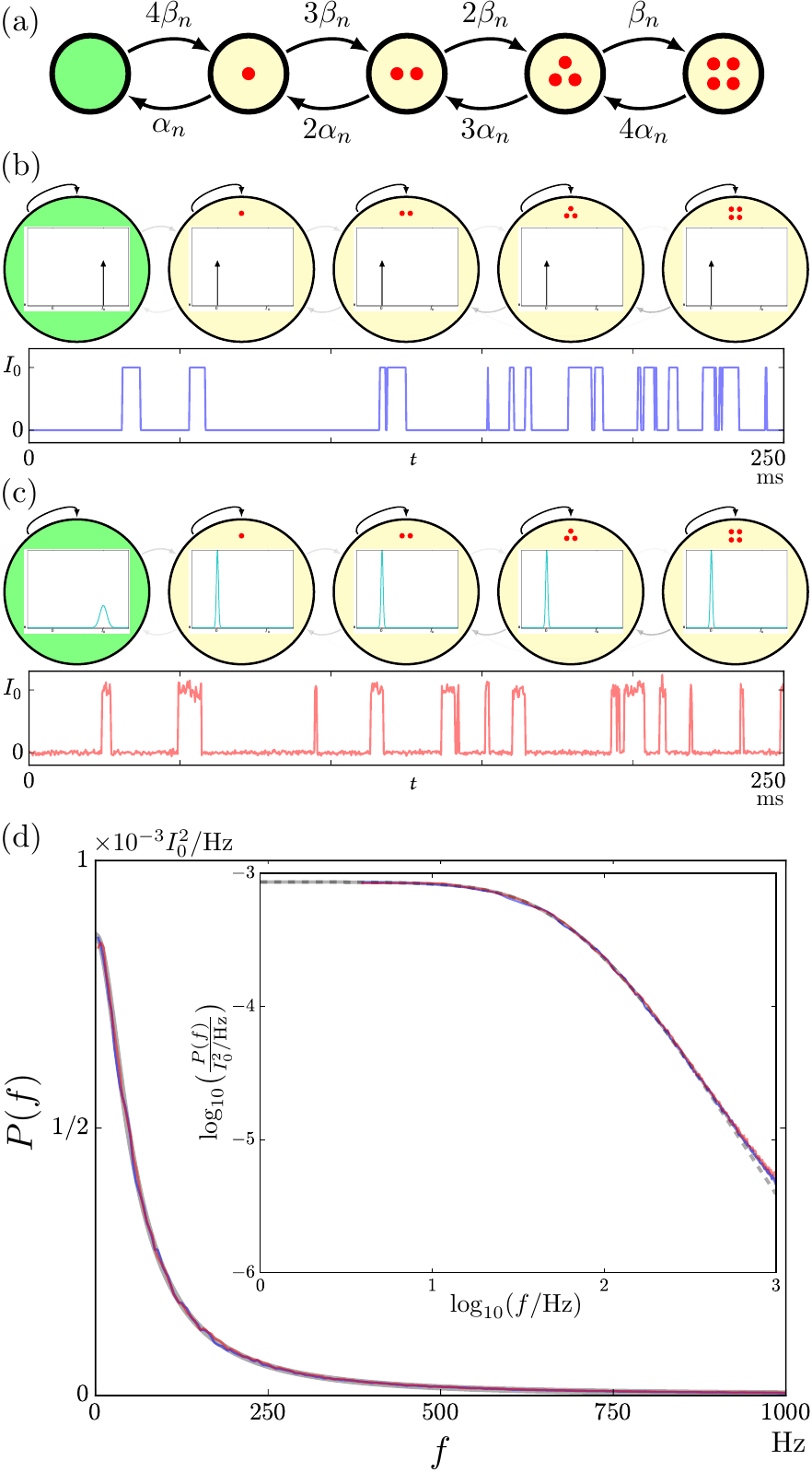}
\end{center}
\caption{Biophysical application of Thm.~\ref{thm:PSDequivalence}: Power
	spectrum of current through a potassium ion channel does \emph{not} depend
	on the details of the probabilistic current in each channel conformation.
	(a) The continuous-time model of transition rates between conformations of
	the channel. Each state has a different number of activation gates that
	block the channel (from zero to four). Panels (b) and (c) show HMMs and
	representative time series of ion current generated from this
	continuous-time model, at a membrane potential of $v=-40$ mV and sampling
	rate of $f_0 = 4$ kHz. (b) Binary output. (c) Continuous-valued output,
	representing both measurement noise and current fluctuations. (d) The power
	spectrum is shown analytically (thick gray) and numerically (thin blue for
	binary model (b); thin red for continuous-valued model (c)). The numerical
	power spectra were each calculated from a simulated time series of length
	$2^{20}$ using the Welch method, performing FFTs on segments of length
	$2^{10}$. The inset log-log plot shows $\sim$constant behavior at low
	frequency, $\sim1/f^2$ behavior at high frequencies, and the effect of
	finite sampling rate at very high frequencies.
	}
\label{fig:PotassiumThm1demo}
\end{figure}

To contribute to these issues concretely, let's consider current fluctuations
in voltage-gated potassium ion channels. Figure~\ref{fig:PotassiumThm1demo}
illustrates an important biophysical application of
Thm.~\ref{thm:PSDequivalence}: The power spectrum of current through a
voltage-gated K$^+$ channel is invariant to mean-preserving changes in the
ion-current PDFs for each channel conformation. We demonstrate this for a
particular physiologically-motivated model of gating kinetics; see
App.~\ref{sec:PotassiumApp} for details. However, it must also hold for any
model of potassium ion current, which may include many hidden open conformations
and electronic states, so long as the output is conditionally-IID in each
hidden state.

Figure \ref{fig:PotassiumThm1demo}(a) shows the continuous-time model of
transition rates between conformations. Each of these conformations has a
different number of activation gates blocking the channel: from zero in the
open state (leftmost, green), through four. Current only flows in the open
state, so the dynamics of K$^+$ current is non-Markovian, as is well known.
The average current in each state is either $I_0$ or $0$, depending on whether
the channel is in an open or closed conformation, respectively. The
Hodgkin--Huxley parameters $\alpha_n$ and $\beta_n$ are voltage-dependent rates
of an individual gate opening or closing. Experiments on ion channels are
typically performed at a fixed membrane voltage~\cite{Hodg52, Fish73, Siwy02}.
With fixed voltage and sampling rate, the continuous-time model generates a
simple discrete-time HMM. (Time-varying voltages produce more complicated HMMs.)

Figures \ref{fig:PotassiumThm1demo}(b) and (c) correspond to a fixed membrane
potential of $v = -40$ mV, with potassium current sampled every $\tau_0 = 250$
microseconds. Panels (b) and (c) each show a HMM and a randomly sampled time
series. For visualization of the HMMs, the opacity of the directed edges is a
simple concave function of the transition probability. (See
App.~\ref{sec:PotassiumApp} for the exact form of the rate matrix and
transition matrix.)

Previous analyses considered the power spectrum from a binary output model 
similar to (b)~\cite{Hill72, Stev72}. Yet with both measurement noise and
current fluctuations, a continuous-valued model like (c) better represents the
stochastic process observed in experiments. Nevertheless, our
Thm.~\ref{thm:PSDequivalence} asserts that both of these models produce exactly
the same power spectrum, up to a frequency-independent offset. Moreover, for
continuous-time processes, this offset vanishes as the sampling rate increases.

We can state this more precisely as a general corollary of
Thm.~\ref{thm:PSDequivalence}.

{\Cor \label{cor:VanishingOffsetInContinuousTime} 
For any two HMMs whose transition matrix comes from the same continuous-time
generator (via $e^{\tau_0 G}$): if the two models have the same average output
in each state, then their power spectra differ only by a frequency-independent
offset: $$P^{(\mathcal{M}')}(\omega) - P^{(\mathcal{M})}(\omega) = \bigl(
\braket{|x|^2}_{\mathcal{M}'} -  \braket{|x|^2}_{\mathcal{M}}  \bigr) / f_0~.$$
For a family of such processes with bounded variance of the instantaneous
observable, this offset must approach zero as the sampling rate increases.
}

{\ProCor 
This follows immediately from Thm.~\ref{thm:PSDequivalence} when we treat $f_0$
explicitly. (Recall that $f_0$ was set to unity in the discrete-time case.)
}

Nevertheless, small constant offsets can be observed between the empirical
power spectra whenever a finite sampling frequency is used.

Figure \ref{fig:PotassiumThm1demo}(d) shows that the power spectra from model
(b) and model (c) are indeed the same, up to a very small constant offset of
$\bigl( \stationary_\text{open} \sigma_\text{open}^2 + (1 -
\stationary_\text{open}) \sigma_\text{closed}^2  \bigr) / f_0 \approx 1.8
\times 10^{-7} I_0^2 / \text{Hz}$.

The power spectrum in the continuous-time limit of $f_0 \to \infty$ is derived
in App.~\ref{sec:PotassiumApp}. The analytic curve (dashed thick gray) is shown
in the log-log inset of Fig.~\ref{fig:PotassiumThm1demo}(d). It is flat at low
frequencies and falls off as $1/f^2$ at high frequencies. However, these
processes are sampled at a finite rate of $f_0 = 4$ kHz. The analytic curve for
the expected \emph{empirical} power spectrum (from model (b)) is shown in thick
gray in the log-log inset. It deviates from the continuous-time model's $1/f^2$
behavior but matches the numerical power spectra extremely well up to
arbitrarily high frequencies. This whitening of empirical power spectra at high
frequencies is predicted by Eq.~\eqref{eq:PcwFromResolvent}.

Our results suggest that observed $1/f$ noise is likely due to non-IID current
fluctuations in the channel's open conformation. This conclusion is at odds
with the conclusion of Ref.~\cite{Siwy02}, but is consistent with
theoretical~\cite{Hill72, Stev72} and experimental~\cite{Fish73} observations
in much earlier work, where the Lorentzian-like power spectrum of the channel's
conductance fluctuations appears to be additive to the $1/f$ flicker noise
background.

Despite $70$ or more years of ongoing investigation and great advances,
potassium ion-channel conduction is still not fully understood \cite{Kasa16,
Hou17}. Fortunately, the analytic results here can help---they can be applied
to evaluate the power spectrum from any proposed model and so aid in bridging
theory to experiment. To make genuine progress, these models will necessarily
be more complicated, including transitions between distinct electronic
conduction states in the channel's open conformation. On the one hand, the
results emphasized that power spectra are indifferent to several stochastic
features of alternative models. Yet, on the other, the relationship between
power spectra and eigenvalues of the rate matrix immediately tells us much
about which models can be ruled out based on nontrivial features of observed
power spectra.
}

\section{Structure in Noise?}
\label{sec:SeekingStructureInNoise}

{ \color{blue} 
Surely leveraging predictions to exclude alternative mechanisms is a central
strategy in physical science, but isn't there a direct way to discover
structure in apparent noise? One approach immediately suggests itself. We first
reflect on, and further develop, the theory of higher-order spectra---which
maintain much of the familiarity and convenience of power spectra. However,
enumerating and interpreting higher-order spectra in general is difficult.  Not
the least reason for this is that the number of possible spectral descriptions
multiplies combinatorially. Or, sometimes the motivating questions are more
pointed. In these cases, it is often more incisive to develop an
information-theoretic probe of statistical interdependencies.

The ultimate goal, though, in using any of these tools is constructing
a testable model that generates the observed features of interest. In the
deterministic case familiar in classical physics, this is synonymous with the
learning the equations of motion. In open complex systems with noise and many
layers of feedback, this may instead take the form of a hidden-state
model---whose input-dependent time-evolution operator generalizes the
deterministic equations of motion. By directly expressing mechanisms,
developing such models allows thoughtful reflection on assumptions,
generalizations, and interventions.
}

{ \color{blue} 
\subsection{Polyspectra}

Higher-order spectra---often simply \emph{polyspectra}---are a natural next
step to detecting structure beyond the pairwise correlations conveyed by power
spectra~\cite{Bril65, Coll98}. As we will show, polyspectra are not the ultimate
answer to structure detection, however they certainly are a tool that
practitioners should be aware of. The following derives new analytical
expressions for polyspectra useful for both experimentalists and theoreticians.
In emphasizing properties already implicit in the foregoing, the analysis
reveals that polyspectra too are blind to predictable structure in processes.

Following Ref.~\cite{Bril65}, we introduce a general formulation for
polyspectra that implicates expectation values---such as, $\braket{g_0(X_{t_0})
g_1(X_{t_1}) g_2(X_{t_2})}$---of time-displaced functions of the observables.
As part of the generalization, let $g_k \colon \Abet \to \mathbb{C}$ be any
function taking observables to complex numbers. If $\Abet$ is an abstract
set---representing, say, observing colors yellow or red $\Abet = \{ \text{y} ,
\text{r} \}$---the $g_k$ functions allow a polyspectral analysis that is not be
possible otherwise.

Consider the \emph{$(g_0, \dots , g_K)$-polyspectrum}:
\begin{align}
S_{g_0 , \dots , g_K}  (\omega_1, \dots , \omega_K) = 
  \lim_{N \to \infty} \frac{1}{N}
    \Biggl\langle 
     \prod_{k=0}^K \widetilde{g_k}^{(N)}(\omega_k) \!
    \Biggr\rangle
	,
\label{eq:Polyspectra_def}
\end{align}
where $\omega_0 \equiv - \sum_{k=1}^K \omega_k $ and:
\begin{align}
\widetilde{g}^{(N)}(\omega) \equiv \sum_{t=1}^N g(X_t) e^{-i \omega t}
  ~.
\label{eq:NFT}
\end{align}

Although challenging to interpret in full generality, in principle polyspectra
provide a window into a process' high-order nonlinear dependencies.
Equation~\eqref{eq:Polyspectra_def} says that polyspectra are the expected
products of Fourier components---components that, in practice, can be obtained
from the FFT. Given the FFT's well-known computational efficiency, polyspectra
are an especially appealing probe of higher-order structure.

Many special cases of the $(g_0, \dots , g_K)$-polyspectrum have been well
studied. For example, $S_{\overline{X},X}(\omega) = P(\omega)$ is the power
spectrum; $S_{\overline{X},Y}(\omega)$ is the cross-spectrum discussed in
App.~\ref{sec:Crossspectra}; $S_{\overline{X},X,X}(\omega_1, \omega_2)$ is the
\emph{moment bispectrum}; $S_{\overline{X},X,X,X}(\omega_1, \omega_2,
\omega_3)$ is the \emph{moment trispectrum};
$S_{\overline{X-\braket{X}},X-\braket{X},X-\braket{X}}(\omega_1, \omega_2)$ is
the \emph{cumulant bispectrum}; and so on. The following, in contrast,
addresses $(g_0, \dots , g_K)$-polyspectra generally. 

Combining Eqs.~\eqref{eq:Polyspectra_def} and \eqref{eq:NFT} yields:
\begin{align}
& S_{g_0 , \dots , g_K}  (\omega_1, \dots , \omega_K) \nonumber \\
  & = 
  \lim_{N \to \infty} \frac{1}{N} 
    \sum_{t_0 = 1}^N  \dots \sum_{t_K = 1}^N
      \Bigl\langle  \prod_{k=0}^{K} g_k(X_{ t_k}) \Bigr\rangle  
      \prod_{k=0}^K e^{-i \omega_k t_k}
    .
\label{eq:Polyspectra_expanded_def}
\end{align}
Thus, the $(g_0, \dots , g_K)$-polyspectrum is closely related to the
expectations $\Bigl\langle \prod_{k=0}^{K} g_k(X_{ t_k}) \Bigr\rangle$, as
suggested. And, crucially, the expectation values can be calculated exactly
from any hidden-state model. Unraveling this exact relationship gives new
insight into what the polyspectrum conveys about a process.

The time variables $(t_k)_{k=0}^K$ in Eq.~\eqref{eq:Polyspectra_expanded_def}
are not necessarily time-ordered by the index $k$. Moreover, time variables may
coincide; i.e., it is possible to have $t_j = t_k$ for $j \neq k$. To remove
these complications, one can work with a reduced and time-ordered collection of
time variables $(t_k')_{k=0}^\kappa$ such that $t_{k}' > t_{k-1}'$, where
$\kappa + 1 = \bigl| \{ t_k \}_{k=0}^K \bigr| \leq K+1$ is the number of
distinct values of the time variables.  These time-ordered variables are
defined recursively via $t_0' = \min \bigl( \{ t_k \}_{k=0}^K \bigr)$ and
$t_\ell' = \min \bigl( \{ t_k \}_{k=0}^K \setminus \{ t_k' \}_{k=0}^{\ell-1}
\bigr)$.

The original time variables $(t_k)_{k=0}^K$ induce a function $\alpha \colon \{
0, 1, \dots K \} \to \{ 0, 1, \dots \kappa \}$ that compresses and time-orders
the indices, such that $t_k = t_{\alpha(k)}'$. Although $\alpha$ does not
generally have a unique inverse, we define $\alpha^{-1}(\ell) = \bigl\{ k \in
\{ 0, 1, \dots K \} : \alpha(k) = \ell \bigr\}$ to be the set of indices that
map to $\ell$.

For HMMs, we can then express the expectations in Eq.~\eqref{eq:Polyspectra_expanded_def} as:
\begin{align}
\Bigl\langle \prod_{k=0}^{K} & g_k(X_{ t_k}) \Bigr\rangle  
  \Bigl\langle \prod_{\ell=0}^{\kappa}
  g_{\alpha^{-1}(\ell)}(X_{ t_\ell'}) \Bigr\rangle  \nonumber \\
  & = \text{tr} \Bigl(  \ket{\one}
\bra{\pi} \Omega_{g_{\alpha^{-1}(0)}} \prod_{\ell=1}^{\kappa}
T^{t_{\ell}' - t_{\ell-1}'} \Omega_{g_{\alpha^{-1}(\ell)}}
\Bigr)
  ~,
\label{eq:OrderedExpectationsFromHMM}
\end{align}
where tr$(\cdot)$ denotes the trace, the product on the right maintains time ordering 
$g_{\alpha^{-1}(\ell)}(x) \equiv \prod_{k \in \alpha^{-1}(\ell)} g_k(x)$,
and
we used the generalized average-observation matrices:
\begin{align}
\Omega_g \equiv 
\sum_{s \in \SSet} \braket{g(X)}_{\pdf( X | s )} \ket{ s } \bra{ s } ~.
\label{eq:ExpectedObsMatrix}
\end{align}

Note that the summations over all time variables in
Eq.~\eqref{eq:Polyspectra_expanded_def} induce all possible functions $\alpha$
that permute and compress the indices. And, within each compressed
time-ordering, all possible values of the indices consistent with that ordering
are summed over. To enumerate all possible compressed time-orderings, it is
useful to explicitly introduce the set $\mathbb{F}_{K}^{(\kappa)}$ of all
surjective functions mapping $\{ 0, 1, \dots K \}$ onto $\{ 0, 1, \dots \kappa
\}$. For HMMs, we can then express the expectations \footnote{Equation
(\ref{eq:Polyspectra_shuffled_def}) is a useful expression for time-ordered
correlators of observables. This contrasts with \emph{out-of-time-order
correlators} (OTOCs) that recently received renewed interest for investigating
temporal correlations. OTOCs are theoretically interesting as they can quantify
information scrambling in thermodynamics, chaotic systems, small quantum
systems, and black holes~\cite{Swin18a}. Experimentally obtaining OTOCs
requires sufficient experimental finesse to effectively reverse time---and this
is simply not an option for most systems. Correlates based on time-ordered
observables therefore remain prominent for most systems.}
in Eq.~\eqref{eq:Polyspectra_expanded_def} as:
\begin{widetext}
\begin{align}
S_{g_0 , \dots , g_K}  (\omega_1, \dots , \omega_K) 
  = \lim_{N \to \infty} \frac{1}{N} 
  \sum_{\kappa = 0}^{K}
  \sum_{\alpha \in \mathbb{F}_{K}^{(\kappa)}} 
    \sum_{t_0' = 1}^{N-\kappa}  
    \sum_{t_1' = t_0' +1}^{N-\kappa + 1}
    \dots \!  \sum_{t_\kappa' = t_{\kappa-1}' + 1}^{N}
      \Bigl\langle  \prod_{\ell=0}^{\kappa} g_{\alpha^{-1}(\ell)}(X_{ t_\ell'}) \Bigr\rangle  
      \prod_{\ell=0}^\kappa e^{-i \omega_{\alpha^{-1}(\ell)} t_\ell'}  
    ~,
\label{eq:Polyspectra_shuffled_def}
\end{align}
where 
$\omega_{\alpha^{-1}(\ell)} \equiv \sum_{k \in \alpha^{-1}(\ell)} \omega_k$.

Leveraging Eq.~\eqref{eq:OrderedExpectationsFromHMM},
App.~\ref{sec:AnalyticPolyspectra} shows that
Eq.~\eqref{eq:Polyspectra_shuffled_def} yields the closed-form expression for
the continuous part of the $(g_0, \dots , g_K)$-polyspectrum:
\begin{align}
S_{g_0 , \dots , g_K}  (\omega_1, \dots , \omega_K) 
  = \sum_{\kappa = 0}^{K}
  \sum_{\alpha \in \mathbb{F}_K^{(\kappa)}} \!\!
     \bra{\stationary} \Omega_{ g_{\alpha^{-1}(0)}}  \!
\Bigl( 
\prod_{\ell = 1}^{\kappa}   T
   \bigl( I /  \mathfrak{z}_{\ell:\kappa}^{(\alpha)}  - T  \bigr)^{-1} 
   \Omega_{g_{\alpha^{-1}(\ell)}}
\Bigr) \!
\ket{\one}           
    ~,
\label{eq:Polyspectra_closed_form}
\end{align}
where $\mathfrak{z}_{\ell:\kappa}^{(\alpha)} \equiv \prod_{k = \ell }^\kappa
\mathfrak{z}_{k}^{(\alpha)} = e^{-i \sum_{k = \ell}^{\kappa}
\omega_{\alpha^{-1}(k)} }$.
\end{widetext}

We see that the $(g_0, \dots , g_K)$-polyspectrum sandwiches up-to $K$
resolvents of the time evolution operator $T$, with each resolvent separated by
average-observation matrices. The resolvents couple the chain of observation
matrices, and the polyspectrum reports their average interaction over arbitrary
displacements. 

Using Eq.~\eqref{eq:ResolventPartialFractExpansion} to express the resolvent
$\bigl( I /  \mathfrak{z}_{\ell:\kappa}^{(\alpha)}  - T  \bigr)^{-1} $ in terms
of $T$'s eigenvalues and spectral projection operators, we again see that the
eigenspectrum of the time evolution operator directly controls the polyspectrum
of the stochastic process. Appendix~\ref{sec:EigenPolySpectra} discusses this
further.

Note, too, that there are contributions to the discrete part of the $(g_0,
\dots , g_K)$-polyspectrum wherever $1  /  \mathfrak{z}_{k:\kappa}^{(\alpha)}
\in \Lambda_T$. This coincides with $\kappa$-wise products of $T$'s eigenvalues
on the unit circle. Moreover, this coincides with eigenvalues of
$\bigotimes_{k=1}^\kappa T$---the tensor product of the transition matrix by
itself $\kappa$ times---that lie on the unit circle. The tensor product
indicates that polyspectra reflect transition matrix properties that unfold
over time.

It is useful to probe several special cases of the $(g_0, \dots , g_K)$-polyspectrum. Consider, first, the $(\overline{X}, X)$-polyspectrum, $S_{\overline{X}, X}(\omega_1)$, which is simply the power spectrum $P(\omega_1)$. In this case, $K=1$. So, we must consider the functions contained in $\mathbb{F}_1^{(0)} = $ \Fonezero and $\mathbb{F}_1^{(1)} = $ \Foneone. For the compressive function $\alpha = $ \FonezeroA,
we obtain $\alpha^{-1}(0) = \{ 0, 1 \}$, yielding:
\begin{align*}
\Omega_{g_{\alpha^{-1}(0) }} & = \Omega_{g_{ \{ 0, 1 \} }} \\
  & = \Omega_{|X|^2} \\
  & = \sum_{s \in \SSet} \braket{|X|^2}_{\pdf( X | s )} \ket{ s } \bra{ s }
  ~.
\end{align*}
The $(\kappa=0)$-contribution to the power spectrum is thus:
\begin{align*}
\bra{\stationary} \Omega_{|X|^2} \ket{\one} = 
\sum_{s \in \SSet} \braket{|X|^2}_{\pdf( X | s )} \braket{ \stationary | s }
= \braket{ | x |^2 }
  ~,
\end{align*}
which is indeed the first term in Eq.~\eqref{eq:PcwFromResolvent}.
The $(\kappa=1)$-contribution to the power spectrum is:
\begin{align*}
\sum_{\alpha \in \mathbb{F}_1^{(1)}} \!
     \bra{\stationary} \Omega_{ g_{\alpha^{-1}(0)}}    T
   \bigl( e^{i \omega_{\alpha^{-1}(1)}} I  - T  \bigr)^{-1} 
   \Omega_{g_{\alpha^{-1}(1)}} \!
\ket{\one}
  ~,
\end{align*}
where it should be recalled that $\omega_0 = -\omega_1$.
Plugging in the identity and swap functions of $\mathbb{F}_1^{(1)}$,
this becomes:
\begin{align*}
2 \text{Re} 
\bra{\stationary} \Omega_{ \overline{X}}    T
   \bigl( e^{i \omega_{1}} I  - T  \bigr)^{-1} 
   \Omega_{X} \!
\ket{\one} 
  ~,
\end{align*}
which is indeed the last term of Eq.~\eqref{eq:PcwFromResolvent}.

Appendix~\ref{sec:PolyExamples} gives a similar analysis of the cumulant bispectrum. Analogous to Cor.~\ref{cor:PSD_from_ZeroMean_pdfs}, we find in Thm.~\ref{pro:Bispectrum_from_ZeroMean_pdfs} of App.~\ref{sec:PolyExamples} that:
\begin{quote}
\emph{The cumulant bispectrum is completely flat for any process generated by a HMM with the same average output $\braket{X}_{\pdf(X|s)} = \braket{x}$ from each
hidden state.}
\end{quote}
This serves as a stark warning against over-reliance on any particular
polyspectrum: Structure and interdependence will be missed and it is
challenging to predict with which polyspectra this will happen.

Can polyspectra overcome the shortcomings of power spectra and avoid the
inherent pitfalls? Only indirectly. For example, the cumulant
bispectrum---often championed as the next-step tool for detecting
nonlinearities in a process~\cite{Rose65, Niki87, Coll98, Petr99}---is
completely flat for the example process from Fig.~\ref{fig:Cor1demo} for all
values of the transition parameter $p \in [0,1]$. That is, the cumulant
bispectrum tells us no more than the power spectrum. Yet the moment bispectrum
should be useful in this case, if one only knew how to interpret it.
Specifically, and more simply, if one is sharp enough to use (in fact, guess)
$g(X) = X^2$, then the change in observable reveals the process' structure
through the single-frequency $S_{X^2, X^2}(\omega_1)$ polyspectrum.

Such guesswork is inescapable and, more to the point, reveals a fundamental
problem: If a process' structure is unknown a priori, there is no guarantee
that the structure will be revealed, even after an infinite number of
higher-order polyspectra have been inspected. Generically, it is not clear
\emph{which} set of polyspectra to use to detect structure. Fortunately,
information theory and model reconstruction both provide more principled
approaches to extracting a process' statistical dependencies~\cite{Crut12a,
Riec18a}.
}

\subsection{Becoming Informed}
\label{sec:BecomingInformed}

A more systematic and direct method for exploring beyond-pairwise correlations
in stationary stochastic processes is through the sequence of \emph{myopic
entropy rates}~\cite{Shan48a,Junc79,Crut83a,Crut01a,Riec18a,Riec18b}:
\begin{align*}
h_L = \H(X_L | X_{1} X_{2} \dots X_{L-1}) ~,
\end{align*}
with $h_1 = \H(X_1)$. For example, the RRXOR process has $h_1 = h_2 = \log |
\Abet | = 1$ bit/symbol---it appears as random as possible when considering
symbols individually or in pairs. Structure is unveiled, though, for $L \geq 3$
when $h_L < 1$. That is, progressively longer Markov-order-$L$ approximations of
the infinite-Markov-order process reveal progressively more of its hidden
structure.

In fact, $h_L$'s convergence reflects how structure is hidden in the stochastic
process \cite{Crut01a}. As $L \to \infty$, $h_L$ approaches the process'
\emph{Shannon entropy rate} $h$---the irreducible randomness per symbol after
all orders of correlation have been taken into account. Notably, the
accumulation of the excess myopic entropy $\sum_{L=1}^\infty ( h_L - h ) =
\EE$---the \emph{excess entropy}---quantifies the total mutual information
between the past and future of a process: $\EE = \I(\dots , X_{-1},
X_0 \, ; \, X_1, X_2, \dots)$. So, while $\I(X_0; X_\tau) = 0$ for
all $\tau > 0$ for the RRXOR process, the past and future are nevertheless
correlated since $\EE > 0$. And, the convergence to predictability can be
viewed in the frequency domain through the excess-entropy spectrum introduced
in Ref.~\cite{Riec18b}. Taken together, this suggests that myopic entropy rates
serve well to identify hidden structure beyond pairwise correlation. They show
how predictability improves as progressively longer historical context is used.

However, correlations are not always restricted to contiguous blocks.
Therefore, there can be pairwise correlations among distant observables while
$h_2=0$. Moreover, the myopic entropy rates as defined above are restricted to
stationary processes. Consequently, despite their utility, myopic entropies
are not ideal for direct indication of $L$-way correlation in the most general
setting.

A more direct indicator of $L$-way correlation is found in the \emph{dependence
function} $D_L$, which quantifies the maximal uniquely-$L$-way correlation that
exists in a process. We say a set $\boldsymbol{\chi}$ of random variables is
\emph{fully correlated} if all constituent random variables inform all of the
others; that is, if:
\begin{align*}
\H(X | \boldsymbol{\chi} \setminus \{ X, X' \} ) & - 
\H(X | \boldsymbol{\chi} \setminus \{ X \} ) 
  \\
  & = \I \bigl( X \, ; \, X' \, | \boldsymbol{\chi}  \setminus \{ X, X' \}
  \bigr) \\
  & > 0
  ~,
\end{align*}
for all $X, X' \in  \boldsymbol{\chi}$. A process is then $L$-way correlated if it has a set of $L$ random variables that are fully correlated. One way to quantify this $L$-way correlation is through the following dependence function:
\begin{align*}
D_L & \equiv 
\sup_{ \bigl\{ \boldsymbol{\chi} \subset  \{ X_t \}_t : \, | \boldsymbol{\chi} | = L \bigr\} } 
\min_{ X, X' \in \boldsymbol{\chi} }
\I \bigl( X \, ; \, X' \, | \boldsymbol{\chi}  \setminus \{ X, X' \}  \bigr)
  ~.
\end{align*}
defined here only for $L \geq 2$. $L$-way dependence is nonzero if and only if
there are novel $L$-way contributions to a process' total correlation. Note
that dependence can be applied to nonstationary processes and processes of
finite duration.

Consider, as a simple example of noncontiguous dependencies, the process
consisting of two interlaced RRXOR processes with unambiguous phase,
which arose from measurement of an entangled quantum system in Sec.~\ref{sec:FWN_fromEntanglement}.
Explicitly:
\begin{align*}
X_{6n}   & = \text{XOR}(X_{6n-4}, X_{6n - 2}) ~\text{and}\\
X_{6n-1} & = \text{XOR}(X_{6n-5}, X_{6n - 3})
  ~, 
\end{align*}
whereas $X_{6n-5}$, $X_{6n-4}$, $X_{6n-3}$, and $X_{6n-2}$ are all generated
from a uniform distribution for all $n \in \{ 1, 2, \dots \}$. Joint
probabilities over contiguous variables are completely uncorrelated and as
random as possible, up until a block-length of five. Let's treat the example
as a stationary process: Calculating probabilities from word frequencies in a
single realization, with the implicit assumption of stationarity, effectively
inducing random phase. Then, we find full randomness in the myopic entropy
rates up to block length five: $h_L = \log | \Abet | = 1$ bit for $1 \leq
L < 5$. Then, finally, a reduction in apparent entropy occurs at $h_5$, after
which $h_{L} < h_{L - 1}$ for $L \geq 5$. Notably, $h_3$ reflects maximal
randomness within its purview. Whereas, the process actually has three-way, but
no lower-order dependencies. This yields $D_1 = D_2 = 0$ and $D_3 > 0$. With
known phase, we would have $D_3 = 1$ bit.

However, when the process is unknown and only a single realization is available
for analysis, probabilities can be inferred only from motifs of random-variable
clusters. For example, estimating $\Pr(X_{t-2}, X_t, X_{t+2})$ \emph{as if} the
process were stationary, leads to finding $0 < \widetilde{D}_3 < 1$, where
$\widetilde{D}_L$ denotes approximating the dependence function assuming
stationarity and testing a limited set of motifs. Usefully, $\widetilde{D}_L$
sets a lower bound on $D_L$. So, nonzero $\widetilde{D}_L$ implies $L$-way
dependence. Curiously, the assumption of stationarity induces $\widetilde{D}_L
> 0$ for all $L \geq 3$; reminiscent of how $h_L - h_{L-1} > 0$ for all $L \geq
3$ for the RRXOR process with ambiguous phase. In each case, these higher-order
correlations correspond to the observer's ability to resolve phase ambiguity.

The dependence function seems to fulfill its desired role of identifying
high-order correlations that cannot be explained by lower-order phenomena.
Taking a step back, though, we might question the whole endeavor. Can a single
model-free signal-analysis method ever reliably detect information processing
and thus complex structure in the world around us? We clearly ousted power
spectra for this task. Nevertheless, our arguments here lend support to an
affirmative answer, but at the cost of more nuanced and computationally
intensive techniques. What is the range of validity of the informational
measures discussed above? Can they be entrusted with finding structure in the
noise?

First, it should be noted that Shannon entropy is only fully justifiable for
alphabets $\Abet$ of countable cardinality. So, apparently continuous
observables must be partitioned into measurable sets to apply the informational
measures like the myopic entropy rates and the dependencies $D_L$. Nevertheless,
quantum physics suggests that even very large and apparently continuous systems
are, in principle, always represented in a countable basis. Practically too,
measurement devices only have a finite precision, so observations are
discretized in practice anyway. Therefore, Shannon entropies (like the myopic
entropy rates and the dependencies) can be applied in principle. 

Second, a likely more-severe challenge arises from limitations built into
information theory itself. Specifically, there are multiway statistical
dependencies that are missed by all joint and conditional entropies and all
mutual and conditional mutual informations \cite{Jame15a,Jame16a}.

Finally, a third and practical challenge arises from limited data:
reliable estimates of probabilities are not always available. Model building
offers the strongest response to this challenge. Generative models inferred
from low-order statistics sometimes encapsulate predictions of rare events
\cite{Youn93a}. And, at least, they give a prediction for high-order
statistics. Testing these predictions against observation allows refining one's
model and discovering new structure.

\section{Conclusion}
\label{sec:Conclusion}

Our investigation began with the modest task of showing how to calculate the
correlation function and power spectrum given a signal's generator. To this
end, we briefly introduced hidden Markov models as signal generators and then
used the linear-operator techniques of Ref.~\cite{Riec18c} to calculate their
autocorrelation and power spectra in closed-form. This led to several lessons.
First, we saw that the power spectrum is a direct fingerprint of the resolvent
of the model's time-evolution operator, analyzed along the unit circle. Second,
spectrally decomposing the not-necessarily-diagonalizable time evolution
operator, we discovered the range of qualitative behaviors that can be
exhibited by autocorrelation functions and power spectra. Third, contributions
from eigenvalues on the unit circle had to be extracted and dealt with
separately. Contributions from eigenvalues on the unit circle correspond to
Dirac $\delta$-functions---the analog of Bragg reflections in diffraction. Whereas,
eigen-contributions from inside the unit circle correspond to diffuse peaks,
which become sharper for eigenvalues closer to the unit circle. Finally, we
found that nondiagonalizable eigenmodes yield qualitatively different line
profiles than their diagonalizable counterparts.

These first results incisively answer the challenges raised by Ruelle--Pollicott
resonance theory about the possible relationship between complex eigenvalues of
time-evolution operators and the correlation and power spectra of
observables~\cite{Poll85,Ruel86,Gasp05}. In short, we provided the exact
relationship between the time-evolution operator and the correlation functions
and power spectra, as well as the possible behavior modes of each. The result
is a deeper theoretical understanding and constructive calculational methods.
These complement early investigations that experimentally delivered meromorphic
power spectra from chaotic dynamical systems~\cite{Holm77,Farm80}.

Accordingly, our findings bear on modern applications of Ruelle--Pollicott
resonance theory. These applications are leading, for example, to better
understanding of sensitivities in climate models~\cite{Chek14} and the dynamics
of open quantum systems via their correspondence to classical chaotic dynamical
systems~\cite{Panc00, Garc04}. Our results provide full analytical
correspondence between observed correlation and the spectral properties of
nonunitary models. Our approach also bears on Koopman operator theory and its
applications, which has received a new wave of attention due to the success of
recent data-driven algorithms~\cite{Budi12}. However, our results also clarify
that resonances discovered via pairwise correlation are generically an
insufficient representation of the spectral features of such nonnormal
dynamics. This emphasizes that the full spectral representation of the
effective nonnormal dynamics~\cite{Riec18c}, generically inaccessible via
pairwise correlation, is worth pursuing. Success in this will immediately yield
predictions about many complex systems of interest.

The most surprising and more immediate finding, though, is that temporal
structure can fully evade detection by power spectra. Arbitrarily sophisticated
processes can have exactly flat power spectra and so masquerade as white noise.
Accordingly, we called such processes \emph{fraudulent white noise} processes.
Theorem~\ref{thm:PSDequivalence} and Cor.~\ref{cor:PSD_from_ZeroMean_pdfs}
characterized the many ways that structure can be hidden from power spectra.
And, ultimately, Thm.~\ref{thm:GenFlatPS} addressed the more general condition
for fraudulent white noise, in which the generated time-series could be
input-dependent and nonstationary.

{ \color{blue}
We showed that fraudulent white noise and the degeneracy of power spectra have
important physical implications. We found that fraudulent white noise arises
from sequential measurements of entangled quantum systems. Moreover, the
generation of high-order structure and the complete absence of pairwise
structure occurred despite the fact that these quantum states resulted from a
simple sequence of pairwise interactions. Beyond quantum physics, our results
on the degeneracy of power spectra have consequences throughout the sciences.
We derived new results on the degeneracy of diffraction patterns and showed
how the entire contents of the present work can be encoded in a crystal with a
flat diffraction pattern. We then leveraged our results to comment on a
longstanding debate about $1/f$ noise in biomolecular ion channels.
}

We started out noting that, on the one hand, divergent correlation length often
heralds the emergence of new types of order. And, on the other, that pairwise
correlation is generically identified as the structure in random systems.
However, we showed that there is often rich structure even in the absence of
pairwise correlations. What types of order are we \emph{failing} to predict due
to an historical emphasis on pairwise correlations? Complex systems surely
exhibit emergent structure beyond the reach of pairwise statistics. There is
almost surely more functionally-relevant brain activity available in EEGs
beyond what is reported in their power spectra. Perhaps, however, we should
consider beyond-pairwise structure for even simple generators of structure. For
example, cosmological models could be more thoroughly tested against structure
in the CMB beyond what is contained in the two-point angular correlation
functions.

Having diagnosed the structures inaccessible via power spectra, we discussed how to detect beyond-pairwise structure. {\color{blue} We obtained a closed-form expression for all polyspectra, but showed that higher-order spectra are also completely flat in some cases where structure should have been apparent.}
In response, we introduced the dependence function to detect any $L$-way
correlations for any $L$. We also stressed the importance of model building
whenever possible. In particular, it can help anticipate and perhaps avoid
not-yet encountered catastrophes, which are often a byproduct of the high
interconnectivity of complex socio-economic systems \cite{Crut09b}. Model
building, beyond pure signal analysis, is key in this---it allows us to
discover new mechanisms in nature.

This all said, nature still keeps us in the dark. We showed that the
correlations in a message can be shifted to arbitrarily-high orders of
correlation. The result is that, for finite length messages, statistical
inference can be made effectively impossible regardless of one's
sophistication. Nature herself employs this technique whenever we observe an
increase in entropy---giving the impression of randomness generated, when it is
only ever structure hidden in inaccessibly-obscure high-order correlations.
Waking up to the true hues of reality---prying open the black box, dispelling
apparent white noise---continues to require new theory and new experimentation.

\epigraph{... it is clearly wise to learn what a procedure really seems to be
telling us about.}
{John Tukey, \emph{The Future of Data Analysis}, 1962 \cite[p. 60]{Tuke62a}}

\section*{Acknowledgments}

The authors thank Alec Boyd and Dowman Varn for insightful discussions. JPC thanks the Santa Fe Institute and the California Institute of Technology and the authors together thank the Telluride Science Research Center for their hospitality during visits. This material is based upon work supported by, or in part by, the U. S. Army Research Laboratory and the U. S. Army Research Office under contracts W911NF-13-1-0390 and W911NF-18-1-0028.

\appendix

\section{Diffraction patterns as power spectra}
\label{sec:DPsAsPSs}

Diffraction patterns are used extensively to infer material structure from the
scattering of, for example, an incident x-ray beam
\cite{Guin63a,Warr69a,Welb85a,Ashc76a,Welb14a}. Generally, consider $\vec{r}
\in \mathbb{R}^\text{d}$ to be a vector in d-dimensional real space. The spatial
arrangement of elastic scatterers is given by the scatterers' density
$f(\vec{r})$. Ideally, we wish to recover $f(\vec{r})$ from our diffraction
experiments, which provide measured intensities. However, far-field patterns of
diffracted intensity yield only $I_\text{diff}(\vec{q}) = c |
F(\vec{q}) |^2$, where $F(\vec{q}) = \int_{{\mathbb{R}}^\text{d}} f(\vec{r})
e^{-i \vec{q} \cdot \vec{r}} \, d^\text{d} \vec{r}$ is the d-dimensional
Fourier transform of $f(\vec{r})$, $c$ is some constant, and $\vec{q} = 2 \pi (\vec{k}_\text{out} - \vec{k}_\text{in})$ 
is the scattering vector that quantifies the change in the incident wave vector. In other words,
$F(\vec{q})$'s phase information is lost when only intensity is measured.
This is known as the `phase problem'~\cite{Tayl10}.
The x-ray beam's expected diffracted intensity is proportional to $\Braket{
\left| F(\vec{q}) \right|^2 }$, which is the d-dimensional
generalization of a power spectrum. However, it is also interesting to relate
the d-dimensional diffraction pattern, along a curve in reciprocal space, to
the more familiar one-dimensional power spectrum.

For a given scattering vector $\vec{q}$, decompose $\vec{r} =
\vec{r}_{\|} + \vec{r}_{\bot}$, where $\vec{r}_{\|} \equiv (\vec{r} \cdot
\widehat{q}) \widehat{q}$ and $\widehat{q} =
\vec{q} / | \vec{q} |$. Then, let $\mu_{\bot}( \vec{r}_{\|} )$
be the accumulated density within the entire cross-sectional plane
perpendicular to and uniquely identified by $\vec{r}_{\|} $; i.e., $\mu_{\bot}(
\vec{r}_{\|} ) \equiv \int_{{\mathbb{R}}^{\text{d}-1}} f( \vec{r}_{\|} +
\vec{r}_{\bot} ) \, d^{\text{d}-1} \vec{r}_{\bot}$. We then find that in general: 
\begin{align}
 I_\text{diff}(\vec{q}) = c \left|  
 \int_{{\mathbb{R}}} \mu_{\bot}( \vec{r}_{\|} )
 e^{-i q r_{\|} } \, d  r_{\|}
  \right|^2 ~.
\end{align}
In particular, we see that the diffraction pattern along any line $\vec{q}
= q \widehat{q}$ (with varying $q$ but fixed $\widehat{q}$)
is the power spectrum of the net magnitude of scatterers within sequential cross
sections of real space perpendicular to $\widehat{q}$.

For molecular or crystalline structures, the net scatterer density may often be
well-approximated by a superposition of more elementary densities $f(\vec{r}) =
\sum_{j} f_j(\vec{r} - \vec{r}_j)$.  
If we partition the real space occupied by the material
into $N$ layers of thickness $\tau_0$, stacked along a particular direction $\hat{\ell}$, 
then we obtain the alternative expression:
\begin{align}
I_\text{diff}(\vec{q}) = c \left|  
  \sum_{n = 1}^{N} F^{(n)}(\vec{q}) e^{-i \omega  n }
  \right|^2
  ~,
\label{eq:DPasPS_v2}
\end{align}
where $\omega = \tau_0 \vec{q} \cdot \hat{\ell}$ is ($2 \pi $ times) 
the change in wavenumber per layer in the stacking direction.
In such cases, the \emph{layer form factors} are:
\begin{align*}
F^{(n)}(\vec{q}) \equiv 
\sum_{j \in n^\text{th} \text{ layer}} F_j(\vec{q}) e^{-i ( \vec{q} \cdot \vec{r}_j - n \omega )}
  ~,
\end{align*}
where the ``$n^\text{th}$ layer'' is the set of indices 
$\{ j : n \tau_0 \leq \braket{\vec{r}_j} \cdot \hat{\ell} < (n+1) \tau_0 \}$ 
for the elementary constituents typically contained in the layer.
And the
\emph{atomic form factor}
\begin{align*}
F_j(\vec{q}) = \int_{{\mathbb{R}}^\text{d}} f_j(\vec{r}) e^{-i \vec{q} \cdot
\vec{r}} \, d^\text{d} \vec{r} 
\end{align*}
is the d-dimensional Fourier transform of $f_j(\vec{r})$. 
As a result, 
we see that the expected diffraction pattern can always be written as the power spectrum
of layer form factors:
\begin{align}
\braket{I_\text{diff}(\vec{q})} = c N P(\omega) 
= c  \Braket{ \left|  
  \sum_{n = 1}^{N} X_n e^{-i \omega  n }
  \right|^2 } ~,
\end{align}
with $X_n = F^{(n)}(\vec{q}) \in \mathbb{C}$
as the layer form factor of the $n^\text{th}$ layer of the material.

The frequency-dependence of $F^{(n)}(\vec{q})$ is often factored out to
`correct' the diffraction pattern, so that only the structure of
interest---features due to the stacking
sequence---remains~\cite{Prin04,Wool97}.

{\color{blue}
\subsection{From fraudulent white noise to Debye--Waller theory}
\label{sec:FWN2DWT}

It is important to recognize that the elementary positions $\{ \vec{r}_j \}_j$
are random variables, since thermal motion---and even quantum uncertainty at
zero temperature---can significantly displace them from their average value.
Indeed, the observed diffraction pattern is not consistent with evaluating $\{
\vec{r}_j \}_j$ at their average values. This is because the expected value of
a structure factor is not the same as the structure factor evaluated at the
expected value of elementary positions. Nevertheless, the observed diffraction
pattern \emph{is} consistent with $\braket{|F(\vec{q})|^2}$, where the
averaging over realizations induces the proper thermal (and
quantum-uncertainty) averaging. However, the thermal averaging appears unwieldy
in the general case. Fortunately, we can leverage our Theorems
\ref{thm:PSDequivalence} and \ref{thm:SamePS} to rigorously recover the
simplifications of Debye--Waller theory in our setting of randomly stacked
structures. 

Suppose there is a hidden-state model $\mathcal{M}(\vec{m}) = \bigl( \SSet,
\Abet, \mathcal{P}, \{ T_t(\vec{m}) \}_t , \boldsymbol{\mu}_1 \bigr)$ that
generates the correct statistics of the layer form factors in the
material---taking the stochastic stacking process, thermal motion, and quantum
uncertainty into account. Theorems \ref{thm:PSDequivalence} and
\ref{thm:SamePS} imply that the diffraction pattern will be the same (up to a
constant offset) if we instead consider the much simpler hidden-state model
$\mathcal{M}'(\vec{m}) = \bigl( \SSet, \mathcal{B}, \mathcal{Q}, \{
T_t(\vec{m}) \}_t , \boldsymbol{\mu}_1 \bigr)$ that outputs only the
\emph{expected} layer form factor from each latent state.

Each of the \emph{expected} layer form factors $b \in \mathcal{B}$ can be
expressed as:
\begin{align*}
b = 
\braket{X}_{\pdf(X|s \in \SSet_b)} &= \!\!
\sum_{j \in \text{type-$b$ layer}} \!\!\!\!\! F_j(\vec{q}) \braket{ e^{-i  \vec{q} \cdot \vec{r}_j  } }  \\
&= \!\!
\sum_{j \in \text{type-$b$ layer}} \!\!\!\!\! F_j(\vec{q}) e^{-i  \vec{q} \cdot \braket{\vec{r}_j}  } 
D_j(\vec{q})
  .
\end{align*}
Notably, 
\begin{align*}
D_j(\vec{q}) & \equiv
\bigl\langle  e^{-i \vec{q} \cdot \left( \vec{r}_j - \braket{\vec{r}_j} \right) } \bigr\rangle  \\
& \approx e^{-\tfrac{1}{6} \sigma_{\vec{r}_j}^2 q^2} 
\end{align*}
is exactly the Debye--Waller factor for an elementary scattering site of type
$j$~\cite{Trea91}. The variance $\sigma_{\vec{r}_j}^2$ scales as $\kB T$ at
high temperatures (via the equipartition theorem), although it is still
nonzero as $T \to 0$ due to zero-point energy. 

In the case that the Debye--Waller factors from all scattering sites are the same (i.e., $D_j(\vec{q}) = D(\vec{q})$), the thermal averaging over positions
does \emph{not} broaden the diffraction pattern at all. Rather, the
Debye--Waller factor only suppresses the diffracted intensity at large
scattering magnitudes by an approximately Gaussian envelope (centered at
$\vec{q}=0$).

In contrast, \emph{thermal broadening}---expected of spectral lines throughout
the domains of physics---is due to a Doppler effect from the \emph{velocity} of
the elementary scatterers (rather than their random positions). This induces a
Gaussian convolution on the otherwise Lorentzian line profile. Whereas the
Debye--Waller factor is important, thermal broadening is not a significant
source of line broadening for X-ray diffraction~\cite{Balz92}.

\subsection{Close-packed structures}
\label{sec:ClosePackedFormFactors}

Recall that each layer of a close-packed structure is a two-dimensional
hexagonal close-packed lattice. The diffracted intensity will thus only be
nonzero at scattering vectors that satisfy the Laue condition for allowed
reflections from the two-dimensional crystal:
\begin{align}
\vec{q} - (\vec{q} \cdot \hat{\ell}) \hat{\ell} = \vec{G}
~,
\label{eq:2dLaue}
\end{align}
where $\vec{G}$ is in the set of reciprocal lattice vectors of the 2-D hexagonal lattice.

For close-packed structures, there are only three types of layers, differing
only via relative displacements of 1/3 of a lattice translation vector
$\vec{t}$ in the plane of the layer~\cite{Hend42}. As a result, if type-$A$
layers have an expected layer form factor of:
\begin{align*}
A 
= \!\!\!\!
\sum_{j \in \text{type-$A$ layer}} \!\!\!\!  D_j(\vec{q}) F_j(\vec{q}) e^{-i  \vec{q} \cdot \braket{\vec{r}_j}  } 
~,
\end{align*}
then type-$B$ layers will have an expected form factor of:
\begin{align*}
B
&=  \!\!\!\!
\sum_{j \in \text{type-$A$ layer}} \!\!\!\! D_j(\vec{q}) F_j(\vec{q}) e^{-i  \vec{q} \cdot ( \braket{\vec{r}_j} - \vec{t}/3)  } 
= 
e^{i  \vec{q} \cdot \vec{t}/3 } A
~,
\end{align*}
and type-$C$ layers will have an expected form factor of:
\begin{align*}
C
&= \!\!\!\!
\sum_{j \in \text{type-$A$ layer}} \!\!\!\! D_j(\vec{q}) F_j(\vec{q}) e^{-i  \vec{q} \cdot ( \braket{\vec{r}_j} + \vec{t}/3)  } 
= 
e^{-i  \vec{q} \cdot \vec{t}/3 } A
~.
\end{align*}
However, due to the periodic crystallinity in two dimensions, $A$ is only
nonzero when Eq.~\eqref{eq:2dLaue} is satisfied. By definition of the
reciprocal lattice, $\vec{G} \cdot \vec{t} = 2 \pi m $ with $m \in \mathbb{Z}$.
Hence, for all values of the scattering vector $\vec{q}$ where the expected
layer form factors are nonzero, the expected layer form factors are related by:
\begin{align*}
B = e^{i 2 \pi / 3} A  
\quad \text{and} \quad 
C = e^{-i 2 \pi / 3} A  ~.
\end{align*}
}

\section{Autocorrelation for processes generated by autonomous HMMs}
\label{sec:ACFforHMMs}

Let's derive the autocorrelation function in general and in closed form for the
class of autonomous HMMs introduced in the main body. Helpfully, for particular
models, the expressions become analytic in terms of the model parameters.

Directly calculating, we find that the autocorrelation function, for $\tau >
0$, for any such HMM is:
\begin{widetext}
\begin{align*}
\gamma(\tau) 
&= \Braket{ \, \overline{X_t} X_{t + \tau}} \nonumber \\
&= \int_{x \in \Abet} \int_{x' \in \Abet} \overline{x} x' \pdf(X_0 = x, X_\tau = x') \, dx \, dx' \\
&= \sum_{s \in \SSet} \sum_{s' \in \SSet} \int_{x \in \Abet} \int_{x' \in
\Abet} \overline{x} x' \pdf(X_0 = x, X_\tau = x', \St_0 = s, \St_\tau = s') \, dx \, dx' \\
&= \sum_{s \in \SSet} \sum_{s' \in \SSet} \int_{x \in \Abet} \int_{x' \in \Abet} \overline{x} x' 
\Pr(\St_0 = s, \St_\tau = s') \, \pdf(X_0 = x | \St_0 = s ) \, \pdf( X_\tau = x' | \St_\tau = s') \, dx \, dx' \\
&= \sum_{s \in \SSet} \sum_{s' \in \SSet}  
\braket{\stationary | s } \bra{ s } T^{\tau} \ket{ s' } \braket{ s' | \one} \, 
\Bigl( \int_{x \in \Abet} \overline{x} \, \pdf( x | s ) \, dx \Bigr)
\, \Bigl( \int_{x' \in \Abet} x' \, \pdf( x' | s' ) \, dx' \Bigr) \\
&= 
\bra{\stationary} \Bigl( \sum_{s \in \SSet} \braket{\overline{X}}_{\pdf( X | s )} \ket{ s } \bra{ s } \Bigr) T^{\tau} 
\Bigl( \sum_{s' \in \SSet} \braket{X}_{\pdf( X | s' )} \ket{ s' } \bra{ s' } \Bigr) \ket{\one} 
~,
\end{align*}
where the integrals are written in a form meant to be easily accessible but
should generally be interpreted as Lebesgue integrals. In the above derivation,
note that:
\begin{align*}
\pdf (X_0 = x, X_\tau = x', \St_0 = s, \St_\tau = s')
  = \Pr(\St_0 = s, \St_\tau = s')
  \pdf(X_0 = x, X_\tau = x' | \St_0 = s, \St_\tau = s')
\end{align*}
holds by definition of conditional probability. The decomposition of:
\begin{align*}
\pdf(X_0 = x, X_\tau = x' | \St_0 = s, \St_\tau = s')
  = \pdf(X_0 = x | \St_0 = s ) \pdf( X_\tau = x' | \St_\tau = s') 
\end{align*}
\end{widetext}
for $\tau \neq 0$ follows from the conditional independence in the relevant
Bayesian network shown in Fig.~\ref{fig:HMM_BayesNet}. Moreover, the equality:
\begin{align*}
\Pr(\St_0 = s, \St_\tau = s')
  = \braket{\stationary | s }
  \bra{ s } T^{\tau} \ket{ s' } \braket{ s' | \one}
\end{align*}
can be derived by marginalizing over all possible intervening state sequences.
We can use the hidden-state basis, where $\ket{ s }$ is the column vector of
all $0$s except for a $1$ at the index corresponding to state $s$, while $\bra{
s }$ is simply its transpose. This yields a natural decomposition of the
identity operator: $I = \sum_{s \in \SSet} \ket{ s } \bra{ s }$.

Since the autocorrelation is a Hermitian function---i.e., $\gamma(-\tau) =
\overline{\gamma}(\tau)$---and:
\begin{align*}
\gamma(0) & = \bigl\langle \left| X \right|^2 \bigr\rangle_{\stationary(X)} \\
          & = \bra{\stationary}
		  \sum_{s \in \SSet} \bigl\langle \left| X \right|^2
		  \bigr\rangle_{\pdf(X|s)} \ket{ s }
  ~,
\end{align*}
we find the full autocorrelation function is given by:
\begin{align}
\gamma(\tau) & = 
\begin{cases}
\bra{\stationary} \Omega \, T^{|\tau| } \, \overline{\Omega} \ket{\one} & \text{if } \tau \leq 1 \\
\bigl\langle \left| x \right|^2 \bigr\rangle & \text{if } \tau = 0 \\
\bra{\stationary} \overline{\Omega} \, T^{|\tau| } \, \Omega \ket{\one} & \text{if } \tau \geq 1
\end{cases}
~,
\label{eq:ExplicitCorrFnForm_inApp}
\end{align}
where 
$\Omega$ is the $|\SSet|$-by-$|\SSet|$ matrix defined by:
\begin{align*}
\Omega 
= \sum_{s \in \SSet} \braket{X}_{\pdf( X | s )} \ket{ s } \bra{ s } ~.
\end{align*}
The $\Omega$ matrix simply places state-conditioned average outputs along its diagonal.

To better understand the range of possible behaviors of autocorrelation, we can
go a step further. In particular, we employ the general spectral decomposition
of $T^\tau$ derived in Ref.~\cite{Riec18c} for nonnormal and potentially
nondiagonalizable operators:
\begin{align}
T^\tau &= 
  \Bigl[ \sum_{m=0}^{\nu_0 - 1} \delta_{\tau, m} T_{0,m} \Bigr]
  \! + \!\!\!\! \sum_{\lambda \in \Lambda_T \setminus \{ 0 \} } \sum_{m=0}^{\nu_\lambda - 1} 
    \binom{\tau}{m}
    \lambda^{\tau - m}
    T_{\lambda, m}
  ,
\label{eq:SpectralDecompOfPowersOfT}
\end{align}
where $\binom{\tau}{m}$ is the generalized binomial coefficient:
\begin{align*}
\binom{\tau}{m} &= \frac{1}{m!} \prod_{n=1}^m (\tau - n + 1)
  ~,
\end{align*}
with $\binom{\tau}{0} = 1$.
As briefly summarized in Sec.~\ref{sec:ApparentStructure}, $\Lambda_T$ is the
set of $T$'s eigenvalues while $T_\lambda$ is the spectral projection operator
associated with the eigenvalue $\lambda$. Recall that $\nu_\lambda$ is the
\emph{index} of the eigenvalue $\lambda$, i.e., the size of the largest Jordan
block associated with $\lambda$, and $T_{\lambda, m} = T_\lambda (T- \lambda I
)^m$. Substituting Eq.~\eqref{eq:SpectralDecompOfPowersOfT} into
Eq.~\eqref{eq:ExplicitCorrFnForm_inApp} yields:
\begin{align*}
\gamma(\tau) &= 
\Bigl[ \sum_{m=1}^{\nu_0 - 1} \delta_{\tau, m}  \bra{\stationary}
\overline{\Omega} \, T_{0,m}  \, \Omega \ket{\one}  \Bigr] \\
  & \qquad + \sum_{\lambda \in \Lambda_T \setminus \{ 0 \} } \sum_{m=0}^{\nu_\lambda - 1} 
    \binom{\tau}{m}
    \lambda^{\tau - m}
     \bra{\stationary} \overline{\Omega} \, T_{\lambda,m}  \, \Omega \ket{\one} 
  ~,
\end{align*}
for $\tau > 0$.

It is significant that the zero eigenvalue contributes a qualitatively distinct
ephemeral behavior to the autocorrelation while $| \tau | < \nu_0$. All other
eigenmodes contribute products of polynomials times decaying exponentials in
$\tau$. When $T$ is diagonalizable, the autocorrelation is simply a sum of
decaying exponentials.


\section{Analytical power spectra}
\label{sec:PSDderivation}

The following derives both the continuous and discrete part of the power
spectrum for HMM-generated processes. The development parallels that in
Ref.~\cite{Riec14b}, although that derivation was restricted to the special
case of diffraction patterns from Mealy (i.e., edge-emitting) HMMs with
countable alphabets. In contrast, the following derives analytical expressions
for the power spectrum of any stochastic process generated by an HMM. Notably,
it also allows uncountably infinite alphabets. Also, it is developed for Moore
(i.e., state-emitting) HMMs---although Mealy and Moore HMMs are
class-equivalent and can be easily transformed from one to the other.

\subsection{Diffuse Spectra}

Recall Eq.~\eqref{eq:PSDfromACF}:
\begin{align*}
P(\omega) = \lim_{N \to \infty} \tfrac{1}{N} \sum_{\tau=-N}^{N} \bigl( N - \left| \tau \right| \bigr) \gamma(\tau) e^{-i \omega \tau}
  ~,
\end{align*}
and Eq.~\eqref{eq:ExplicitCorrFnForm}'s explicit expression for the correlation function:
\begin{align*}
\gamma(\tau) & = 
\begin{cases}
\bra{\stationary} \Omega \, T^{|\tau| } \, \overline{\Omega} \ket{\one} & \text{if } \tau \leq 1 \\
\bigl\langle \left| x \right|^2 \bigr\rangle & \text{if } \tau = 0 \\
\bra{\stationary} \overline{\Omega} \, T^{|\tau| } \, \Omega \ket{\one} & \text{if } \tau \geq 1
\end{cases}
~.
\end{align*}
From these we can rewrite the power spectrum directly in terms of the
generating HMM's transition matrix:
\begin{align}
P(\omega) 
&= \bigl\langle \left| x \right|^2 \bigr\rangle + \nonumber \\
  & \qquad \lim_{N \to \infty} \frac{2}{N}  \, \text{Re} 
  \sum_{\tau=1}^{N} \bigl( N -  \tau \bigr) \bra{\stationary} \overline{\Omega}
  \, T^{\tau } \, \Omega \ket{\one}  e^{-i \omega \tau} \nonumber \\
&= \bigl\langle \left| x \right|^2 \bigr\rangle +  \nonumber \\
  & \qquad \lim_{N \to \infty} \frac{2}{N}  \, \text{Re} 
   \bra{\stationary} \overline{\Omega} \, \Bigl(
  \sum_{\tau=1}^{N} \bigl( N -  \tau \bigr) T^{\tau } e^{-i \omega \tau} 
  \Bigr)  \, \Omega \ket{\one} 
    ~.
\label{eq:SwallowedSummation}
\end{align}
We used the fact that $z + \overline{z} = 2 \text{Re} (z)$ for any $z \in
\mathbb{C}$. For convenience, we introduce the variable $\mathfrak{z} \equiv
e^{-i \omega}$. We then note that the summation splits:
\begin{align*}
\sum_{\tau=1}^{N} \bigl( N -  \tau \bigr) T^{\tau } e^{-i \omega \tau} 
&= 
N \sum_{\tau=1}^N (\mathfrak{z} T)^\tau - \sum_{\tau=1}^{N} \tau (\mathfrak{z} T)^\tau
    ~.
\end{align*}
For positive integer $N$, it is always true that:
\begin{align*}
(I - \mathfrak{z} T) \sum_{\tau=1}^{N}  ( \mathfrak{z} T)^{\tau }  
&= 
\mathfrak{z} T - \mathfrak{z}^{N+1} T^{N+1}
    ~,
\end{align*}
and:
\begin{align*}
(I - \mathfrak{z} T) \sum_{\tau=1}^{N}  \tau ( \mathfrak{z} T)^{\tau }  
&= 
- N \mathfrak{z}^{N+1} T^{N+1} + \sum_{\tau=1}^{N}  ( \mathfrak{z} T)^{\tau }
    ~.
\end{align*}
Hence, whenever $I - \mathfrak{z} T$ is invertible (i.e., whenever $e^{i
\omega} \notin \Lambda_T$), we have:
\begin{align*}
\sum_{\tau=1}^{N} ( \mathfrak{z} T)^{\tau }  
  &= (I - \mathfrak{z} T)^{-1}
 \bigl(  \mathfrak{z} T - \mathfrak{z}^{N+1} T^{N+1} \bigr)
    ~,
\end{align*}
and:
\begin{widetext}
\begin{align*}
\sum_{\tau=1}^{N}  \tau ( \mathfrak{z} T)^{\tau }  
  & = 
(I - \mathfrak{z} T)^{-1}
\Bigl( - N \mathfrak{z}^{N+1} T^{N+1} + 
    (I - \mathfrak{z} T)^{-1}
       \bigl(  \mathfrak{z} T - \mathfrak{z}^{N+1} T^{N+1} \bigr) \Bigr)
    ~.
\end{align*}

Together, this yields:
\begin{align*}
\sum_{\tau=1}^{N} \bigl( N -  \tau \bigr) T^{\tau } e^{-i \omega \tau} 
&= 
N \sum_{\tau=1}^N (\mathfrak{z} T)^\tau - \sum_{\tau=1}^{N} \tau (\mathfrak{z} T)^\tau \\
&= 
N (I - \mathfrak{z} T)^{-1} \bigl( \mathfrak{z} T - \mathfrak{z}^{N+1} T^{N+1} + \mathfrak{z}^{N+1} T^{N+1}  \bigr)
-  (I - \mathfrak{z} T)^{-2} \bigl( \mathfrak{z} T - \mathfrak{z}^{N+1} T^{N+1}  \bigr) \\
&= 
N T (\mathfrak{z}^{-1} I -  T)^{-1} 
-  (I - \mathfrak{z} T)^{-2} \bigl( \mathfrak{z} T - \mathfrak{z}^{N+1} T^{N+1}  \bigr) 
    ~.
\end{align*}
Noting that $(\mathfrak{z}^{-1} I -  T)^{-1} = (e^{i \omega} I - T)^{-1}$, 
this implies that the continuous (i.e., diffuse) part of the power spectrum becomes:
\begin{align}
P_\text{c}(\omega) 
&= \bigl\langle \left| x \right|^2 \bigr\rangle + 
\lim_{N \to \infty} \frac{2}{N}  \, \text{Re} 
   \bra{\stationary} \overline{\Omega} \, \Bigl(
  \sum_{\tau=1}^{N} \bigl( N -  \tau \bigr) T^{\tau } e^{-i \omega \tau} 
  \Bigr)  \, \Omega \ket{\one} \nonumber \\
&=   
\bigl\langle \left| x \right|^2 \bigr\rangle + 
\lim_{N \to \infty} \frac{2}{N}  \, \text{Re} 
   \bra{\stationary} \overline{\Omega} \, \Bigl(
  N T (\mathfrak{z}^{-1} I -  T)^{-1} 
-  (I - \mathfrak{z} T)^{-2} \bigl( \mathfrak{z} T - \mathfrak{z}^{N+1} T^{N+1}  \bigr) 
  \Bigr)  \, \Omega \ket{\one}  \nonumber \\
&=   
\bigl\langle \left| x \right|^2 \bigr\rangle + 
2  \, \text{Re} 
   \bra{\stationary} \overline{\Omega} \, 
  T (\mathfrak{z}^{-1} I -  T)^{-1} 
  \, \Omega \ket{\one} 
- \lim_{N \to \infty} \frac{2}{N}  
   \, \text{Re} 
   \bra{\stationary} \overline{\Omega} \, 
   (I - \mathfrak{z} T)^{-2} \bigl( \mathfrak{z} T - \mathfrak{z}^{N+1} T^{N+1}  \bigr) 
    \, \Omega \ket{\one} 
\label{eq:Pc_wN}
   \\
&=   
\bigl\langle \left| x \right|^2 \bigr\rangle + 
2  \, \text{Re} 
   \bra{\stationary} \overline{\Omega} \, 
  T ( e^{i \omega} I -  T)^{-1} 
 \, \Omega \ket{\one} 
 \label{eq:Pc_final}
    ~.
\end{align}
\end{widetext}
Equation~\eqref{eq:Pc_final} is the principle result, yielding the continuous
part of the power spectrum in closed form. However, it is also worth noting
that Eq.~\eqref{eq:Pc_wN} (without the $N \to \infty$ limit yet being taken)
provides the exact result for the expected periodogram from finite length-$N$
samples.

\subsection{Discrete Spectra}

The transition dynamic's eigenvalues $\Lambda_{\rho(T)} = \bigl\{ \lambda \in
\Lambda_T: | \lambda |  = 1 \bigr\}$ on the unit circle are responsible for a
power spectrum's Dirac $\delta$-functions. In the physical context of diffraction
patterns, these $\delta$-functions are the familiar Bragg reflections. For finite length-$N$
samples, eigenvalues on the unit circle give rise to Dirichlet kernels. As $N
\to \infty$, the analysis simplifies since the Dirichlet kernels converge to
$\delta$-functions.

The following derives the exact form of the $\delta$-function contributions,
showing how their presence and integrated magnitude can be calculated directly
from the stochastic transition dynamic. Recall that the spectral projection
operator $T_{\lambda, 0}$ associated with the eigenvalue $\lambda$ can be
defined as the residue of $(z I - T)^{-1}$ as $z  \to \lambda$:
\begin{align*}
T_{\lambda, 0} = \tfrac{1}{2 \pi i} \oint_{C_\lambda} \bigl( zI - T \bigr)^{-1} \, dz ~.
\end{align*}
The spectral companion operators are:
\begin{align*}
T_{\lambda, m} = T_{\lambda, 0} (T - \lambda I)^m 
  ~,
\end{align*}
with the useful property that 
$T_{\lambda, m} T_{\zeta, n} = \delta_{\lambda, \zeta} T_{\lambda, m+n} $ and
$T_{\lambda, m} = \boldsymbol{0}$ for $m \geq \nu_\lambda$.
The index $\nu_\lambda$ of the eigenvalue $\lambda$ is the size of the largest Jordan block associated with $\lambda$.

The Perron--Frobenius theorem guarantees that all eigenvalues on the unit circle
have an index of one: i.e., $\nu_\lambda = 1$ for all $\lambda \in
\Lambda_{\rho(T)}$. This means that the algebraic and geometric multiplicities
of these eigenvalues coincide and they are all associated with diagonalizable
subspaces.

Taking advantage of the index-one nature of the eigenvalues on the unit circle, 
and using the shorthand $T_\lambda \equiv T_{\lambda, 0}$ for the spectral
projection operators, we define:
\begin{align*}
\Theta \equiv \sum_{\lambda \in \Lambda_{\rho(T)} } \lambda  T_\lambda 
\end{align*}
and 
\begin{align*}
F \equiv T - \Theta ~.
\end{align*}
We then consider how the spectral decomposition of $T^\tau$ splits into contributions from these two independent components:
From Ref.~\cite{Riec18c}, and employing the simplifying notation that $0^{\tau-m} = \delta_{\tau-m, 0}$,
we find:
\begin{align*}
T^\tau
& = \sum_{\lambda \in \Lambda_T } 
		\sum_{m=0}^{\nu_\lambda - 1}  
		\lambda^{\tau-m}  \binom{\tau}{m} 
		T_{\lambda, m}  \\
&= \Bigl( \sum_{\lambda \in \Lambda_{\rho(T)} }  
		\lambda^{\tau} 
		T_{\lambda}  \Bigr)
	+ \Bigl( \sum_{\lambda \in \Lambda_T \setminus \Lambda_{\rho(T)} } 
		\sum_{m=0}^{\nu_\lambda - 1}  
		\lambda^{\tau-m}  \binom{\tau}{m} 
		T_{\lambda, m}  \Bigr)
		\\
&= \Theta^\tau + F^\tau	
  ~,	
\end{align*}
where $\binom{\tau}{m} = \frac{1}{m!} \prod_{n=1}^m (\tau-n+1)$ is the
generalized binomial coefficient. 

As the sequence length $N \to \infty$, the summation over $\tau$ in Eq.~\eqref{eq:SwallowedSummation} divided by the sequence length becomes:
\begin{align}
\lim_{N \to \infty} \sum_{\tau = 1}^N & \frac{N-\tau}{N} T^\tau e^{-i \omega
\tau} \nonumber \\
  & = \sum_{\tau = 1}^\infty T^\tau e^{-i \omega \tau } \nonumber \\
  & = \Bigl( \sum_{\tau = 1}^\infty \Theta^\tau e^{-i \omega \tau } \Bigr) + \Bigl( \sum_{\tau = 1}^\infty F^\tau e^{-i \omega \tau } \Bigr)  ~.
\label{eq:SplitSum}
\end{align}
In Eq.~\eqref{eq:SplitSum}, only the summation involving $\Theta$ is capable of contributing $\delta$-functions.
Expanding that sum yields:
\begin{align}
\sum_{\tau = 1}^\infty & \Theta^\tau e^{-i \omega \tau } \nonumber \\
  &= \sum_{\lambda \in \Lambda_{\rho(T)}} T_\lambda \sum_{\tau = 1}^{\infty} (
  \lambda e^{-i \omega})^\tau \nonumber \\
  &= \sum_{\lambda \in \Lambda_{\rho(T)}} T_\lambda \Bigl( -1 + \sum_{\tau =
  0}^{\infty} e^{i ( \omega_\lambda - \omega ) \tau} \Bigr) \nonumber \\
  &= \!\! \sum_{\lambda \in \Lambda_{\rho(T)}} \!\!\!\!  T_\lambda 
      \Bigl( 
       \frac{-1}{1-e^{i (\omega - \omega_\lambda)}} + \!\!\!\!  \sum_{k =
	   -\infty}^\infty \!\!\!\!  \pi \, \delta( \omega - \omega_\lambda + 2 \pi k )
      \Bigr)
  ,
\label{eq:DeltaSum}
\end{align}
where $\omega_\lambda$ is related to $\lambda$ by $\lambda = e^{i \omega_\lambda}$.
The last line is obtained using well-known properties of the discrete-time Fourier transform~\cite{Oppe14}.

From Eqs.~(\ref{eq:SwallowedSummation}), (\ref{eq:SplitSum}), and
(\ref{eq:DeltaSum}), we find that the potential $\delta$-function at
$\omega_\lambda$ (and its $2\pi$-periodic offsets) has integrated magnitude:
\begin{widetext}
\begin{align}
\Delta_\lambda 
  & \equiv \lim_{\epsilon \to 0} \int_{\omega_\lambda -
  \epsilon}^{\omega_\lambda + \epsilon} P(\omega) \, d \omega \nonumber \\
  &= \lim_{\epsilon \to 0} \int_{\omega_\lambda - \epsilon}^{\omega_\lambda + \epsilon}
  2  \, \text{Re} 
   \bra{\stationary} \overline{\Omega} \, \Bigl(
   \lim_{N \to \infty} 
  \sum_{\tau=1}^{N} \frac{ N -  \tau }{N}  T^{\tau } e^{-i \omega \tau} 
  \Bigr)  \, \Omega \ket{\one} 
  \, d \omega \nonumber \\
  &= \lim_{\epsilon \to 0} \int_{\omega_\lambda - \epsilon}^{\omega_\lambda + \epsilon}
  2  \, \text{Re} 
   \bra{\stationary} \overline{\Omega} \, \Bigl(
   \sum_{\tau = 1}^\infty \Theta^\tau e^{-i \omega \tau } 
  \Bigr)  \, \Omega \ket{\one} 
  \, d \omega \nonumber \\
  &= \lim_{\epsilon \to 0} \int_{\omega_\lambda - \epsilon}^{\omega_\lambda + \epsilon}
  2  \, \text{Re} 
   \bra{\stationary} \overline{\Omega} \, 
   \sum_{\zeta \in \Lambda_{\rho(T)}} T_\zeta 
      \Bigl( 
       \frac{-1}{1-e^{i (\omega - \omega_\zeta)}} + \sum_{k = -\infty}^\infty \pi \, \delta( \omega - \omega_\zeta + 2 \pi k )
      \Bigr)
     \, \Omega \ket{\one} 
  \, d \omega \nonumber \\
  &= 
  2 \pi \, \text{Re} 
   \bra{\stationary} \overline{\Omega} \, 
   T_\lambda
     \, \Omega \ket{\one} \,
     \lim_{\epsilon \to 0} \int_{\omega_\lambda - \epsilon}^{\omega_\lambda + \epsilon}
      \delta( \omega - \omega_\lambda )
  \, d \omega \nonumber \\
  &= 
  2 \pi \, \text{Re} 
   \bra{\stationary} \overline{\Omega} \, 
   T_\lambda
     \, \Omega \ket{\one} ~.
\label{eq:IntMagnitude}
\end{align}
\end{widetext}

Finally, from Eq.~\eqref{eq:IntMagnitude} and the $2 \pi$-periodicity of the power spectrum, we obtain 
the full discrete (i.e., $\delta$-function) contribution to the power spectrum:
\begin{align}
P_\text{d}(\omega) &= 
  \!\!\!\!  \sum_{\lambda \in \Lambda_{\rho(T)}}
  \!\! 2 \pi \,
  \text{Re} \bra{\stationary} \overline{\Omega} \, T_\lambda \Omega \ket{\one}
 \!\! \sum_{k = -\infty}^{\infty} 
   \!\! \delta( \omega - \omega_\lambda + 2 \pi k) 
 .
\label{eq:PdwDerived}
\end{align}

{ \color{blue} 

\section{A new condition for $1/f$ noise}
\label{sec:OneOverfApp}

Here we obtain a sufficient condition for $1/f$ noise.

Eq.~\eqref{eq:ContTimePSD} gave the general formula for power spectra from
continuous-time processes:
\begin{align*}
P_\text{c}(f) = 
\sum_{\lambda \in \Lambda_G} \sum_{m = 0}^{\nu_\lambda - 1}
  2 \, \text{Re} \frac{
  \bra{\stationary} \overline{\Omega} \, G_{\lambda, m} \Omega \ket{\one}}{(i 2 \pi f - \lambda)^{m+1}}
  ~.
\end{align*}

We restrict attention to diagonalizable transition rate operators. To simplify
notation, we relabel the spectral intensity as $c_\lambda \equiv
\bra{\stationary} \overline{\Omega} \, G_{\lambda, 0} \Omega \ket{\one}$.
Recall the following.

{\textbf{Definition \ref{def:UniformDim}. }}
\emph{An observable continuous-time process has \emph{\DHD} if:
\begin{enumerate}
\item its generator of time evolution $G$ is diagonalizable and has $N+1$
	evenly spaced eigenvalues along the real line $\Lambda_G = \{ -n a
	\}_{n=0}^N$  for some $a>0$, and 
\item its spectral intensity fades with increasing frequency according to
	$c_{-na} = c / n$ for $n \geq 1$ and some $c \in \mathbb{R}$.
\end{enumerate}
}

We will show that any process with \DHD\ produces $1/f$ noise over a frequency
bandwidth proportional to $N$.

For a process with \DHD, the power spectrum simplifies considerably to:
\begin{align}
P_\text{c}(f) 
&= 
\sum_{\lambda \in \Lambda_G} 
  2 \, \text{Re} \frac{
  c_\lambda }{i 2 \pi f - \lambda} \nonumber \\
&= 
\sum_{n=1}^N 
  2 \, \text{Re} \frac{ c / n
   }{i 2 \pi f + n a} \nonumber \\
&= 
\frac{2 c}{a}
\sum_{n=1}^N 
 \frac{
  1 }{ n^2 + \bigl( \tfrac{2 \pi f}{a} \bigr)^2 }       
  ~.
\label{eq:MainOneOverfEq}
\end{align}

By considering various limits, we see that Eq.~\eqref{eq:MainOneOverfEq}
leads to nearly perfect $1/f$ noise over a significant bandwidth.

\subsection{Constant spectrum for $ f \ll a / 2 \pi$}

If $2 \pi f \ll a$, then $1+ \bigl( \frac{2 \pi f}{ na } \bigr)^2 \approx 1$ for all $n \geq 1$. Accordingly:
\begin{align*}
P(f)
&= 
\frac{2 c}{a} \sum_{n=1}^N 
 \frac{1}{ n^2 \Bigl[ 1 + \bigl( \tfrac{2 \pi f}{ n a} \bigr)^2 \Bigr] } \\
& \approx    
\frac{2 c}{a} \sum_{n=1}^N 
 \frac{ 1 }{ n^2 } \; =  \frac{2 c}{a} H_{N,2} \\
& \to \frac{c \pi^2}{ 3 a} \qquad \quad \text{as } N \to \infty
  ~,
\end{align*}
where $H_{N, 2} = \sum_{n=1}^N \tfrac{1}{n^2}$ is a generalized harmonic number.
Notably, $H_{N, 2} \to \pi^2 / 6$ as $N \to \infty$.

\subsection{$1/f^2$ spectrum for $f \gg N a / 2 \pi$}

If $2 \pi f \gg N a$, then $2 \pi f \gg n a$ and $ 1 + \bigl( \tfrac{n a}{2 \pi f} \bigr)^2 \approx 1$ for all $n \leq N$. Accordingly:
\begin{align*}
P(f) &= \frac{2 c}{a} \sum_{n=1}^N 
 \frac{
  1 }{ \bigl( \frac{2 \pi f}{ a} \bigr)^2  \Bigl[ 1 + \bigl( \tfrac{n a}{2 \pi f} \bigr)^2 \Bigr] } 
\\
  & \approx \frac{a c }{ 2 \pi^2 f^2}
  \sum_{n=1}^N  1 \; =   \frac{c a N }{ 2 \pi^2 f^2}
  ~.
\end{align*}  

\subsection{$1/f$ spectrum for $\frac{a}{2 \pi} \ll f \ll \frac{N a }{ 2 \pi }$ }

If $2 \pi f \ll N a $, then $2 \pi f \ll n a $ and $1 + \bigl( \tfrac{2 \pi f}{
n a} \bigr)^2 \approx 1$ for any $n \geq N$. Then:
\begin{align*}
P(f) &= 
\tfrac{2 c}{a} \Biggl[  \biggl(
  \sum_{n=1}^\infty
    \tfrac{ 1 }{ n^2 + \bigl( \tfrac{2 \pi f}{a} \bigr)^2 }  \biggr)
- \biggl(
   \sum_{n=N+1}^\infty  
      \tfrac{ 1 }{ n^2 \Bigl[ 1 + \bigl( \tfrac{2 \pi f}{ n a} \bigr)^2 \Bigr] } 
\biggr)
  \Biggr] \\
& \approx
\tfrac{2 c}{a} \Biggl[  \biggl(
  \sum_{n=1}^\infty
    \frac{ 1 }{ n^2 + \bigl( \tfrac{2 \pi f}{a} \bigr)^2 }  \biggr)
- \biggl(
   \sum_{n=N+1}^\infty  
      \frac{ 1 }{ n^2} 
\biggr)
  \Biggr] \\
& =
\tfrac{2 c}{a} \Biggl[  \biggl(
  \sum_{n=1}^\infty
    \frac{ 1 }{ n^2 + \bigl( \tfrac{2 \pi f}{a} \bigr)^2 }  \biggr)
- \bigl(
  \tfrac{\pi^2}{6} - H_{N, 2} 
\bigr)
  \Biggr] ~.
\end{align*}
With the identity
\begin{align*}
\sum_{n=1}^\infty
    \frac{ 1 }{ n^2 + \bigl( \tfrac{2 \pi f}{a} \bigr)^2 }  
    = \frac{ a \coth(2 \pi^2 f / a)}{ 4 f } - \frac{1}{ 2 \bigl( \tfrac{2 \pi f}{a} \bigr)^2 }
  ~,
\end{align*}
this yields:
\begin{align*}
P(f) \approx \frac{c \coth(2 \pi^2 f / a)}{ 2 f} - \frac{ac}{(2 \pi f)^2}  
  - \tfrac{2 c}{a} \bigl(  \tfrac{\pi^2}{6} - H_{N, 2}  \bigr)
\end{align*}
for $2 \pi f \ll N a $.

For $f > \frac{a}{2 \pi^2}$, the hyperbolic cotangent $\coth(2 \pi^2 f / a)$
quickly converges to unity. Hence, for $\frac{a}{2 \pi^2} \ll f \ll \frac{N
a}{2 \pi}$, the power spectrum is well approximated by:
\begin{align*}
P(f) & \approx
\frac{c }{ 2 f} \Bigl( 1 - \frac{a}{2 \pi^2 f} \Bigr)  - \tfrac{2 c}{a} \bigl(  \tfrac{\pi^2}{6} - H_{N, 2}  \bigr)
\nonumber \\
& \approx \frac{c }{ 2 f} - \tfrac{2 c}{a} \bigl(  \tfrac{\pi^2}{6} - H_{N, 2}  \bigr)
  ~.
\end{align*}

Moreover, $ \tfrac{\pi^2}{6} - H_{N, 2} \to 0 $ as $N \to \infty$.

\subsection{Combining the regimes}

We showed that any process with \DHD\ has three distinctive regimes in its
power spectrum: nearly constant for very low frequency, $1/f$ decay over a
broad bandwidth, and $1/f^2$ decay at very large frequencies.

The transition frequencies between these three behavior regimes is found
more specifically by looking for the crossover frequencies---$f^*$ where the
constant and $1/f$ approximations meet, and $f^{**}$ where the $1/f$ and
$1/f^2$ approximations meet.

The first transition frequency $f^*$, from constant to $1/f$ behavior,
satisfies $\frac{2 c}{a} H_{N, 2} = \frac{c }{ 2 f^*} - \tfrac{2 c}{a} \bigl(
\tfrac{\pi^2}{6} - H_{N, 2}  \bigr)$. We find that:
\begin{align}
f^* = \frac{3 a}{2 \pi^2} ~. 
\end{align}

The second transition frequency $f^{**}$, from $1/f$ behavior to $1/f^2$
behavior, satisfies:
\begin{align*}
\frac{c }{ 2 f^{**}} - \tfrac{2 c}{a} \bigl(  \tfrac{\pi^2}{6} - H_{N, 2}  \bigr) = \frac{a c N}{2 \pi^2 (f^{**})^2}
  ~.
\end{align*}
We find that:
\begin{align*}
f^{**} = \frac{a}{8 \bigl(  \tfrac{\pi^2}{6} - H_{N, 2}  \bigr) } \biggl( 1 - \sqrt{1 - \tfrac{8}{3} N \bigl(  1 - \tfrac{ 6 H_{N, 2} }{\pi^2}  \bigr) } \, \biggr)
  ~.
\end{align*}
This exact expression for $f^{**}$ can be expanded in terms of the small
parameter:
\begin{align*}
\epsilon = \frac{8}{3}  N \bigl( 1 - \tfrac{ 6 H_{N, 2} }{\pi^2} \bigr)
\end{align*}
such that:  
\begin{align*}
f^{**} 
&= 
\frac{aN}{ 2 \pi^2 \epsilon } \Bigl[ 1 - (1 - \epsilon)^{1/2} \Bigr] \\
&=
\frac{aN}{ 2 \pi^2} \sum_{k=1}^\infty {1/2 \choose k } (-\epsilon)^{k-1} \\
&= 
\frac{aN}{ 4 \pi^2} \Bigl( 1 - \tfrac{1}{4} \epsilon + \tfrac{1}{8} \epsilon^2 - \mathcal{O}(\epsilon^3) \Bigr)
  ~,
\end{align*}
that, to first order, yields the approximation
$f^{**} \approx \frac{aN}{ 4 \pi^2} $.

Altogether, this leads to:
\begin{align*}
P(f) \approx
\begin{cases}
\frac{2 c}{a} H_{N,2}  & \text{if } f < \frac{3 a}{ 2 \pi^2} \\
\frac{c}{2 f} - \tfrac{2 c}{a} \bigl(  \tfrac{\pi^2}{6} - H_{N, 2}  \bigr)   & \text{if } \frac{3 a}{ 2 \pi^2} < f <  f^{**} \\
\frac{c a N}{ 2 \pi^2 f^2} & \text{if } f >  f^{**} 
\end{cases}
\end{align*}
or, more simply:
\begin{align*}
P(f) \sim
\begin{cases}
\text{constant} & \text{if } f < \frac{3 a}{ 2 \pi^2} \\
1/f  & \text{if } \frac{3 a}{ 2 \pi^2} < f \lesssim \frac{a N}{ 4 \pi^2} \\
1/{f^2} & \text{if } f \gtrsim  \frac{a N}{ 4 \pi^2}
\end{cases} ~.
\end{align*}
}

{ \color{blue} 

\section{Brownian Noise}
\label{sec:AppendixOnDeterministicTransduction}

Here, we show how to recover the power spectrum of Brownian motion using the
tools of Sec.~\ref{sec:DeterministicTransduction}. This simple example
indicates how to leverage the tools more generally to analyze the power spectra
of more sophisticated Langevin-type differential equations that can transduce
arbitrarily sophisticated noise models.

Each spatial dimension of a Brownian trajectory behaves independently and
simply integrates white noise. In the discrete-time case, the fundamental
equation for Brownian noise is:
\begin{align}
Y_t - Y_{t-1} =  X_t
  ~,
\label{eq:BrownianInApp}
\end{align}
where $X_t$ is a Gaussian white noise
of variance  $\sigma^2 = 2 D \tau_0$, where  $D$ is the diffusion coefficient,
which implies $P_{XX}(\omega) = \sigma^2 / f_0 = 2 D / f_0^2$.
Equation~\eqref{eq:BrownianInApp} corresponds to $\mathscr{P}(\mathcal{D}) =
\mathcal{D}^0 - \mathcal{D}$ and $\mathscr{Q}(\mathcal{D}) =  \mathcal{D}^0$,
which leads to: 
\begin{align}
 | H_{X \to Y} (\omega) |^2 = \frac{1}{| 1 - e^{i \omega} |^2 } = \frac{1}{ 2 \bigl( 1 - \cos( \omega ) \bigr) } 
\end{align}
and
\begin{align}
P_{YY}(\omega) 
&=  \frac{2 D / f_0^2}{ 2 \bigl( 1 - \cos( \omega ) \bigr) } \\
&=  \frac{D}{2 \pi^2 f^2 \left[ 1 - \frac{\pi^2}{3}
  \left(\frac{f}{f_0}\right)^2
  + \mathcal{O}\left( \left(\frac{f}{f_0}\right)^4 \right) \right] } 
  \label{eq:FiniteFrequencyBrownianApprox} \\
  & \to  \frac{D}{2 \pi^2 f^2 } \quad \text{ as } \tfrac{f}{f_0} \to 0
  ~.
\end{align}
The last line gives the limiting power spectrum of Brownian noise in the continuous-time case, where it is well-known that $P_{YY}(f) \propto 1/f^2$.

It is worth noting that, at finite sampling frequency, the experimentally or
numerically obtained power spectrum deviates significantly from the $1/f^2$
spectrum as $f \to f_0/2$, according to
Eq.~\eqref{eq:FiniteFrequencyBrownianApprox}. This could lead to misidentifying
$1/f^\alpha$ noise.



}

\section{Proof of Lemma 1}
\label{sec:Lemma1Proof}

Recall {\textbf{ Lemma 1}}:

\emph{Any stochastic process (not necessarily stationary) with the
\emph{Single-Condition-Independent Property} (SCIP):
\begin{align*}
\Pr(X_t | X_{t'} = x) & = \Pr(X_t) \\
  & = \Pr(X_{t'})
  ~,
\end{align*}
for all $x \in \Abet$ and all $t \neq t'$, generates a flat power spectrum,
mimicking white noise.
}

{\ProLem 
For any such process, $\Pr(X_t)$ is the stationary distribution
$\boldsymbol{\mu}_X$ of the instantaneous observable under the stochastic
dynamic.  Moreover, SCIP means that the joint probability of any
two observations decomposes:
\begin{align*}
\Pr(X_t = x, X_{t+\tau} = x') & = \Pr(X_{t+\tau} = x' | X_t = x) \Pr(X_t = x) \\
  & = \Pr(X_{t+\tau} = x' ) \Pr(X_t = x) \\
  & = \boldsymbol{\mu}_X( x' ) \boldsymbol{\mu}_X( x)
  ~.
\end{align*}
Substituting $\boldsymbol{\mu}_X( x' ) \boldsymbol{\mu}_X( x)$ for $\Pr(X_t =
x, X_{t+\tau} = x')$ in the autocorrelation definition of
Eq.~\eqref{eq:AutocorrDef} immediately implies that SCIP processes have
$\tau$-independent pairwise correlation $\gamma(\tau) = \left| \braket{x}
\right|^2$ for $\tau \neq 0$. The power spectrum is thus flat over all
frequencies, except possibly with a $\delta$-function at $\omega=0$.
}

\section{Proof of Theorem 2}
\label{sec:Thm2Proof}

We define the set $\Xi$ of average outputs exhibited by the
states: $\Xi \equiv \bigcup_{s \in \SSet} \bigl\{ \braket{x}_{\pdf(X | s)}
\bigr\}$. Furthermore, we define $\SSet_\xi \subset \SSet$ as the set of states
that all exhibit the same average output $\xi \in \Xi$. Explicitly, $\SSet_\xi
\equiv \{ s \in \SSet :  \braket{x}_{\pdf(X | s)} = \xi \}$. 

Recall {\textbf{Theorem \ref{thm:GenFlatPS}}}:

\emph{
Let $\{ X_t \}_t$ be a stochastic process generated by a hidden-state
model $\mathcal{M}(\vec{m})$. 
$X_t$ is the random variable for the observable at
time $t$, and $\St_t$ is the random variable for the hidden state at time $t$.
Such processes have \emph{constant autocorrelation and a flat power spectrum}
if:
\begin{align}
\Pr(\St_{t+\tau} \in \SSet_{\xi'} | \St_t \in \SSet_\xi )
  & = \Pr(\St_{t+\tau} \in \SSet_{\xi'} ) \nonumber \\
  & = \Pr(\St_{t} \in \SSet_{\xi'} )
  ~,
\label{eq:ThmConditionApp}
\end{align}
for all separations $\tau > 0$, for all $t \in \mathcal{T}$, and for all $\xi, \xi' \in \Xi$.
}

{\ProThe 
Starting from Eq.~\eqref{eq:AutocorrAsAvgAvgs}, we find the autocorrelation for
all such processes (for $\tau \geq 1$):
\begin{align}
\gamma(\tau) 
  & = \big\langle  \braket{\overline{x}}_{\pdf(X | \St_t)}
  \braket{x}_{\pdf(X | \St_{t+\tau})}
  \big\rangle_{\Pr(\St_t, \St_{t+\tau}) } \nonumber \\
  & = \sum_{s, s' \in \SSet}
\Pr(\St_t = s, \St_{t+\tau} = s')
  \braket{\overline{x}}_{\pdf(X | s)} \braket{x}_{\pdf(X | s')} \nonumber \\
  & = \sum_{\xi, \xi' \in \Xi}
  \Pr(\St_t \in \SSet_\xi , \St_{t+\tau} \in \SSet_{\xi'})
  \, \overline{\xi}  \, \xi'  \nonumber \\
  & = \sum_{\xi \in \Xi}
  \Pr(\St_t \in \SSet_\xi)
  \overline{\xi}  \nonumber \\
  & \qquad \times \sum_{\xi' \in \Xi}
  \Pr(\St_{t+\tau} \in \SSet_{\xi'} | \St_t \in \SSet_\xi )
  \xi'  
  ~.
\label{eq:ParsedXis}
\end{align}
Combining Eq.~\eqref{eq:ThmConditionApp} and Eq.~\eqref{eq:ParsedXis},
we see that: 
\begin{align*}
\gamma(\tau) 
  & = \sum_{\xi \in \Xi} \Pr(\St_t \in \SSet_\xi)  \overline{\xi}
  \sum_{\xi' \in \Xi}  \Pr(\St_{t+\tau} \in \SSet_{\xi'}  )  \xi' \\
  & = \left| \braket{\xi} \right|^2
  ~,
\end{align*}
which is a constant. 
With the same reasoning, we likewise find that $\gamma(\tau) = \left| \braket{\xi} \right|^2$ for $\tau \leq -1$.
The autocorrelation function is thus
$\gamma(\tau) = |\braket{ \xi }|^2 + c \delta_{\tau, 0}$,
where $c \equiv \gamma(0) - |\braket{ \xi }|^2$ is a constant.
Thus, the power spectrum is flat, \emph{if} 
Eq.~(\ref{eq:ThmConditionApp}) holds.
}

\section{Proof of Theorem 3}
\label{sec:ProofOfSamePSD}

{\color{blue}
Recall {\textbf{Theorem \ref{thm:SamePS}}}: 
\emph{Let $\{ X_t \}_t$ and $\{ Y_t \}_t$ be two stochastic processes generated
by any of the hidden-state models $\mathcal{M}(\vec{m})$ discussed above,
including autonomous HMMs and input-dependent generators, $X_t$ and $Y_t$ the
random variables for the observables at time $t$, and $\St_t \in \SSet$ and
$\mathcal{R}_t \in \boldsymbol{\mathcal{R}}$ the random variables for the
respective hidden states at time $t$. These processes have \emph{identical
power spectra}, up to a constant offset, if:
\begin{align*}
\Pr(\St_t \in \SSet_\xi , \St_{t+\tau} \in \SSet_{\xi'} )
  & = \Pr(\mathcal{R}_t \in \boldsymbol{\mathcal{R}}_\xi , \mathcal{R}_{t+\tau} \in \boldsymbol{\mathcal{R}}_{\xi'} )
  ~,
\end{align*}
for all separations $\tau > 0$, for all $t \in \mathcal{T}$, and for all $\xi,
\xi' \in \Xi$, which is the set of average outputs emitted by the states.
}

{\textbf{Proof.}}
\emph{Let $\gamma(\tau)$ be the autocorrelation function for the first process
$\{ X_t \}_t$, and let $\gamma'(\tau)$ be the autocorrelation function for the
second process $\{ Y_t \}_t$. Assume:
\begin{align*}
\Pr(\St_t \in \SSet_\xi , \St_{t+\tau} \in \SSet_{\xi'} )
  = \Pr(\mathcal{R}_t \in \boldsymbol{\mathcal{R}}_\xi , \mathcal{R}_{t+\tau} \in \boldsymbol{\mathcal{R}}_{\xi'} )
\end{align*}
for all separations $\tau > 0$, for all $t \in \mathcal{T}$, and for all $\xi, \xi' \in \Xi$.
Then, starting from Eq.~\eqref{eq:AutocorrAsAvgAvgs}, we find the
autocorrelation for the first process (for $\tau \geq 1$):
\begin{align*}
\gamma(\tau) 
  & = \big\langle  \braket{\overline{x}}_{\pdf(X | \St_t)}
  \braket{x}_{\pdf(X | \St_{t+\tau})}
  \big\rangle_{\Pr(\St_t, \St_{t+\tau}) } \\
  & = \sum_{s, s' \in \SSet}
\Pr(\St_t = s, \St_{t+\tau} = s')
  \braket{\overline{x}}_{\pdf(X | s)} \braket{x}_{\pdf(X | s')} \\
  & = \sum_{\xi, \xi' \in \Xi}
  \Pr(\St_t \in \SSet_\xi , \St_{t+\tau} \in \SSet_{\xi'})
  \, \overline{\xi}  \, \xi'  \\
& = \sum_{\xi, \xi' \in \Xi}
\Pr(\mathcal{R}_t \in \boldsymbol{\mathcal{R}}_\xi , \mathcal{R}_{t+\tau} \in \boldsymbol{\mathcal{R}}_{\xi'} )  
 \, \overline{\xi}  \, \xi'  \\
   & = \sum_{r, r' \in \boldsymbol{\mathcal{R}}}
\Pr(\mathcal{R}_t = r, \mathcal{R}_{t+\tau} = r')
  \braket{\overline{x}}_{\pdf(X | r)} \braket{x}_{\pdf(X | r')} \\
  & = \big\langle  \braket{\overline{x}}_{\pdf(X | \mathcal{R}_t)}
  \braket{x}_{\pdf(X | \mathcal{R}_{t+\tau})}
  \big\rangle_{\Pr(\mathcal{R}_t, \mathcal{R}_{t+\tau}) } \\
  & = \gamma'(\tau)  
  ~.
\end{align*}
With the same reasoning, we find that $\gamma(\tau) = \gamma'(\tau)$ for $\tau
\leq -1$. Hence, the autocorrelations for the two processes agree everywhere
except possibly at $\tau=0$.
}

\emph{Define the constant $c \equiv \gamma(0) - \gamma'(0)$. The
autocorrelation functions for the two processes are then related by
$\gamma(\tau) = \gamma'(\tau) + c \delta_{\tau, 0}$ for all $\tau$. It then
follows that the power spectrum of the processes differ at most by a constant
offset.}
}

{ \color{blue}

\section{Diffraction patterns of chaotic crystals from HMMs}
\label{sec:DPsFromHMMs}

Let's analyze two examples of HMM-designed chaotic crystals.

\subsection{Example One}

Consider a $p$-parametrized stochastic process for the stacking of layers of a
close-packed structure. The stochastic stacking process is described by a HMM,
where the transition matrix and average-observation matrix are:
\begin{align*}
T = 
\begin{bmatrix}
0 & 1 & 0 \\
p & 0 & 1-p \\
1 & 0 & 0
\end{bmatrix}
\quad \text{and } \quad
\Omega =  
\begin{bmatrix}
A & 0 & 0 \\
0 & B & 0 \\
0 & 0 & C
\end{bmatrix}
  ~,
\end{align*}
respectively. For $p=1$, we recover the deterministic period-$2$ hcp structure.
For $p=0$, we recover the deterministic period-$3$ ccp structure. For other
values of $p$, the structure is described by a stochastic stacking process. 

For any $p$, the eigenvalues of the transition matrix are 
$\Lambda_T = \bigl\{ 1, - \frac{1}{2} \pm \sqrt{p - \frac{3}{4}} \bigr\}$.
The transition matrix is diagonalizable unless $p=3/4$, where it becomes nondiagonalizable.

We aim to calculate the diffracted intensity for any $p$ in closed form via
Eq.~\eqref{eq:PcwFromDecomposedResolvent}:
\begin{align*}
P_\text{c}(\omega) = \bigl\langle \left| x \right|^2 \bigr\rangle
  + \sum_{\lambda \in \Lambda_T} \sum_{m = 0}^{\nu_\lambda - 1}
  2 \, \text{Re}
  \frac{\bra{\stationary} \overline{\Omega} \, T \, T_{\lambda, m} \Omega \ket{\one}}{(e^{i \omega} - \lambda)^{m+1}}
\end{align*} 
and
Eq.~\eqref{eq:PdwFromResolvent}:
\begin{align*}
P_\text{d}(\omega) &= \!\!\! \sum_{k = -\infty}^{\infty} 
  \!\! \sum_{\lambda \in \Lambda_T \atop |\lambda| = 1}
  \!\! 2 \pi  \, \delta( \omega \!-\! \omega_\lambda \!+\! 2 \pi k) 
 \, \text{Re} \bra{\stationary} \overline{\Omega} \, T_\lambda \Omega \ket{\one}
 .
\end{align*}
We note that $\braket{ |x|^2 } = \braket{ |\psi|^2 } =  |\psi|^2$.

For $p \neq 3/4$, the continuous spectrum simplifies to:
\begin{align*}
P_\text{c}(\omega) = |\psi|^2
  + \sum_{\lambda \in \Lambda_T} 
  2 \, \text{Re}
  \frac{\lambda \bra{\stationary} \overline{\Omega}  \, T_{\lambda} \Omega \ket{\one}}{e^{i \omega} - \lambda} ~,
\end{align*} 
and
each spectral projection operator is given by
$T_\lambda = \ket{\lambda} \bra{\lambda}$,
with:
\begin{align*}
\bra{\lambda} & = \frac{1}{3 \lambda^2 - p} \begin{bmatrix} \lambda & 1 &
\lambda^2 - p \end{bmatrix} \text{ and }\\
\ket{\lambda} & = \begin{bmatrix} \lambda & \lambda^2 & 1 \end{bmatrix}^\top
  ~,
\end{align*}
where $\top$ denotes transposition. Recall that the stationary distribution is
the left eigenvector $\bra{\stationary} = \bra{1} = \frac{1}{3-p}
\begin{bmatrix} 1 & 1 & 1-p\end{bmatrix}$.

From these elements, we can calculate the spectral intensity:
\begin{widetext}
\begin{align}
\bra{\stationary} \overline{\Omega}  \, T_{\lambda} \Omega \ket{\one} 
  & = \frac{|\psi|^2}{(3-p)(3 \lambda^2 - p)} 
  \begin{bmatrix} 1 & 1 & 1-p \end{bmatrix}
  \begin{bmatrix} 
      1 & \, & \, \\
      \, & e^{-i 2 \pi / 3} & \, \\
      \, & \, & e^{i 2 \pi / 3}
  \end{bmatrix}
  \begin{bmatrix} \lambda \\ \lambda^2 \\ 1 \end{bmatrix}
  \begin{bmatrix} \lambda & 1 & \lambda^2 - p \end{bmatrix}
  \begin{bmatrix} 
      1 & \, & \, \\
      \, & e^{i 2 \pi / 3} & \, \\
      \, & \, & e^{-i 2 \pi / 3}
  \end{bmatrix}
  \begin{bmatrix} 1 \\ 1 \\ 1 \end{bmatrix} \nonumber \\
  & =   \frac{|\psi|^2}{(3-p)(3 \lambda^2 - p)} 
  \Bigl( \lambda + \lambda^2 e^{-i 2 \pi / 3} + (1-p) e^{i 2 \pi / 3} \Bigr)
  \Bigl( \lambda + e^{i 2 \pi / 3} + (\lambda^2-p) e^{-i 2 \pi / 3} \Bigr)
\label{eq:SpectralIntensityExample}  
\end{align}
for any $\lambda \in \Lambda_T$ and for any $p \neq 3/4$.
\end{widetext}

For $\lambda=1$, Eq.~\eqref{eq:SpectralIntensityExample} reduces to:
\begin{align}
\bra{\stationary} \overline{\Omega}  \, T_{1} \Omega \ket{\one} 
  & = \frac{ p^2 |\psi|^2}{(3-p)^2} 
  ~,
\label{eq:UnityIntensityExampleApp}
\end{align}
where we have used the identity $1+e^{i 2 \pi / 3} + e^{-i 2 \pi / 3} = 0$.
Equation~\eqref{eq:UnityIntensityExampleApp} is in fact valid for any $p \in
[0, 1]$.

For $p \in (0, 1)$, transition matrix $T$ only has one eigenvalue on the unit
circle, so the discrete (Bragg) spectrum has a single contribution from the
eigenvalue of unity:
\begin{align}
P_\text{d}(\omega) &= \frac{ 2 \pi p^2 |\psi|^2 }{(3-p)^2}  \! \sum_{k = -\infty}^{\infty} 
  \!  \delta( \omega \!+\! 2 \pi k) ~
 .
 \label{eq:UnityBraggExampleApp}
\end{align}
Although not resulting from a deterministic periodicity, this Bragg reflection
can nevertheless be regarded as a result of spatial periodicity in
probabilistic behavior.

In fact, for any $p>0$, the top-left panel of Fig.~\ref{fig:Crystal1Example}
shows that orientations $A$ and $B$ are more common than orientation $C$.
However, Eq.~\eqref{eq:UnityBraggExampleApp} survives a cyclic permutation of
the alphabet (i.e., $A \mapsto B$, $B \mapsto C$, and $C \mapsto A$). So, this
Bragg reflection persists even in multi-crystalline materials---where each
component chaotic crystal, with its own absolute orientation, is stacked
according to either the process in Fig.~\ref{fig:Crystal1Example} or one of its
cyclic permutations.

\begin{figure}[h]
\begin{center}
\includegraphics[width=0.98\columnwidth]{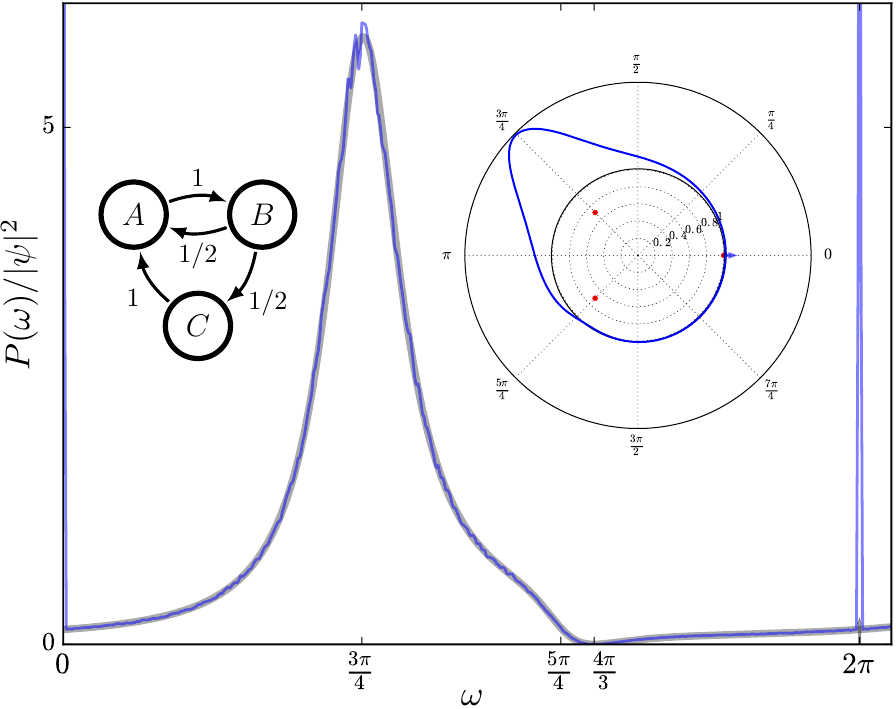}
\end{center}
\caption{Example One stochastic stacking process at $p=1/2$ (left inset) and 
	its diffraction pattern. (Main) Numerical diffraction pattern (thin blue
	line) generated from a sampled stacking sequence of $2^{20}$ layers, using
	the Welch method to calculate the power spectrum on subsamples of length
	$2^9$. It closely matches the thick gray line, which is the analytic
	solution for the diffracted intensity. (Right inset) HMM stacking process
	diffraction pattern and eigenvalues, as a coronal spectrogram.
	}
\label{fig:Crystal1at_0pt5p}
\end{figure}

There is a diffuse contribution to the power spectrum for all $p \in (0, 1)$.
For $p \in (0, 3/4) \cup (3/4, 1)$, this contribution is:
\begin{align*}
P_\text{c}(\omega) = |\psi|^2 \Bigl(  1 - \frac{ p^2 }{(3-p)^2}  \Bigr) +  \!\!
  \sum_{\lambda \in \Lambda_T \setminus \{ 1 \} } \!\!\!\!
  2 \, \text{Re}
  \frac{ \bra{\stationary} \overline{\Omega}  \, T_{\lambda} \Omega \ket{\one}}{e^{i \omega} / \lambda - 1} 
  ~.
\end{align*}
Expanding this via Eq.~\eqref{eq:SpectralIntensityExample} initially appears
unwieldy, but the expressions can be simplified as soon as one recognizes that
$\lambda^2 = p - 1 - \lambda$ for $\lambda \in \Lambda_T \setminus \{ 1 \}$.
Further simplification leverages the properties of the eigenvalues in the
distinct regimes of $p > 3/4$ and $p < 3/4$. For $p > 3/4$, all eigenvalues
have distinct real values. For $p < 3/4$, the two nonunity eigenvalues are
complex conjugate pairs and, accordingly, have the same real part (Re$(\lambda)
= -1/2$) and the same magnitude ($| \lambda | = \sqrt{1-p}$), with angular
frequencies $\omega_\lambda = \pi \pm \arctan(\sqrt{3-4p})$.

Figure~\ref{fig:Crystal1at_0pt5p} shows the ``corrected'' diffraction pattern
$P(\omega) / |\psi|^2$ for $p=1/2$. There is a Bragg reflection at $\omega = 2
\pi n$ (for all $n \in \mathbb{Z}$) due to the eigenvalue of unity. The
nonunity eigenvalues $\lambda = -1/2 \pm i 1/2$ appear at angular frequencies
$\omega_\lambda \in \{ 3 \pi / 4 , \, 5 \pi / 4\}$. The Lorentzian line profile
contributed at $\omega_\lambda = 3 \pi / 4$ is prominent. There is a local
feature around $\omega_\lambda = 5 \pi / 4$, but it is more nuanced since it is
not a peak in the diffracted intensity. Rather, the contribution from the
eigenvalue at $\omega_\lambda = 5 \pi / 4$ primarily \emph{depresses} the
diffraction pattern around it, which allows for the zero at $\omega = 4 \pi /
3$. Diffracted intensity is forbidden at $4 \pi / 3$ since the stochastic
process does not allow for a full anti-cyclic sequence of layers; i.e., $C B
A$, $BAC$, and $ACB$ are all forbidden sequences.

For $p=1$, the continuous spectrum vanishes while the discrete spectrum picks
up another Bragg reflection at $\omega = \pi$ with intensity $\bra{\stationary}
\overline{\Omega}  \, T_{-1} \Omega \ket{\one} = \frac{3}{4} | \psi |^2$,
yielding the diffraction pattern for a 2H hcp crystal:
\begin{align*}
P(\omega) \!  & =
2 \pi | \psi |^2 \! \sum_{k=-\infty}^{\infty} \! \bigl[ \frac{1}{4}  \delta( \omega \!+\! 2 \pi k) + \frac{3}{4}  \delta( \omega \!-\! \pi  \!+\! 2 \pi k) \bigr]
.
\end{align*}

Similarly, for $p = 0$, the continuous spectrum vanishes while the discrete
spectrum picks up a Bragg reflection at $\omega = 2 \pi / 3$ with intensity $
\bra{\stationary} \overline{\Omega}  \, T_{e^{i 2 \pi / 3}} \Omega \ket{\one} =
| \psi |^2$, yielding the diffraction pattern for a 3C$^+$ ccp crystal:
\begin{align*}
P(\omega) \!  &=
  2 \pi | \psi |^2 \! \sum_{k=-\infty}^{\infty}
  \! \delta( \omega \!-\! 2 \pi / 3 \!+\! 2 \pi k) 
~.
\end{align*}
Notice from Eq.~\eqref{eq:UnityIntensityExampleApp} that the former Bragg
reflection at $\omega=0$ has vanished at $p=0$.

For $p = 3/4$, the transition matrix is nondiagonalizable. Since the spectral
projection operators always sum to the identity, we can calculate $T_{-1/2}$
easily via $T_{-1/2} = I - \ket{\one} \bra{\stationary}$, with
$\bra{\stationary} = \frac{1}{9} \begin{bmatrix} 4 & 4 & 1 \end{bmatrix}$,
which yields:
\begin{align*}
T_{-1/2} = \frac{1}{9} 
    \begin{bmatrix}
      5 & -4 & -1 \\
      -4 & 5 & -1 \\
      -4 & -4 & 8
    \end{bmatrix} ~.
\end{align*}
The spectral companion operator $T_{-1/2, 1}$ is then found as:
\begin{align*}
T_{-1/2, 1} 
  & = T_{-1/2} (T + \frac{1}{2} I ) \\
  & = \frac{1}{12} 
    \begin{bmatrix}
      -2 \\
      1 \\
      4
    \end{bmatrix} 
    \begin{bmatrix}
      1 & -2 & 1 
    \end{bmatrix} 
    ~.
\end{align*}

To obtain the diffracted intensity, we calculate:
\begin{align*}
\bra{\stationary} \overline{\Omega} T_{-1/2} \Omega \ket{\one} = 8 |\psi|^2 /9
\end{align*}
and
\begin{align*}
\bra{\stationary} \overline{\Omega} 
T_{-1/2, 1} \Omega \ket{\one} = \frac{ |\psi|^2 }{3} e^{i 2 \pi / 3}
  ~.
\end{align*}
We then leverage the fact that $T T_{-1/2, 1} = -\frac{1}{2} T_{-1/2, 1}$
and $T T_{-1/2} = -\frac{1}{2} T_{-1/2} + T_{-1/2, 1}$ to calculate:
\begin{align*}
\bra{\stationary} \overline{\Omega} T T_{-1/2, 1}
	\Omega \ket{\one} = -\frac{ |\psi|^2 }{6} e^{i 2 \pi / 3}
\end{align*}
and:
\begin{align*}
\bra{\stationary} \overline{\Omega} T T_{-1/2}
\Omega \ket{\one} = (-\frac{4}{9} + \frac{1}{3} e^{i 2 \pi / 3})  |\psi|^2
  ~.
\end{align*}
Finally, this yields the nondiagonalizable power spectrum at $p=3/4$:
\begin{align*}
\frac{P_\text{c}(\omega) }{ |\psi|^2 } = \frac{8}{9} 
    - \frac{1}{3} \text{Re} \frac{ e^{i 2 \pi / 3} }{ (e^{i \omega} + \frac{1}{2})^2 } 
    + \frac{2}{3} \text{Re} \frac{ e^{i 2 \pi / 3} - \frac{4}{3} }{ e^{i \omega} + \frac{1}{2} }
	~.
\end{align*}


\subsection{Example Two}
\label{sec:ChaoticCrystalEx2App}

Here, we analyze a generalization of the second chaotic crystal discussed in
the main body. For any $q$, the transition matrix and average-observation
matrix are:
\begin{align*}
T = 
\begin{bmatrix}
0 & 1-q & q \\
1 & 0 & 0 \\
1 & 0 & 0
\end{bmatrix}
\quad \text{and } \quad
\Omega =  
\begin{bmatrix}
A & 0 & 0 \\
0 & B & 0 \\
0 & 0 & C
\end{bmatrix}
  ~,
\end{align*}
respectively.
The transition matrix  eigenvalues are 
$\Lambda_T = \bigl\{ 0, \pm 1 \bigr\}$, independent of $q$.

Each spectral projection operator is given by $T_\lambda = \ket{\lambda}
\bra{\lambda}$, with:
\begin{align*}
\bra{\lambda} = \frac{1}{3 \lambda - 1}
	\begin{bmatrix} \lambda & \lambda^2 - q & q \end{bmatrix}
\end{align*}
and:
\begin{align*}
\ket{\lambda} = 
  \begin{bmatrix} \lambda & 1 & (\lambda^2 + q-1)/q \end{bmatrix}^\top
  ~.
\end{align*}
Recall that the stationary distribution is the left eigenvector
$\bra{\stationary} = \bra{1} = \frac{1}{2} \begin{bmatrix} 1 & 1-q & q
\end{bmatrix}$. From these elements, we calculate $\bra{\stationary}
\overline{\Omega}  \, T_{\lambda} \Omega \ket{\one}$ and the power spectrum
analytically as a function of the transition parameter $q$. In particular:
\begin{align*}
\bra{\stationary} \overline{\Omega}  \, T_{1} \Omega \ket{\one}
  = \frac{1}{4}(3 q^2 - 3q + 1) |\psi|^2
\end{align*}
and:
\begin{align*}
\bra{\stationary} \overline{\Omega}  \, T_{-1} \Omega \ket{\one}
  = \frac{3}{4}(q^2 - q + 1) |\psi|^2
  ~.
\end{align*}
The net power spectrum thus consists of a flat ``white noise'' component:
\begin{align*}
P_\text{c}(\omega) &= \frac{ 3 }{2}  q ( 1 - q ) |\psi|^2 ~
\end{align*}
in addition to two Bragg reflections per $2 \pi $ of angular frequency bandwidth
\begin{align*}
P_\text{d}(\omega) = \frac{ \pi |\psi|^2 }{2}  \! \sum_{k = -\infty}^{\infty} 
  \! 
  & \bigl[ (3 q^2 - 3q + 1) \delta( \omega \!+\! 2 \pi k) \\
  & + 3(q^2 - q + 1)  \delta( \omega \!-\! \pi \!+\! 2 \pi k) \bigr] ~
 .
\end{align*}
}

{ \color{blue} 

\section{Potassium ion channel: Details}
\label{sec:PotassiumApp}

This section lays out the details for the voltage-gated potassium ion channel,
as an input-dependent transition rate matrix for partially-observable
conformational states---i.e., a continuous-time input-dependent HMM. Potassium
ion channels are embedded in neural membranes and, together with sodium ion
channels, are critical to generating and propagating action potentials that
transmit and process information throughout the brain.

\begin{figure}[h]
\begin{center}
\includegraphics[width=0.95\columnwidth]{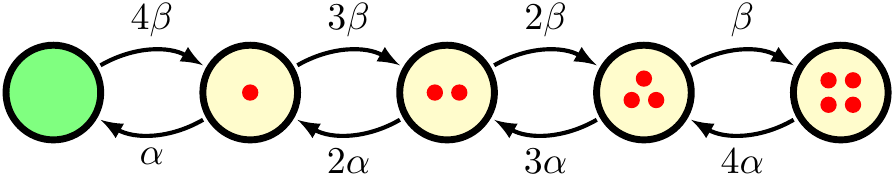}
\end{center}
\caption{Voltage-dependent continuous-time Markov chain specifying the
	transition rates between the conformational states of the K$^+$ channel.
	$\alpha$ and $\beta$ are voltage-dependent transition rates. Only the empty
	(green) state conducts current. The other states have between one and four
	activation gates (indicated by the number of red dots) blocking the
	channel. This model is thus an input-dependent continuous-time HMM for
	potassium ion current through the channel.
	}
\label{fig:PotassiumChannel}
\end{figure}

What are the dynamics and power spectra of potassium current flowing through
the channel? Only one of the five conformational states corresponds to an open
channel where current can flow. The other states are distinguished by the
number of activation gates closing the channels (from one to four), but
observation of the current does not allow for direct observation of these
conformational states. Nevertheless, the dynamics among these states influence
the statistical properties of the current. In particular, the current is
non-Markovian and exhibits a nonexponential distribution of closure durations.

The transition structure between conformational states of the K$^+$ channel is
depicted in Fig.~\ref{fig:PotassiumChannel}. The Hodgkin--Huxley model's
voltage-dependent transition rates $\alpha$ and $\beta$---often denoted
$\alpha_n$ and $\beta_n$---describe the probability that an activation gate
opens or closes (respectively) at a given voltage:
\begin{align*}
\alpha = \frac{(v + 55)/100 \text{ms}}{1 - e^{ -(v + 55)/10 } }
\quad
\text{and}
\quad
\beta = \frac{1}{8 \text{ms}} e^{-(v+65) / 80} ~,
\end{align*}
where $v$ is the voltage (in mV) across the membrane~\cite{Hodg52, Daya05}.
The voltage-dependent transition rate matrix can be written explicitly as:
\begin{align*}
&
G^{(\SSet \to \SSet | v)} \equiv \\
&
\begin{bmatrix}
- 4 \beta &	4 \beta 	&	0	&	0	&	0 	\\
\alpha	&	- (\alpha + 3 \beta) 	&	3 \beta &	0 &	0 \\
0			&	2 \alpha	& - (2\alpha + 2\beta) & 2 \beta & 0 \\
0			&	0			&	3 \alpha 	&	- (3 \alpha + \beta) & \beta \\
0			&	0			&	0 	&	4 \alpha &	- 4 \alpha 	
\end{bmatrix}
~.
\end{align*}

The average current through a single channel is binary---either $0$ or $I_0$.
Appreciable current only flows in the open conformation. In the open
conformation, $I_0 = g_0 (v - V_\text{K})$, where $g_0$ is the conductance of
an open K$^+$ channel and $V_\text{K}$ is the Nernst potential for potassium.

The rate matrix eigenvalues are $\Lambda_G = \{ - n ( \alpha + \beta)
\}_{n=0}^4$.

Applying Eq.~\eqref{eq:ContTimePSD} the power spectrum at a fixed voltage is:
\begin{align*}
P_\text{c}(f) 
&= 
\sum_{\lambda \in \Lambda_G} \sum_{m = 0}^{\nu_\lambda - 1}
  2 \, \text{Re} \frac{
  \bra{\stationary} \overline{\Omega} \, G_{\lambda, m} \Omega \ket{\one}}{(i 2 \pi f - \lambda)^{m+1}} 
\\
&= \sum_{n=0}^4  
  2 \, \text{Re} \frac{
  \bra{\stationary} \overline{\Omega} \, G_{- n (\alpha + \beta)} \Omega \ket{\one}}{i 2 \pi f + n ( \alpha + \beta ) }
\\
&= 2 I_0^2 \braket{ \stationary | \text{open} } \sum_{n=0}^4  
  \text{Re} \frac{
  \bra{ \text{open} }   G_{- n (\alpha + \beta)}  \ket{ \text{open} }}{i 2 \pi f + n ( \alpha + \beta ) }   \\
&= 2 I_0^2 \braket{ \stationary | \text{open} } \sum_{n=1}^4  
  \frac{
  \bra{ \text{open} }   G_{- n (\alpha + \beta)}  \ket{ \text{open} }}{ n ( \alpha + \beta ) \Bigl[ 1 + \Bigl( \frac{2 \pi f }{ n (\alpha + \beta) }\Bigr)^2 \Bigr]}  
  ~.
\end{align*}

For convenience, define the opening bias $\psi \equiv \alpha / \beta$ as the
ratio between an individual gate's rates of opening versus closing. The
spectral projection operators $\{ G_{- n (\alpha + \beta)} \}_n$ are simple
analytic functions of $\psi$. We then find the open state's overlap with the
spectral projection operators:
\begin{align*}
\bra{ \text{open} }   G_{- n (\alpha + \beta)}  \ket{ \text{open} } 
= 
{4 \choose n} \frac{\psi^{4-n}}{ (1+\psi)^{4} } ~.
\end{align*}
By setting $n=0$, this expression also yields the stationary probability of the open state: 
$\braket{ \stationary | \text{open} } = \bigl( \frac{\psi}{1+\psi} \bigr)^4$.

The power spectrum for potassium current is:
\begin{align}
P_\text{c}(f) 
&= \frac{I_0^2}{\pi}  \Bigl( \frac{\psi }{1 + \psi} \Bigr)^8  \sum_{n=1}^4  
  \frac{  { 4 \choose n }
  \psi^{-n} }{ 
  n w \Bigl( 1 + \bigl( \frac{ f }{ n w }\bigr)^2 \Bigr) }  
  ~,
\label{eq:PotassiumSpectrum}  
\end{align}
where $w \equiv \alpha + \beta / 2 \pi$. Each nonzero eigenmode contributes a
Lorentzian profile to the power spectrum, each with a different crossover
frequency $f_\text{c} = nw$ depending on $n$. This spread of crossover
frequencies smooths the transition between the flat power spectrum at low
frequencies and the $1/f^2$ spectrum at high frequencies. At $v=-40$ mV, the
crossover frequency of the net spectrum is $f_\text{c}^\text{net} \approx 3w$.

In fact, a power spectral signature of this general form has been
experimentally observed above the $1/f$ background noise \cite{Fish73}. That
said, the empirically observed crossover frequency suggests that the model is
not a complete description of the ion channel dynamics. 

Equation~\eqref{eq:PotassiumSpectrum}, derived from the rate matrix's spectral
properties, agrees with the much earlier results calculated via alternative
methods in Refs.~\cite{Stev72, Hill72}. For ease of comparison with those
references, note that the Hodgkin--Huxley parameter $n_\infty$ is related to
$\psi$ via $n_\infty = \psi / (1+\psi)$.

In voltage-clamped experiments, a common neurophysiological measurement
technique, the voltage is held fixed. Then, the finite-duration transition
matrix is simply $T = e^{\tau_0 G}$, where $\tau_0$ is the duration between
measurements. Since K$^+$ current (rather than conformational states) is
measured, this gives the transition matrix of a HMM for the observed current.
The finite sampling rate associated with such a discrete-time HMM allows exact
predicting the expected \emph{empirical} spectrum. At high frequencies, this
deviates from the continuous-time spectrum as the latter implicitly assumes an
infinite sampling rate.
}

\section{Cross-correlation and spectral densities}
\label{sec:Crossspectra}

Cross-correlation and cross-spectral densities are often important in
applications~\cite{Bend80, Bend11}. These may be especially useful when
analyzing input--output processes, to characterize the correlation of input and
output, or to characterize the correlation between different aspects of the
output. Our results can be easily extended to address these quantities.

Using an HMM that describes the \emph{joint} stochastic process of two
observables $(x, y) \in \Abet$, it is straightforward to generalize our
developments to \emph{cross-correlation} $\gamma_{XY}(\tau)$:
\begin{align*}
\gamma_{XY}(\tau) = \braket{\overline{X_t} Y_{t+\tau}} 
\end{align*}
(rather than necessarily \emph{auto}correlation
$\gamma = \gamma_{XX}$) and the associated \emph{cross-spectral densities}
$P_{XY}(\omega)$: 
\begin{align*}
P_{XY}(\omega) 
&= \lim_{N \to \infty} \tfrac{1}{N} \Braket{ \Bigl( \sum_{t=1}^{N}  \overline{X_t} e^{i \omega t} \Bigr) \,  \Bigl( \sum_{t=1}^{N} Y_t e^{-i \omega t} \Bigr) } \\
&=  \lim_{N \to \infty} \tfrac{1}{N} \sum_{\tau=-N}^{N} \bigl( N - \left| \tau \right| \bigr) \gamma_{XY}(\tau) e^{-i \omega \tau}
\end{align*}
of distinct observables $x \in \mathcal{X}$ and $y \in
\mathcal{Y}$. The individual stochastic processes for each observable by itself
can simply be obtained by marginalizing over the other observable.

Explicitly, the expressions take the form:
\begin{align*}
\gamma_{XY}(\tau) & = 
\begin{cases}
\bra{\stationary} \Omega_Y \, T^{|\tau| } \, \overline{\Omega}_X \ket{\one} & \text{if } \tau \leq 1 \\
\bigl\langle  \overline{X}_t Y_t \bigr\rangle & \text{if } \tau = 0 \\
\bra{\stationary} \overline{\Omega}_X \, T^{|\tau| } \, \Omega_Y \ket{\one} & \text{if } \tau \geq 1
\end{cases}
~,
\end{align*}
where:
\begin{align*}
\Omega_Y 
= \sum_{s \in \SSet} \braket{Y}_{\pdf( X, Y | s )} \ket{ s } \bra{ s } ~,
\end{align*}
and:
\begin{align*}
\bigl\langle  \overline{X}_t Y_t \bigr\rangle = 
\sum_{s \in \SSet} \braket{ \stationary | s } \braket{ \overline{X} Y }_{\pdf(X, Y | s)}
~.
\end{align*}
Moreover, the continuous part of the cross-spectral density is given by:
\begin{align*}
P_{XY \text{c}}(\omega) 
= \bigl\langle  \overline{X}_t Y_t \bigr\rangle
 & + \bra{\stationary} \overline{\Omega}_X \, T \, \bigl( e^{i \omega} I - T \bigr)^{-1} \Omega_Y \ket{\one} \\
 & + \bra{\stationary} \Omega_Y \, T \, \bigl( e^{- i \omega} I - T \bigr)^{-1} \overline{\Omega}_X \ket{\one} 
 ~.
\end{align*}
And so on.

\section{Pairwise mutual information example}
\label{sec:PairwiseMI}

For the process generated by the HMM given in Fig.~\ref{fig:Cor1demo}, taking
the limit of ever-narrower Gaussians in the state-conditioned PDFs, so that we
work with pairs of $\delta$-functions, then the process becomes Markovian and the
pairwise mutual information can be calculated exactly:
\begin{align}
\I(X_0 & ; X_\tau) = H (X_0) - H (X_\tau | X_0 ) \nonumber \\
&= H (X_0 , \St_0) - H (X_\tau, \St_\tau | X_0, \St_0 ) \nonumber \\
&= H (\St_0) + \H(X_0 | \St_0) - H (X_\tau, \St_\tau | \St_0 ) \nonumber \\
&= H (\St_0) + \H(X_0 | \St_0) 
    - H ( \St_\tau | \St_0 ) - H (X_\tau | \St_\tau ) \nonumber \\
&= H (\St_0) - H ( \St_\tau | \St_0 ) \nonumber \\
&= H (\stationary) - \sum_{s \in \SSet} \stationary(s) H ( \St_\tau | \St_0 = s
) \nonumber \\
&= H (\stationary) - \sum_{s \in \SSet} \stationary(s) H \bigl( \bra{s} T^\tau
\bigr) \nonumber \\
&= H (\stationary) 
    + \! \! \sum_{s, s' \in \SSet} \! \stationary(s)  \bra{s} T^\tau \ket{s'} \log \bra{s} T^\tau \ket{s'} 
,
\label{eq:PairwiseInfo}
\end{align}
where $\stationary = [ 1, 1-p, 1-p, 1-p ] / (4-3p)$.

Continuing, $\bra{s} T^\tau \ket{s'}$ can be calculated via $T$'s spectral
decomposition. Since $T$ is diagonalizable and nondegenerate for all values of
the transition parameter $p$, we find:
\begin{align*}
\bra{s} T^\tau \ket{s'} = \sum_{\lambda \in \Lambda_T} \lambda^\tau \bra{s} T_\lambda \ket{s'}
  ~.
\end{align*}
Moreover:
\begin{align*}
\bra{s} T_1 \ket{s'} & = \braket{ s | \one } \braket{ \stationary | s'} \\
  & = \stationary(s')
  ~,
\end{align*}
so $\bra{s} T^\tau \ket{s'}$ simplifies somewhat to: 
\begin{align*}
\bra{s} T^\tau \ket{s'} =  \stationary(s')  +  \sum_{\lambda \in \Lambda_T \setminus \{ 1 \} } \lambda^\tau \bra{s} T_\lambda \ket{s'}
  ~
\end{align*}

In fact, Eq.~\eqref{eq:PairwiseInfo} is valid for any set of four PDFs we could
have chosen for the example HMM's states, as long as the PDFs all have mutually
exclusive support for the observable output, since this then makes the hidden
state a function of the instantaneous observable.

\begin{figure}[h]
\begin{center}
\includegraphics[width=0.95\columnwidth]{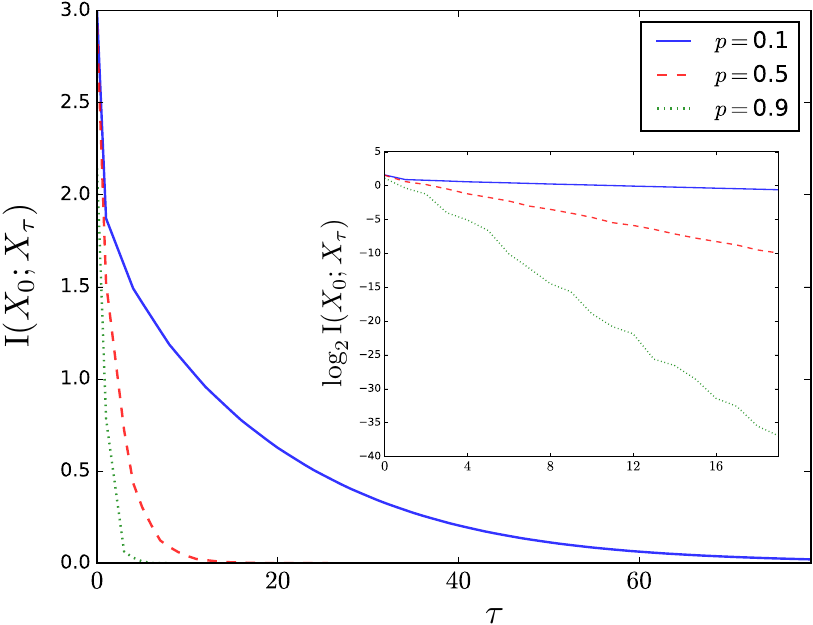}
\end{center}
\caption{Nontrivial pairwise mutual information for the process from Fig.~\ref{fig:Cor1demo} with a flat power spectrum.}
\label{fig:PMI}
\end{figure}

Using the linear algebra of Eq.~\eqref{eq:PairwiseInfo}, 
we calculate the pairwise mutual information and POPI spectrum numerically.
The pairwise mutual informations are shown for $p \in \{ 0.1, 0.5, 0.9\}$
in Fig.~\ref{fig:PMI}.
Reasonably, the loss of information is monotonic over temporal distance.
More surprisingly, the decay of pairwise mutual information is very-nearly exponential as made clear in the inset 
logarithmic plot.

\begin{figure}[h]
\begin{center}
\includegraphics[width=0.95\columnwidth]{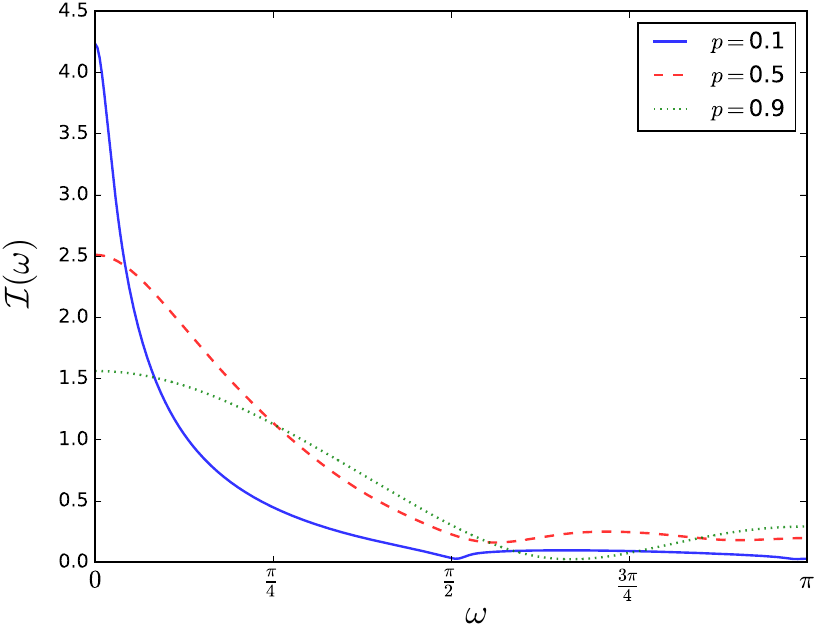}
\end{center}
\caption{Power-of-Pairwise-Information (POPI) spectrum for the process from Fig.~\ref{fig:Cor1demo}.
	}
\label{fig:POPI}
\end{figure}

The POPI spectrum, 
which can be rewritten for a wide-sense stationary process as:
\begin{align*}
\mathcal{I}(\omega) = \lim_{N \to \infty} 2 \sum_{\tau=1}^N \cos(\omega \tau) \I (X_0; X_\tau) ~,
\end{align*}
is shown for these same $p$-values in Fig.~\ref{fig:POPI}.
The POPI spectrum was approximated by truncating the summation 
of modulated pairwise mutual informations at a sufficiently large separation 
of $\tau = 2000$.

\section{Measurement Feedback Models} 
\label{sec:MealyForQuantum}

Let's now turn to describe an alternative set of possibly-input-dependent
models, which may be more convenient for describing certain phenomena. For
example, they are more natural for describing measured quantum systems. They
also reduce to the canonical models used in \emph{computational
mechanics}~\cite{Crut12a, Barn15} after a number of simplifying assumptions.

After introducing them, we show that Thm.~\ref{thm:GenFlatPS} applies to them
as well as to the other model types discussed in the main body. In this way, we
extend the theory of fraudulent white noise to these models as well.

The models we consider generate observable behavior \emph{during transitions}
between states, rather than in the states themselves. This is a natural
approach in the quantum setting since measurement feedback changes the state of
the quantum system with dependence on the measurement outcome. For projective
measurements, measurement fully defines the new state, but for the much more
general class of quantum measurements described by \emph{positive operator
valued measures} (POVMs), the measurement outcome plays a more nuanced role in
updating the state. More generally, edge-emitting models can be natural
descriptors of complex systems with control and feedback. And, fittingly,
edge-emitting models have been used elsewhere as well. For instance, they
appear extensively in computer science and computational mechanics---the latter
of which spans the study of natural computation in physical systems and the
minimal resources required for prediction.

\subsection{Measurement Feedback Models}
\label{sec:MFM}

\newcommand{\mfHMM}{mfHMM}
\newcommand{\mfHMMs}{mfHMMs}

Here we introduce Measurement Feedback Models (MFMs)
$\mathcal{M}_\text{MFM}(\vec{m})$,
which are input-dependent generators of an
observable output process $\{ X_t \}_{t \in \mathcal{T} }$. As before, the
lengths and alphabets of the inputs and outputs need not be commensurate. The
output is generated via $\mathcal{M}_\text{MFM}(\vec{m}) = \bigl( \SSet, \Abet, \{
T_t^{(x)}(\vec{m}) \}_{t \in \mathcal{T}, x \in \Abet } , \boldsymbol{\mu}_1
\bigr)$, where $\SSet$ is the countable set of hidden states, $\Abet$ is the
alphabet of observables, and $ \boldsymbol{\mu}_1$ is the initial distribution
over hidden states. For a given $t$ and $x$, the matrix elements $\bra{s}
T_t^{(x)}(\vec{m}) \ket{s'}$ provide the probability density of transitioning
from state $s$ to $s'$ while emitting the observable $x$; that is:
\begin{align*}
\bra{s} T_t^{(x)}(\vec{m}) \ket{s'} = \pdf_{\vec{m}} ( X_{t+1} = x, \St_{t+1} = s'  | \St_t = s)
  ~.
\end{align*}
where $\pdf_{\vec{m}} $ is the probability density (\emph{induced} by
$\vec{m}$) of the labeled transition. The symbol-labeled transition matrices
$\{ T_t^{(x)}(\vec{m}) \}_{t \in \mathcal{T}, x \in \Abet }$ yield the net
state-to-state transition probabilities when marginalizing over all possible
observations:
\begin{align*}
\int_{x \in \Abet}   T_t^{(x)}(\vec{m})  \, dx = T_t(\vec{m}) 
\end{align*}
where $\bra{s} T_t (\vec{m}) \ket{s'} = \Pr_{\vec{m}} ( \St_{t+1} = s'  | \St_t
= s)$.

\begin{figure}[h]
\begin{center}
\includegraphics[width=0.43\textwidth]{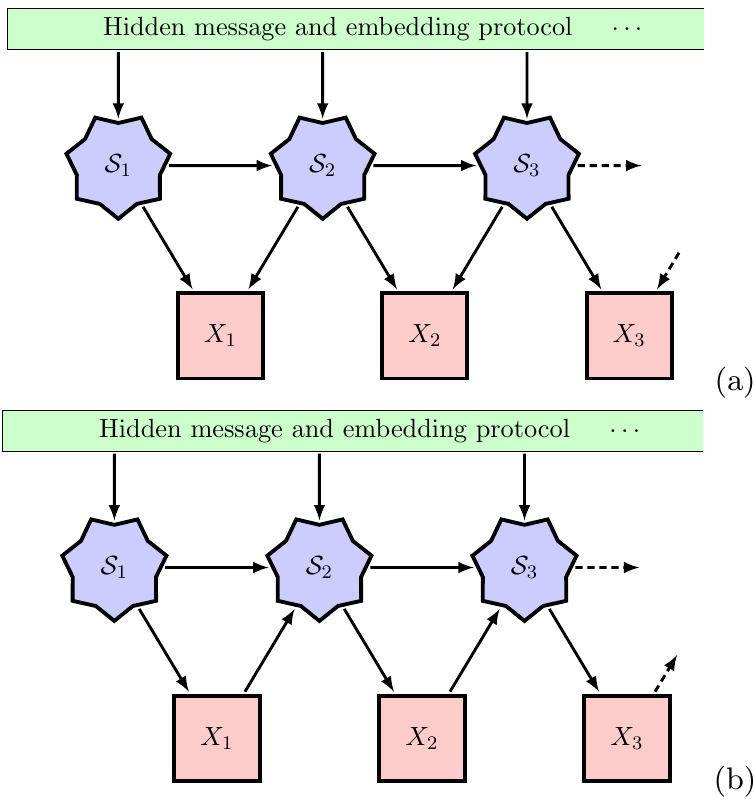}
\end{center}
\caption{Alternative Bayesian networks for measurement feedback models.
  }
\label{fig:AltInputDepGen_BayesNet}
\end{figure}


Figure~\ref{fig:AltInputDepGen_BayesNet} displays two different (but equally valid) 
Bayesian networks
for the decomposition of conditional dependencies among observables and latent states of a MFM. 
Each decomposition suggests a preferred interpretation.
The decomposition of the top panel (a) 
allows identifying a PDF with each directed edge between latent states of a 
measurement feedback model $\mathcal{M}_\text{MFM}(\vec{m})$.
Accordingly, panel (a)
suggests that the transited edge 
determines the probability of the observable.
Whereas, the decomposition of the bottom panel (b)
suggests that the observation determines the probability of the
latent state transition. 
The fact that both decompositions are valid
insists, perhaps surprisingly, that the interpretations have no physical distinction.
The interpretation of causality is ambiguous although 
each calculus of conditional dependencies is reliable.

The measurement feedback models may initially appear rather restrictive when considering the
possibilities of, say, measuring a quantum system in different bases and with
different instruments. However, in principle, the different measurement choices
are incorporated through the different transformations $T_t(\vec{m})$, both
through any pre-determined measurement choices in $\vec{m}$ and through
dynamic-determination via feedback of the measurement outcomes themselves.  

Reference~\cite{Poll18}'s \emph{process tensors} can also be used to model
classical observable processes generated by general quantum dynamics. Although
unnecessarily elaborate for most purposes, process tensors are appealing since
they rigorously incorporate general quantum measurements. Ultimately though,
they, together with a set of ``experiments'' $\vec{m}$, could be mapped onto the
alternative rather-simpler models proposed here, if the goal is only to model
the observable classical output process.

\subsection{Theorem~\ref{thm:GenFlatPS} for Measurement Feedback}
\label{sec:ModThmForAltModels}

The MFM's average-observation matrices are:
\begin{align*}
\Omega_t &=  \int_{x \in \Abet} x \, T_t^{(x)} (\vec{m}) \, dx  ~.
\end{align*}
Notably, they are no longer diagonal in the hidden-state basis. Rather, they
assign to each matrix element the average observation associated with that
transition, multiplied by the probability of the edge being traversed when
conditioned on occupying the outgoing state. That is:
\begin{align}
\bra{s} & \Omega_t \ket{s'} 
  = \int_{x \in \Abet} x \, \pdf_{\vec{m}}( \St_{t+1} = s', X_t = x | \St_t =
  s ) \, dx \nonumber \\
  & =
\Pr_{\vec{m}}(\St_{t+1} = s' | \St_t = s) \nonumber \\
  & \qquad \times \int_{x \in \Abet} x \,
  \pdf_{\vec{m}}( X_t = x | \St_t = s, \St_{t+1} = s') \, dx
  \nonumber \\
&= \bra{s} T_t (\vec{m}) \ket{s'}  \braket{x}_{\pdf_{\vec{m}}(X_t |  \St_t = s, \St_{t+1} = s' )}
  ~.
\label{eq:OmegaEdgeProbRelation}
\end{align}

If the process is wide-sense stationary, then for $\tau > 0$:
\begin{align}
\gamma(\tau) &= \braket{ \boldsymbol{\mu}_t | \overline{\Omega}_t \, T_{t+1:t+\tau}(\vec{m}) \, \Omega_{t+\tau} | \one }
  ~,
\label{eq:AutocorrForAltInputDepModels}
\end{align}
which must be $t$-independent.

For input-independent processes with time-independent transition dynamics---where $T_t^{(x)} (\vec{m}) = T^{(x)}$ and $ \boldsymbol{\mu}_1= \boldsymbol{\pi}$---this simplifies to the
autonomous Mealy-type HMMs with continuous PDFs for the observable associated
with each hidden-state-to-state transition. The autocorrelation function (for $\tau \geq 1$) then
reduces to:
\begin{align*}
\gamma(\tau) &= \braket{ \boldsymbol{\pi} | \overline{\Omega} \, T^{\tau-1} \, \Omega | \one }
  ~,
\end{align*}
while the power spectrum's continuous part is:
\begin{align}
P_\text{c}(\omega) 
= \bigl\langle \left| x \right|^2 \bigr\rangle
 + 2 \, \text{Re} \bra{\stationary} \overline{\Omega} \, \bigl( e^{i \omega} I - T \bigr)^{-1} \Omega \ket{\one} ~.
\label{eq:AltPcwFromResolvent}
\end{align}
Note that this expression lacks $T$, the transition dynamic, when compared to
Eq.~\eqref{eq:PcwFromResolvent}. This follows since $\Omega$ induces a
transition for these Mealy-type HMMs, reducing the number of subsequent
transitions by one.

Let's return to the general setting for autocorrelation given by
Eq.~\eqref{eq:AutocorrForAltInputDepModels} for processes generated by
possibly-input-dependent models. Developing the analog of
Thm.~\ref{thm:GenFlatPS} requires recognizing that the average observation
\emph{on each edge} matters, rather than previously, where the average
observation from each state mattered. For MFMs, constant autocorrelation and
flat power spectrum can again be guaranteed by a rather weak condition: 
The average output of the current edge does not by itself influence the
average output of a future edge.

More explicitly, consider the set of all edges:
\begin{align*}
\boldsymbol{\mathcal{E}}^{(t)} \equiv \bigl\{ (s, s') \in \SSet
\times \SSet : \bra{s} T_t (\vec{m}) \ket{s'} \neq 0 \bigr\}
  ~,
\end{align*}
which are transitions between hidden states that can be traversed at time $t$ with
positive probability. Since outputs occur during edge transitions, we redefine
$\Xi$ as the set of average outputs exhibited by the edges.
Equation~\eqref{eq:OmegaEdgeProbRelation} indicates that the desired definition
is:
\begin{align*}
\Xi \equiv \bigcup_{ t \in \mathcal{T} } \bigcup_{ (s, s') \in \boldsymbol{\mathcal{E}}^{(t)} } \biggl\{  \frac{  \bra{s} \Omega_t \ket{s'}  }{  \bra{s} T_t (\vec{m}) \ket{s'}   }  \biggr\}
  ~.
\end{align*}

Furthermore, we define $\mathcal{E}_t$ to be the random variable for the edge
traversed at time $t$; i.e., $\mathcal{E}_t$ is the joint random variable:
$\mathcal{E}_t = ( \St_t, \St_{t+1} )$. And we define
$\boldsymbol{\mathcal{E}}_{\xi}^{(t)} \subset \boldsymbol{\mathcal{E}}^{(t)} $
as the set of edges (at time $t$) with average output $\xi \in \Xi$: 
\begin{align}
\boldsymbol{\mathcal{E}}_{\xi}^{(t)} 
&\equiv \Bigl\{ (s, s') \in \boldsymbol{\mathcal{E}}^{(t)} :   
\tfrac{  \bra{s} \Omega_t \ket{s'}  }{  \bra{s} T_t (\vec{m}) \ket{s'}   }  = \xi  \Bigr\} ~.
\end{align}
With these in hand, we can state the theorem analogous to Thm.~\ref{thm:GenFlatPS}.

{\The \label{thm:GenFlatPSforAltModels}
Let $\{ X_t \}_t$ be a stochastic process generated by any 
measurement feedback model $\mathcal{M}_\text{MFM}(\vec{m})$,
including autonomous Mealy-type HMMs and input-dependent generators. Such processes have
\emph{constant autocorrelation and a flat power spectrum} if:
\begin{align*}
\Pr(\mathcal{E}_{t+\tau} \in \boldsymbol{\mathcal{E}}_{\xi'}^{(t+\tau)}  | \mathcal{E}_{t} \in \boldsymbol{\mathcal{E}}_{\xi}^{(t)} ) =  \Pr(\mathcal{E}_{t+\tau} \in \boldsymbol{\mathcal{E}}_{\xi'}^{(t+\tau)} )   
\end{align*}
and there exists a constant $c \in \mathbb{C}$ such that:
\begin{align*}
\sum_{\xi \in \Xi} \xi \,  \Pr( \mathcal{E}_{t} \in \boldsymbol{\mathcal{E}}_{\xi}^{(t)} ) = c
  ~,
\end{align*}
for all separations $\tau > 0$, $t \in \mathcal{T}$, and $\xi, \xi' \in \Xi$.
}

{\ProThe 
Starting from Eq.~\eqref{eq:AutocorrForAltInputDepModels}, we find the
autocorrelation for all such processes by calculating:
\begin{widetext}
\begin{align*}
\gamma(\tau) 
&= \braket{ \boldsymbol{\mu}_t | \overline{\Omega}_t \, T_{t+1:t+\tau}(\vec{m}) \, \Omega_{t+\tau} | \one } \\
&= \! \sum_{s,s',s'',s''' \in \SSet} \braket{ \boldsymbol{\mu}_t \ket{s} \bra{s} \overline{\Omega}_t \ket{s'} \bra{s'}  T_{t+1:t+\tau}(\vec{m}) \ket{s''} \bra{s''}  \Omega_{t+\tau} \ket{s'''} \bra{s'''}  \one }   \\
&= \!\!\!\!  
    \sum_{(s,s') \in \boldsymbol{\mathcal{E}}^{(t)} \atop (s',s''') \in \boldsymbol{\mathcal{E}}^{(t+\tau)}  } 
    \Bigl( \tfrac{  \bra{s} \overline{\Omega}_t \ket{s'} }{ \bra{s} T_{t}(\vec{m}) \ket{s'} } \Bigr) \Bigl( \tfrac{ \bra{s''}  \Omega_{t+\tau}  \ket{s'''} }{ \bra{s''}  T_{t+\tau}(\vec{m}) \ket{s'''} } \Bigr)
   \braket{  \boldsymbol{\mu}_t \ket{s}  \bra{s} T_{t}(\vec{m}) \ket{s'} \bra{s'}  T_{t+1:t+\tau}(\vec{m}) \ket{s''} \bra{s''}  T_{t+\tau}(\vec{m}) \ket{s'''} \bra{s'''}  \one } 
 \\
&= \!\!\!\!  
    \sum_{(s,s') \in \boldsymbol{\mathcal{E}}^{(t)} \atop (s',s''') \in \boldsymbol{\mathcal{E}}^{(t+\tau)}  } 
    \Bigl( \frac{  \bra{s} \overline{\Omega}_t \ket{s'} }{ \bra{s} T_{t}(\vec{m}) \ket{s'} } \Bigr) \Bigl( \frac{ \bra{s''}  \Omega_{t+\tau}  \ket{s'''} }{ \bra{s''}  T_{t+\tau}(\vec{m}) \ket{s'''} } \Bigr)
   \Pr(\St_t = s, \St_{t+1} = s', \St_{t+\tau} = s'', \St_{t+\tau+1} = s''')
 \\ 
&= \!\!\!\!  
    \sum_{(s,s') \in \boldsymbol{\mathcal{E}}^{(t)} \atop (s',s''') \in \boldsymbol{\mathcal{E}}^{(t+\tau)}  } 
    \Bigl( \frac{  \bra{s} \overline{\Omega}_t \ket{s'} }{ \bra{s} T_{t}(\vec{m}) \ket{s'} } \Bigr) \Bigl( \frac{ \bra{s''}  \Omega_{t+\tau}  \ket{s'''} }{ \bra{s''}  T_{t+\tau}(\vec{m}) \ket{s'''} } \Bigr)
   \Pr \bigl( \mathcal{E}_t = (s, s'), \mathcal{E}_{t+\tau} = (s'', s''') \bigr)
 \\  
&= 
    \sum_{ \xi, \xi' \in \Xi } 
    \overline{\xi} \xi'
   \Pr \bigl( \mathcal{E}_t \in \boldsymbol{\mathcal{E}}_{\xi}^{(t)}  , \mathcal{E}_{t+\tau} \in \boldsymbol{\mathcal{E}}_{\xi'}^{(t+\tau)}   \bigr)
 \\   
&= 
    \sum_{ \xi \in \Xi } 
    \overline{\xi} 
   \Pr \bigl( \mathcal{E}_t \in \boldsymbol{\mathcal{E}}_{\xi}^{(t)}  \bigr)
   \Bigl[ 
     \sum_{ \xi' \in \Xi } \xi' 
          \Pr \bigl(  \mathcal{E}_{t+\tau} \in \boldsymbol{\mathcal{E}}_{\xi'}^{(t+\tau)} \big| \mathcal{E}_t \in \boldsymbol{\mathcal{E}}_{\xi}^{(t)}  \bigr)
   \Bigr] ~.
\end{align*}
\end{widetext}
Now, suppose that:
\begin{align*}
\Pr(\mathcal{E}_{t+\tau} \in \boldsymbol{\mathcal{E}}_{\xi'}^{(t+\tau)}  | \mathcal{E}_{t} \in \boldsymbol{\mathcal{E}}_{\xi}^{(t)} ) =  \Pr(\mathcal{E}_{t+\tau} \in \boldsymbol{\mathcal{E}}_{\xi'}^{(t+\tau)} )   
\end{align*}
and there exists some constant $c \in \mathbb{C}$ such that:
\begin{align*}
\sum_{\xi \in \Xi} \xi \,  \Pr( \mathcal{E}_{t} \in \boldsymbol{\mathcal{E}}_{\xi}^{(t)} ) = c
  ~,
\end{align*}
for all separations $\tau > 0$, $t \in \mathcal{T}$, and $\xi, \xi' \in \Xi$.
Then, we find:
\begin{align*}
\gamma(\tau) & = \sum_{ \xi \in \Xi } 
    \overline{\xi} 
   \Pr \bigl( \mathcal{E}_t \in \boldsymbol{\mathcal{E}}_{\xi}^{(t)}  \bigr)
   \\
   & \qquad \times
   \Bigl[ 
     \sum_{ \xi' \in \Xi } \xi' 
          \Pr \bigl(  \mathcal{E}_{t+\tau} \in \boldsymbol{\mathcal{E}}_{\xi'}^{(t+\tau)} \big| \mathcal{E}_t \in \boldsymbol{\mathcal{E}}_{\xi}^{(t)}  \bigr)
   \Bigr] \\
  & = \Bigl(
    \sum_{ \xi \in \Xi } 
    \overline{\xi} 
   \Pr \bigl( \mathcal{E}_t \in \boldsymbol{\mathcal{E}}_{\xi}^{(t)}  \bigr)
   \Bigr)
   \Bigl( 
     \sum_{ \xi' \in \Xi } \xi' 
          \Pr \bigl(  \mathcal{E}_{t+\tau}
		  \in \boldsymbol{\mathcal{E}}_{\xi'}^{(t+\tau)}  \bigr)
   \Bigr) \\
&= | c |^2
    ~,
\end{align*}
which is a constant for all separations $\tau > 0$, $t \in \mathcal{T}$, and
$\xi, \xi' \in \Xi$. Finally, a process with stationary low-order statistics
and a flat autocorrelation has a flat power spectrum, as an immediate
consequence of Eq.~\eqref{eq:PSDfromACF}. This proves
Thm.~\ref{thm:GenFlatPSforAltModels}.
}

For the special case of an autonomous HMM that generates observations during
hidden-state-to-state transitions, this condition simplifies significantly.
Specifically, $\Omega_t \to \Omega$ and $T_t(\vec{m}) \to T$ become
$t$-independent, which furthermore means that
$\boldsymbol{\mathcal{E}}_{\xi}^{(t)} \to \boldsymbol{\mathcal{E}}_{\xi}$
becomes $t$-independent. For autonomous wide-sense stationary processes, we
have $\Pr( \mathcal{E}_{t} ) = \Pr( \mathcal{E}_{t + \tau} )$ for all
separations $\tau > 0$ and for all $t \in \mathcal{T}$. It then trivially
follows that $\sum_{\xi \in \Xi} \xi \,  \Pr( \mathcal{E}_{t} \in
\boldsymbol{\mathcal{E}}_{\xi} )$ is constant for all $t \in \mathcal{T}$. So,
the only requirement for an autonomous edge-emitting HMM to produce fraudulent
white noise is that it satisfies the condition:
\begin{align*}
\Pr(\mathcal{E}_{t+\tau} \in \boldsymbol{\mathcal{E}}_{\xi'}  | \mathcal{E}_{t} \in \boldsymbol{\mathcal{E}}_{\xi} ) =  \Pr(\mathcal{E}_{t+\tau} \in \boldsymbol{\mathcal{E}}_{\xi'} )
\end{align*}
for all separations $\tau > 0$, $t \in \mathcal{T}$, and $\xi, \xi' \in \Xi$.

Theorem~\ref{thm:GenFlatPSforAltModels} provides a very general condition for
flat power spectra from measurement feedback models.

\section{Theorem~\ref{thm:GenFlatPS} for time-dependent PDFs}
\label{sec:GenThmForTimeDepPDFs}

Moreover, Thm.~\ref{thm:GenFlatPSforAltModels} suggests how
Thm.~\ref{thm:GenFlatPS} generalizes even further to possibly-input-dependent
hidden-state models with time-dependent PDFs associated with each state. We
will call these \emph{morphing hidden models} (MHMs)
$\mathcal{M}_\text{MHM}(\vec{m})$. MHMs include, as special cases, all models
(Moore-type HMMs and input-dependent generators) considered in the main text.
We employ methods similar to those used in \S~\ref{sec:ModThmForAltModels}.

A MHM is a possibly-input-dependent generator
of an observable output process $\{ X_t \}_{t \in
\mathcal{T}}$.  The output is generated via $\mathcal{M}_\text{MHM}(\vec{m}) =
\bigl( \SSet, \Abet, \{ \mathcal{P}_t(\vec{m}) \}_t, \{ T_t(\vec{m}) \}_t ,
\boldsymbol{\mu}_1 \bigr)$. Here, again, the lengths and alphabets of the
inputs and outputs need not be commensurate. That is, the internal states
$\SSet$ and output alphabet $\Abet$ are static. However, the
hidden-state-to-state transition matrix $ T_t(\vec{m})$---as well as the
state-dependent PDFs $\mathcal{P}_t(\vec{m})$---are time-dependent such that
their values at time $t$ are potentially a function of the full input vector
$\vec{m}$. More specifically, $\mathcal{P}_t(\vec{m})$ is the set of
hidden-state-dependent probability density functions $\pdf_{\vec{m}}( X_t | s
)$ at time $t$. As before, $\boldsymbol{\mu}_1 $ specifies the initial
distribution over hidden states: $\St_1 \sim \boldsymbol{\mu}_1 $.

For such cases, set:
\begin{align*}
\Omega_t 
= \sum_{s \in \SSet} \braket{x}_{\pdf_{\vec{m}}( X_t | s )} \ket{ s } \bra{ s } 
  ~.
\end{align*}
The $\Omega_t$ matrix is time-dependent with the state-conditioned expected outputs along its diagonal.

Since the average state output now varies in time, we must generalize $\Xi$
from its more restricted use in the main text. Specifically, redefine $\Xi$ as
the set of state-dependent average outputs generated throughout time:
\begin{align*}
\Xi \equiv \bigcup_{ t \in \mathcal{T} } \bigcup_{ s \in \SSet} \Bigl\{  \braket{x}_{\pdf_{\vec{m}}( X_t | s )}  \Bigr\}
  ~.
\end{align*}
Furthermore,
we define $\SSet_{\xi}^{(t)} \subset \SSet $ as the set of states (at time $t$) with average 
output $\xi \in \Xi$: 
\begin{align*}
\SSet_{\xi}^{(t)} 
&\equiv \Bigl\{ s \in \SSet :   
\braket{X_t}_{\pdf_{\vec{m}}( X_t | s )} = \xi  \Bigr\} ~.
\end{align*}
Using these, we can state the following theorem, which generalizes
Thm.~\ref{thm:GenFlatPS}.

{\The \label{thm:GenFlatPSforModelswTDepPDFs} 
Let $\{ X_t \}_t$ be a stochastic process generated by 
any morphing hidden model
$\mathcal{M}_\text{MHM}(\vec{m})$. Such processes have
\emph{constant autocorrelation and a flat power spectrum} if:
\begin{align*}
\Pr(\St_{t+\tau} \in \SSet_{\xi'}^{(t+\tau)}  | \St_{t} \in \SSet_{\xi}^{(t)} ) =  \Pr(\St_{t+\tau} \in \SSet_{\xi'}^{(t+\tau)} )
\end{align*}
and there exists a constant $c \in \mathbb{C}$ such that:
\begin{align*}
\sum_{\xi \in \Xi} \xi \,  \Pr( \St_{t} \in \SSet_{\xi}^{(t)} ) = c
  ~,
\end{align*}
for all separations $\tau > 0$, $t \in \mathcal{T}$, and $\xi, \xi' \in \Xi$.
}

{\ProThe
\label{proof:GenFlatPSforInputDepModels_wTdepPDFs}
For the processes under consideration, we find the linear pairwise correlation
(for $\tau \geq 1$) to be:
\begin{widetext}
\begin{align}
\braket{\overline{X_t} X_{t+\tau} }_{\pdf_{\vec{m}}(X_t, X_{t+\tau})} 
&= \braket{ \boldsymbol{\mu}_t | \overline{\Omega}_t \, T_{t:t+\tau}(\vec{m})
\, \Omega_{t + \tau} | \one } \nonumber \\
&= \sum_{s, s' \in \SSet}  \braket{ \boldsymbol{\mu}_t \ket{s} \bra{s}
\overline{\Omega}_t  \ket{s} \bra{s}  T_{t:t+\tau}(\vec{m})  \ket{s'} \bra{s'}
\Omega_{t+\tau}  \ket{s'} \bra{s'}  \one } \nonumber \\
&= \sum_{\xi, \xi' \in \Xi} \overline{\xi} \xi' \sum_{s \in \SSet_{\xi}^{(t)}
\atop s' \in \SSet_{\xi'}^{(t+\tau)}  } \Pr_{\vec{m}}( \St_t = s , \St_{t+\tau}
= s') \nonumber \\
&= \sum_{\xi, \xi' \in \Xi} \overline{\xi} \xi'  \Pr( \St_t \in
\SSet_{\xi}^{(t)} , \St_{t+\tau}  \in \SSet_{\xi'}^{(t+\tau)} ) \nonumber \\
&= \sum_{\xi \in \Xi} \overline{\xi}  \Pr( \St_t \in \SSet_{\xi}^{(t)}  ) 
    \Bigl( \sum_{ \xi' \in \Xi} \xi'    \Pr( \St_{t+\tau}  \in \SSet_{\xi'}^{(t+\tau)} |  \St_t \in \SSet_{\xi}^{(t)})  \Bigr)
  ~.
\label{eq:AutocorrForInputDepModels_wTdepPDFs}
\end{align}
\end{widetext}
Now, suppose that:
\begin{align*}
\Pr(\St_{t+\tau} \in \SSet_{\xi'}^{(t+\tau)}  | \St_{t} \in \SSet_{\xi}^{(t)} ) =  \Pr(\St_{t+\tau} \in \SSet_{\xi'}^{(t+\tau)} )   
\end{align*}
and there exists some constant $c \in \mathbb{C}$ such that:
\begin{align*}
\sum_{\xi \in \Xi} \xi \,  \Pr( \St_{t} \in \SSet_{\xi}^{(t)} ) = c
\end{align*}
for all separations $\tau > 0$, for all $t \in \mathcal{T}$, and for all $\xi,
\xi' \in \Xi$. Then, we find:
\begin{align*}
& \braket{\overline{X_t} X_{t+\tau} }_{\pdf_{\vec{m}}(X_t, X_{t+\tau})} \\
& \quad = \sum_{\xi \in \Xi} \overline{\xi}  \Pr( \St_t \in \SSet_{\xi}^{(t)}  ) 
    \Bigl( \sum_{ \xi' \in \Xi} \! \xi'    \Pr( \St_{t+\tau}  \in \SSet_{\xi'}^{(t+\tau)} |  \St_t \in \SSet_{\xi}^{(t)}) \! \Bigr) \\
& \quad =    
    \Bigl(
        \sum_{\xi \in \Xi} \overline{\xi}  \Pr( \St_t \in \SSet_{\xi}^{(t)}  ) 
    \Bigr)
   \Bigl( 
         \sum_{ \xi' \in \Xi} \xi'    \Pr( \St_{t+\tau}  \in \SSet_{\xi'}^{(t+\tau)} )
   \Bigr) \\
& \quad = | c |^2
  ~.	
\end{align*}
is constant for all $t \in \mathcal{T}, \text{ and } \forall \xi, \xi' \in \Xi$.

That $\braket{\overline{X_t} X_{t+\tau} }_{\pdf_{\vec{m}}(X_t, X_{t+\tau})}$ is
constant verifies that the autocorrelation does not depend on the overall time
shift of the process, so $\braket{\overline{X_t} X_{t+\tau}
}_{\pdf_{\vec{m}}(X_t, X_{t+\tau})} = \gamma(\tau)$. Moreover, $\gamma(\tau)$
is constant. Finally, a process with constant autocorrelation has a flat power
spectrum, as an immediate consequence of Eq.~\eqref{eq:PSDfromACF}. This proves
Thm.~\ref{thm:GenFlatPSforModelswTDepPDFs}.
}

{ \color{blue} 
\section{Analytical Polyspectra}
\label{sec:AnalyticPolyspectra}

This section derives new analytical expressions for polyspectra, revealing
their close relationship with the time-evolution operator's eigenspectrum and
resolvent.

The $(g_0, \dots , g_K)$-polyspectrum is defined as:  
\begin{align}
S_{g_0 , \dots , g_K}  (\omega_1, \dots , \omega_K) 
\equiv 
  \lim_{N \to \infty} \tfrac{1}{N}
    \Biggl\langle 
     \prod_{k=0}^K \widetilde{g_k}^{(N)}(\omega_k) \!
    \Biggr\rangle ,
\label{eq:Polyspectra_def_app}
\end{align}
where $\omega_0 \equiv - \sum_{k=1}^K \omega_k $ and:
\begin{align}
\widetilde{g}^{(N)}(\omega) \equiv \sum_{t=1}^N g(X_t) e^{-i \omega t} ~.
\label{eq:NFT_app}
\end{align}
Each $g_k \colon \Abet \to \mathbb{C}$ can be any function taking observables
to complex numbers.

Combining Eqs.~\eqref{eq:Polyspectra_def_app} and \eqref{eq:NFT_app} yields:
}
\begin{widetext}
{ \color{blue} 
\begin{align}
S_{g_0 , \dots , g_K}  (\omega_1, \dots , \omega_K) 
& = 
  \lim_{N \to \infty} \tfrac{1}{N} 
    \sum_{t_0 = 1}^N  \dots \sum_{t_K = 1}^N
      \Bigl\langle  \prod_{k=0}^{K} g_k(X_{ t_k}) \Bigr\rangle  
      \prod_{k=0}^K e^{-i \omega_k t_k}
    ~.
\label{eq:Polyspectra_expanded_def_app}
\end{align}

The original time variables $(t_k)_{k=0}^K$ induce a function $\alpha \colon \{
0, 1, \dots K \} \to \{ 0, 1, \dots \kappa \}$ that compresses and time-orders
the indices, such that $t_k = t_{\alpha(k)}'$. Since $\alpha$ does not have a
unique inverse, we define the pre-image $\alpha^{-1}(\ell) = \bigl\{ k \in \{
0, 1, \dots K \} : \alpha(k) = \ell \bigr\}$ to be the set of indices that map
to $\ell$.

For HMMs, we then express the expectations in
Eq.~\eqref{eq:Polyspectra_expanded_def_app} as:
\begin{align}
\Bigl\langle \prod_{k=0}^{K} g_k(X_{ t_k}) \Bigr\rangle  
  = \Bigl\langle
  \prod_{\ell=0}^{\kappa} g_{\alpha^{-1}(\ell)}(X_{ t_\ell'}) \Bigr\rangle  
  = \text{tr} \Bigl(  
\ket{\one} \bra{\pi} \Omega_{g_{\alpha^{-1}(0)}} \prod_{\ell=1}^{\kappa} T^{t_{\ell}' - t_{\ell-1}'} \Omega_{g_{\alpha^{-1}(\ell)}}
\label{eq:OrderedExpectationsFromHMM_app}
\Bigr)
  ~,
\end{align}
where tr$(\cdot)$ denotes the trace, the product on the right maintains time
ordering, $g_{\alpha^{-1}(\ell)}(x) \equiv \prod_{k \in \alpha^{-1}(\ell)}
g_k(x)$, and $ \Omega_g \equiv \sum_{s \in \SSet} \braket{g(X)}_{\pdf( X | s )}
\ket{ s } \bra{ s }$.

The summations over all time variables in
Eq.~\eqref{eq:Polyspectra_expanded_def_app} induce all possible functions
$\alpha$ that permute and compress the indices. And, within each compressed
time-ordering, all possible values of the indices consistent with that ordering
are summed over. To enumerate all possible compressed time-orderings, it is
useful to explicitly introduce the set $\mathbb{F}_{K}^{(\kappa)}$ where
$\mathbb{F}_{K}^{(\kappa)}$ is the set of all surjective functions mapping $\{
0, 1, \dots K \}$ onto $\{ 0, 1, \dots \kappa \}$. Then, we can rewrite
Eq.~\eqref{eq:Polyspectra_expanded_def_app} in terms of the new time-ordered
set of variables $(t_0', t_1', \dots t_\kappa')$ where $t_k' - t_{k-1}' > 0$
for all $k > 0$. Dropping the prime on the $t_k'$ variables, we obtain:
\begin{align}
S_{g_0 , \dots , g_K}  (\omega_1, \dots , \omega_K) 
  & = \lim_{N \to \infty} \tfrac{1}{N} \sum_{\kappa = 0}^{K}
  \sum_{\alpha \in \mathbb{F}_K^{(\kappa)}}
    \sum_{t_0 = 1}^{N-\kappa}  \sum_{t_1 = t_0+1}^{N-\kappa + 1}
    \dots \!\!\!  \sum_{t_\kappa = t_{\kappa-1} + 1}^{N}
      \Bigl\langle  \prod_{k=0}^{\kappa} g_{\alpha^{-1}(k)}(X_{ t_k}) \Bigr\rangle  
      \prod_{k=0}^\kappa e^{-i \omega_{\alpha^{-1}(k)} t_k}
    ~,
\label{eq:Polyspectra_shuffled_def_app}
\end{align}
where 
$\omega_{\alpha^{-1}(k)} \equiv \sum_{\ell \in \alpha^{-1}(k)} \omega_\ell$.

The manifest time-ordering in Eq.~\eqref{eq:Polyspectra_shuffled_def_app}
allows us to use Eq.~\eqref{eq:OrderedExpectationsFromHMM_app} for the
expectation. It is convenient to rewrite this as:
\begin{align}
\Bigl\langle  \prod_{k=0}^{\kappa} g_{\alpha^{-1}(k)}(X_{ t_k}) \Bigr\rangle  
&= 
\text{tr} \Biggl(  
\ket{\one} \bra{\pi} \Omega_{ g_{\alpha^{-1}(0)}} T^{-t_0}  \Bigl( \prod_{k=1}^{\kappa - 1} T^{t_k} 
\Omega_{g_{\alpha^{-1}(k)}}
T^{ - t_k} 
\Bigr)
T^{ t_\kappa} 
\Omega_{g_{\alpha^{-1}(\kappa)}}
\Biggr) 
    ~.
\label{eq:GreatExpectations_app}
\end{align}
Technically assumes that the index of the transition matrix is bounded by
$\nu_0(T) \leq t_k - t_{k-1}$. This assumption is valid, for example, if $T$ is
not singular. Otherwise, a slight modification of the derivation is required, 
where the zero eigenspace is treated separately, which we out.

Plugging Eq.~\eqref{eq:GreatExpectations_app} back into
Eq.~\eqref{eq:Polyspectra_shuffled_def_app} consolidates and eventually
eliminates the $t_k$ dependencies, starting with $t_\kappa$. To see this, we
introduce $\mathfrak{z}_k^{(\alpha)} \equiv e^{-i \omega_{\alpha^{-1}(k)} }$
and rearrange terms: 
\begin{align}
& 
S_{g_0 , \dots , g_K}  (\omega_1, \dots , \omega_K) 
\nonumber \\
& = 
  \lim_{N \to \infty} \tfrac{1}{N} 
  \sum_{\kappa = 0}^{K}
  \sum_{\alpha \in \mathbb{F}_K^{(\kappa)}}
    \sum_{t_0 = 1}^{N-\kappa}  
    \sum_{t_1 = t_0+1}^{N-\kappa + 1}
    \! \dots \!\!\!\!  \sum_{t_\kappa = t_{\kappa-1} + 1}^{N}
      \!\text{tr} \Biggl(  
\ket{\one} \bra{\pi} \Omega_{ g_{\alpha^{-1}(0)}} T^{-t_0}  \Bigl( \prod_{k=1}^{\kappa - 1} T^{t_k} 
\Omega_{g_{\alpha^{-1}(k)}}
T^{ - t_k} 
\Bigr)
T^{ t_\kappa} 
\Omega_{g_{\alpha^{-1}(\kappa)}}
\Biggr) 
      \prod_{k=0}^\kappa e^{-i \omega_{\alpha^{-1}(k)} t_k}
  \nonumber \\
&=
  \lim_{N \to \infty} \tfrac{1}{N} 
  \sum_{\kappa = 0}^{K}
  \sum_{\alpha \in \mathbb{F}_K^{(\kappa)}}
    \sum_{t_0 = 1}^{N-\kappa}  
    \sum_{t_1 = t_0+1}^{N-\kappa + 1}
    \! \dots \!\!\!\!  \sum_{t_\kappa = t_{\kappa-1} + 1}^{N}
     \bra{\pi} \Omega_{ g_{\alpha^{-1}(0)}} T^{-t_0}  \Bigl( \prod_{k=1}^{\kappa - 1} 
(\mathfrak{z}_k^{(\alpha)} T)^{t_k} 
\Omega_{g_{\alpha^{-1}(k)}}
 T^{ - t_k} 
\Bigr)
( \mathfrak{z}_\kappa^{(\alpha)} T)^{ t_\kappa} 
\Omega_{g_{\alpha^{-1}(\kappa)}}
\ket{\one} 
      (\mathfrak{z}_0^{(\alpha)})^{t_0}   
  \nonumber \\
&=
  \lim_{N \to \infty} \tfrac{1}{N} 
  \sum_{\kappa = 0}^{K}
  \sum_{\alpha \in \mathbb{F}_K^{(\kappa)}}
    \sum_{t_0 = 1}^{N-\kappa}  
    \dots \!\!\!
    \sum_{t_\ell = t_{\ell-1}+1}^{N-\kappa + \ell}
    \! \dots \!\!\!\!  \sum_{t_{\kappa - 1} = t_{\kappa-2} + 1}^{N-1}
     \bra{\pi} \Omega_{ g_{\alpha^{-1}(0)}} T^{-t_0}  \Bigl( \prod_{k=1}^{\kappa - 1} 
(\mathfrak{z}_k^{(\alpha)} T)^{t_k} 
\Omega_{g_{\alpha^{-1}(k)}}
 T^{ - t_k} 
\Bigr)
  \nonumber \\
  & \qquad \qquad \times 
\Bigl(
\sum_{t_\kappa = t_{\kappa-1} + 1}^{N}
( \mathfrak{z}_\kappa^{(\alpha)} T)^{ t_\kappa} 
\Bigr)
\Omega_{g_{\alpha^{-1}(\kappa)}}
\ket{\one} 
      (\mathfrak{z}_0^{(\alpha)})^{t_0}   
    ~.
\label{eq:Polyspectra_expanded_app}
\end{align}

This results in summations of the form $\sum_{t=a}^b (zT)^t$. It is always true
that $(I - z T) \sum_{t=a}^b (zT)^t = (zT)^a - (zT)^{b+1}$. Hence, when $z^{-1}
\notin \Lambda_T$, the operator $(I - z T)$ can be inverted to yield:
\begin{align*}
\sum_{t=a}^b (zT)^t = (I - z T)^{-1} \bigl( (zT)^a - (zT)^{b+1} \bigr)
  ~.
\end{align*}
The first such summation is:
\begin{align}
\sum_{t_\kappa = t_{\kappa-1} + 1}^{N}
( \mathfrak{z}_\kappa^{(\alpha)} T)^{ t_\kappa} 
&=
 (I - \mathfrak{z}_\kappa^{(\alpha)}  T)^{-1} 
 \bigl( ( \mathfrak{z}_\kappa^{(\alpha)}  T)^{ t_{\kappa-1} + 1} 
 - ( \mathfrak{z}_\kappa^{(\alpha)} T)^{N+1} \bigr) 
  \nonumber \\
&= 
 ( \mathfrak{z}_\kappa^{(\alpha)} T)^{t_{\kappa-1}} 
     T
     \bigl( I / \mathfrak{z}_\kappa^{(\alpha)} - T  \bigr)^{-1}
 -  (I - \mathfrak{z}_\kappa^{(\alpha)}  T)^{-1}
 ( \mathfrak{z}_\kappa^{(\alpha)} T)^{N+1}   
  ~.
\label{eq:InitialSummation2Terms_app}
\end{align}

As $N \to \infty$ the contribution from the rightmost term vanishes. In
particular, since $(e^{-i \omega} T)^N = e^{- i \omega N} T^N$ can be
rewritten as:
\begin{align*}
\sum_{\lambda \in \Lambda_{\rho(T)} } (\lambda / e^{i \omega})^N  T_\lambda 
  ~.
\end{align*}
the contribution from the decaying eigenmodes (with eigenvalue magnitude less
than unity) vanishes as $N \to \infty$. However, for eigenvalues $\lambda$ on
the unit circle, $(\lambda / e^{i \omega})^N$ does not converge for generic
$\omega$ as $N \to \infty$. Therefore, if the polyspectrum is to be
well-behaved in the $N \to \infty$ limit, the contribution from these terms
also must vanish.

The surviving term, leftmost in Eq.~\eqref{eq:InitialSummation2Terms_app},
conveniently has a $T^{t_{\kappa-1}}$ operator on the lefthand side that
cancels with the $T^{- t_{\kappa-1}}$ operator in
Eq.~\eqref{eq:Polyspectra_expanded_app}. In effect for:
\begin{align*}
\Bigl( \prod_{k=1}^{\kappa - 1} 
(\mathfrak{z}_k^{(\alpha)} T)^{t_k} 
\Omega_{g_{\alpha^{-1}(k)}}
 T^{ - t_k} 
\Bigr)
\Bigl(
\sum_{t_\kappa = t_{\kappa-1} + 1}^{N}
( \mathfrak{z}_\kappa^{(\alpha)} T)^{ t_\kappa} 
\Bigr) 
\end{align*}
we substitute:
\begin{align*}
\Bigl( \prod_{k=1}^{\kappa - 2} 
(\mathfrak{z}_k^{(\alpha)} T)^{t_k} 
\Omega_{g_{\alpha^{-1}(k)}}
 T^{ - t_k} 
\Bigr)
( \mathfrak{z}_{\kappa-1:\kappa}^{(\alpha)} T)^{ t_{\kappa-1}} 
\Omega_{g_{\alpha^{-1}(\kappa-1)}}
 T
 \bigl( I / \mathfrak{z}_\kappa^{(\alpha)} - T  \bigr)^{-1} ~,
\end{align*}
where $\mathfrak{z}_{\ell:\kappa}^{(\alpha)} \equiv \prod_{k = \ell }^\kappa
\mathfrak{z}_{k}^{(\alpha)} = e^{-i \sum_{k = \ell}^{\kappa}
\omega_{\alpha^{-1}(k)} }$. The $t_{\kappa-1}$ term can now be summed over in
the same fashion as just done for the $t_{\kappa}$ term. This summation and
annihilation proceeds recursively to yield a surprisingly concise closed-form
solution for any polyspectrum.

To carry out the specified recursion, we note that each new summation is of the
form:
\begin{align*}
\sum_{t_\ell = t_{\ell-1}+1}^{N-\kappa + \ell}
  ( \mathfrak{z}_{\ell:\kappa}^{(\alpha)} T)^{ t_{\ell}} 
&=
 (I -  \mathfrak{z}_{\ell:\kappa}^{(\alpha)}  T)^{-1} 
 \bigl( (  \mathfrak{z}_{\ell:\kappa}^{(\alpha)}  T)^{ t_{\ell-1} + 1} 
 - (  \mathfrak{z}_{\ell:\kappa}^{(\alpha)}  T)^{N-\kappa + \ell+1} \bigr) 
\\
&= 
 (  \mathfrak{z}_{\ell:\kappa}^{(\alpha)}  T)^{t_{\ell-1}}   T
     \bigl( I /  \mathfrak{z}_{\ell:\kappa}^{(\alpha)}  - T  \bigr)^{-1}
 -  (I -  \mathfrak{z}_{\ell:\kappa}^{(\alpha)}   T)^{-1}   (  \mathfrak{z}_{\ell:\kappa}^{(\alpha)}  T)^{N-\kappa + \ell+1}  
\\
& \to_{N \to \infty}
 (  \mathfrak{z}_{\ell:\kappa}^{(\alpha)}  T)^{t_{\ell-1}}   T
     \bigl( I /  \mathfrak{z}_{\ell:\kappa}^{(\alpha)}  - T  \bigr)^{-1} 
  ~.
\end{align*}
This provides the desired annihilation with $T^{-t_{\ell-1}}$, allowing the
recursion. (Again, the contribution of the rightmost term vanishes for generic
$\omega$ in the $N \to \infty$ limit.)

As an intermediate step in this recursive procedure, we obtain:
\begin{align*}
S_{g_0 , \dots , g_K}  (\omega_1, \dots , \omega_K) 
  &=
  \lim_{N \to \infty} \tfrac{1}{N} 
  \sum_{\kappa = 0}^{K}
  \sum_{\alpha \in \mathbb{F}_K^{(\kappa)}}
    \sum_{t_0 = 1}^{N-\kappa}  
    \dots \!\!\!
    \sum_{t_\ell = t_{\ell-1}+1}^{N-\kappa + \ell}
     \bra{\pi} \Omega_{ g_{\alpha^{-1}(0)}} T^{-t_0}  \Bigl( \prod_{k=1}^{\ell} 
(\mathfrak{z}_k^{(\alpha)} T)^{t_k} 
\Omega_{g_{\alpha^{-1}(k)}}
 T^{ - t_k} 
\Bigr)
( \mathfrak{z}_{\ell+1: \kappa}^{(\alpha)} T)^{ t_\ell} 
  \\
  & \qquad \times \Bigl( 
\prod_{k = \ell+1}^{\kappa}   T
   \bigl( I /  \mathfrak{z}_{k:\kappa}^{(\alpha)}  - T  \bigr)^{-1} 
   \Omega_{g_{\alpha^{-1}(k)}}
\Bigr)
\ket{\one} 
      (\mathfrak{z}_0^{(\alpha)})^{t_0}   
\\
&=
  \lim_{N \to \infty} \tfrac{1}{N} 
  \sum_{\kappa = 0}^{K} \!
  \sum_{\alpha \in \mathbb{F}_K^{(\kappa)}}
    \sum_{t_0 = 1}^{N-\kappa}  
    \dots \!\!\!
    \sum_{t_\ell = t_{\ell-1}+1}^{N-\kappa + \ell} \!
     \bra{\pi} \Omega_{ g_{\alpha^{-1}(0)}} T^{-t_0}  \Bigl( \prod_{k=1}^{\ell-1} 
(\mathfrak{z}_k^{(\alpha)} T)^{t_k} 
\Omega_{g_{\alpha^{-1}(k)}}
 T^{ - t_k} 
\Bigr)
( \mathfrak{z}_{\ell: \kappa}^{(\alpha)} T)^{ t_\ell} \Omega_{f_{\alpha^{-1}(\ell)}}
  \\
  & \qquad \times
\Bigl( 
\prod_{k = \ell+1}^{\kappa}    T
   \bigl( I /  \mathfrak{z}_{k:\kappa}^{(\alpha)}  - T  \bigr)^{-1} 
   \Omega_{g_{\alpha^{-1}(k)}}
\Bigr)
\ket{\one} 
      (\mathfrak{z}_0^{(\alpha)})^{t_0}   
\\
&=
  \lim_{N \to \infty} \tfrac{1}{N} 
  \sum_{\kappa = 0}^{K}
  \sum_{\alpha \in \mathbb{F}_K^{(\kappa)}}
    \sum_{t_0 = 1}^{N-\kappa}  
    \dots \!\!\!\!
    \sum_{t_{\ell-1} = t_{\ell-2}+1}^{N-\kappa + \ell - 1} \!\!
     \bra{\pi} \Omega_{ g_{\alpha^{-1}(0)}} T^{-t_0}  \Bigl( \prod_{k=1}^{\ell-1} 
(\mathfrak{z}_k^{(\alpha)} T)^{t_k} 
\Omega_{g_{\alpha^{-1}(k)}}
 T^{ - t_k} 
\Bigr)
( \mathfrak{z}_{\ell: \kappa}^{(\alpha)} T)^{ t_{\ell-1}} 
  \\
  & \qquad \times
\Bigl( 
\prod_{k = \ell}^{\kappa}    T
   \bigl( I /  \mathfrak{z}_{k:\kappa}^{(\alpha)}  - T  \bigr)^{-1} 
   \Omega_{g_{\alpha^{-1}(k)}}
\Bigr)
\ket{\one} 
      (\mathfrak{z}_0^{(\alpha)})^{t_0}         
    ~.
\label{eq:Polyspectra_mid_recursion_app}
\end{align*}

Eventually only the $t_0$ summation remains:
\begin{align*}
S_{g_0 , \dots , g_K}  (\omega_1, \dots , \omega_K) 
  & = \lim_{N \to \infty} \tfrac{1}{N} \sum_{\kappa = 0}^{K} \!
  \sum_{\alpha \in \mathbb{F}_K^{(\kappa)}} \!
    \sum_{t_0 = 1}^{N-\kappa}  
    \sum_{t_1 = t_0+1}^{N-\kappa + 1}
     \bra{\pi} \Omega_{ g_{\alpha^{-1}(0)}} T^{-t_0}  \Bigl( 
(\mathfrak{z}_k^{(\alpha)} T)^{t_1} 
\Omega_{g_{\alpha^{-1}(1)}} T^{ - t_1} \Bigr)
  \\
  & \qquad \times ( \mathfrak{z}_{2: \kappa}^{(\alpha)} T)^{ t_{1}} 
\Bigl( \prod_{k = 2}^{\kappa}    T
   \bigl( I /  \mathfrak{z}_{k:\kappa}^{(\alpha)}  - T  \bigr)^{-1} 
   \Omega_{g_{\alpha^{-1}(k)}}
\Bigr)
\ket{\one} (\mathfrak{z}_0^{(\alpha)})^{t_0}         
\\
  & = \lim_{N \to \infty} \tfrac{1}{N} \sum_{\kappa = 0}^{K}
  \sum_{\alpha \in \mathbb{F}_K^{(\kappa)}}
    \sum_{t_0 = 1}^{N-\kappa}  
     \bra{\pi} \Omega_{ g_{\alpha^{-1}(0)}} T^{-t_0} 
   ( \mathfrak{z}_{1: \kappa}^{(\alpha)} T)^{ t_{0}} 
\Bigl( \prod_{k = 1}^{\kappa}    T
   \bigl( I /  \mathfrak{z}_{k:\kappa}^{(\alpha)}  - T  \bigr)^{-1} 
   \Omega_{g_{\alpha^{-1}(k)}}
\Bigr)
\ket{\one} (\mathfrak{z}_0^{(\alpha)})^{t_0}      
  \\
  & = \lim_{N \to \infty} \tfrac{1}{N} \sum_{\kappa = 0}^{K}
  \sum_{\alpha \in \mathbb{F}_K^{(\kappa)}}
     \bra{\pi} \Omega_{ g_{\alpha^{-1}(0)}} 
\Bigl( \prod_{k = 1}^{\kappa}    T
   \bigl( I /  \mathfrak{z}_{k:\kappa}^{(\alpha)}  - T  \bigr)^{-1} 
   \Omega_{g_{\alpha^{-1}(k)}}
\Bigr)
\ket{\one} 
\sum_{t_0 = 1}^{N-\kappa}  
      (\mathfrak{z}_{0:\kappa}^{(\alpha)})^{t_0}                 
  ~.
\end{align*}

It is now crucial to notice that:
\begin{align*}
\mathfrak{z}_{0:\kappa}^{(\alpha)}
  & = e^{-i \sum_{k = 0}^{\kappa} \omega_{\alpha^{-1}(k)} } \\
  & = e^{-i \sum_{k = 0}^{K} \omega_{k} } \\
  & = e^{i 0} \\
  & = 1
  ~, 
\end{align*}
since $\omega_0 = - \sum_{k=1}^K \omega_k$. Accordingly, the summation over
$t_0$ becomes:
\begin{align*}
\sum_{t_0 = 1}^{N-\kappa}  (\mathfrak{z}_{0:\kappa}^{(\alpha)})^{t_0}      
  & = \sum_{t_0 = 1}^{N-\kappa} 1^{t_0} \\
  & = \sum_{t_0 = 1}^{N-\kappa} 1 \\
  & = N - \kappa
    ~.
\end{align*}
Thus, for finite $K$, the continuous part of the $(g_0, \dots ,
g_K)$-polyspectrum has the closed-form expression:
\begin{align}
S_{g_0 , \dots , g_K}  (\omega_1, \dots , \omega_K) 
  &  = \sum_{\kappa = 0}^{K}
  \sum_{\alpha \in \mathbb{F}_K^{(\kappa)}}
     \bra{\pi} \Omega_{ g_{\alpha^{-1}(0)}} 
\Bigl( \prod_{\ell = 1}^{\kappa}    T
   \bigl( I /  \mathfrak{z}_{\ell:\kappa}^{(\alpha)}  - T  \bigr)^{-1} 
   \Omega_{g_{\alpha^{-1}(\ell)}}
\Bigr)
\ket{\one} \left( \lim_{N \to \infty} \frac{N - \kappa}{N} \right)
  \\
  & = \sum_{\kappa = 0}^{K} \sum_{\alpha \in \mathbb{F}_K^{(\kappa)}}
     \bra{\pi} \Omega_{ g_{\alpha^{-1}(0)}} 
\Bigl( 
\prod_{\ell = 1}^{\kappa}    T
   \bigl( I /  \mathfrak{z}_{\ell:\kappa}^{(\alpha)}  - T  \bigr)^{-1} 
   \Omega_{g_{\alpha^{-1}(\ell)}}
\Bigr)
\ket{\one}           
    ~.
\label{eq:Polyspectra_closed_form_app}
\end{align}

Note that there are possible contributions to the discrete part of the $(g_0,
\dots , g_K)$-polyspectrum wherever $1 / \mathfrak{z}_{k:\kappa}^{(\alpha)}
\in \Lambda_T$. This coincides with $\kappa$-wise products of $T$'s eigenvalues
on the unit circle. Moreover, this coincides with those eigenvalues of
$\bigotimes_{k=1}^\kappa T$ that lie on the unit circle.

That a tensor product appears here tells us that polyspectra reflect properties
of the transition matrix that unfold over time.

\subsection{(Eigen)Spectral expansion of polyspectra}
\label{sec:EigenPolySpectra}

Using Eq.~\eqref{eq:ResolventPartialFractExpansion} to express the resolvent
$\bigl( I /  \mathfrak{z}_{\ell:\kappa}^{(\alpha)}  - T  \bigr)^{-1} $ in terms
of the transition-matrix eigenvalues and spectral projection operators:
\begin{align} 
( I /  \mathfrak{z}_{\ell:\kappa}^{(\alpha)} - T)^{-1}
  & = \sum_{\lambda \in \Lambda_T} \sum_{m = 0}^{\nu_\lambda - 1}
  \frac{1}{(1 / \mathfrak{z}_{\ell:\kappa}^{(\alpha)} - \lambda)^{m+1}}  T_{\lambda,m}
  ~,
\label{eq:ResolventPartialFractExpansion2}
\end{align}
we again see that the time-evolution operator $T$'s eigenspectrum directly
controls the process' polyspectrum. Furthermore, recall that $T =
\sum_{\lambda} \bigl( \lambda T_{\lambda, 0} + T_{\lambda, 1}  \bigr)$ and
$T_{\lambda, m} T_{\zeta, n} = \delta_{\lambda, \zeta} T_{\lambda, m+n}$. With
this we find:
\begin{align}
& S_{g_0 , \dots , g_K}  (\omega_1, \dots , \omega_K)  \nonumber \\
  & \qquad  = \sum_{\kappa = 0}^{K} \sum_{\alpha \in \mathbb{F}_K^{(\kappa)}}
    \sum_{\lambda_1 \in \Lambda_T} \sum_{m_1 = 0}^{\nu_{\lambda_1} - 1}
    \sum_{\lambda_2 \in \Lambda_T} \sum_{m_2 = 0}^{\nu_{\lambda_2} - 1}
    \dots
    \sum_{\lambda_\kappa \in \Lambda_T}
	\sum_{m_\kappa = 0}^{\nu_{\lambda_\kappa} - 1}
  \frac{\bra{\pi} \Omega_{ g_{\alpha^{-1}(0)}} 
\Bigl( \prod_{\ell = 1}^{\kappa}    T
    T_{\lambda_j , m_j}     
   \Omega_{g_{\alpha^{-1}(\ell)}}
\Bigr)
  \ket{\one} }{\prod_{\ell = 1}^{\kappa}
  (1 /  \mathfrak{z}_{\ell:\kappa}^{(\alpha)} - \lambda_j)^{m_j+1}}        
  .
\label{eq:Polyspectra_spectral_closed_form_app}
\end{align}
For a diagonalizable transition matrix $T$, this reduces to:
\begin{align}
S_{g_0 , \dots , g_K}  (\omega_1, \dots , \omega_K) 
&=
  \sum_{\kappa = 0}^{K}
  \sum_{\alpha \in \mathbb{F}_K^{(\kappa)}}
    \sum_{\lambda_1 \in \Lambda_T} 
    \sum_{\lambda_2 \in \Lambda_T} 
    \dots
    \sum_{\lambda_\kappa \in \Lambda_T} 
  \frac{\bra{\pi} \Omega_{ g_{\alpha^{-1}(0)}} 
\Bigl( 
\prod_{\ell = 1}^{\kappa}    T
    T_{\lambda_j }     
   \Omega_{g_{\alpha^{-1}(\ell)}}
\Bigr)
\ket{\one} }{\prod_{\ell = 1}^{\kappa}  (1 /  \mathfrak{z}_{\ell:\kappa}^{(\alpha)} - \lambda_j)}        
    ~.
\label{eq:Polyspectra_diagonable_spectral_closed_form_app}
\end{align}

\subsection{Polyspectra examples}
\label{sec:PolyExamples}

It is instructive to explore several special cases of the $(g_0, \dots ,
g_K)$-polyspectrum. To aid in this, we explicitly construct the surjective
function sets $\mathbb{F}_1^{(0)}$, $\mathbb{F}_1^{(1)}$, $\mathbb{F}_2^{(0)}$,
$\mathbb{F}_2^{(1)}$, and $\mathbb{F}_2^{(2)}$, shown in
Fig.~\ref{fig:FunctionSets}.

\begin{figure}[t]
\begin{center}
\includegraphics[width=0.7\textwidth]{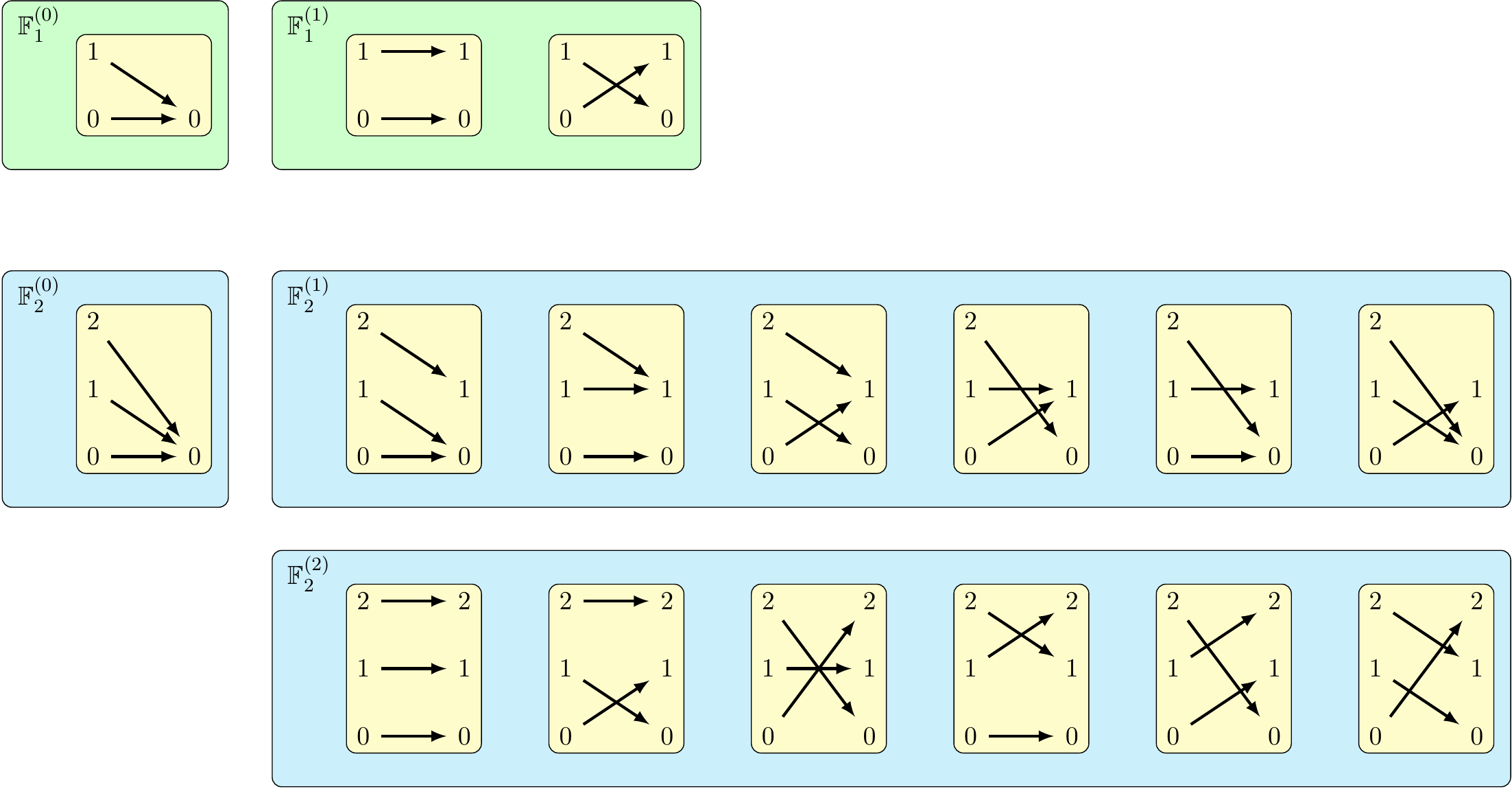}
\end{center}
\caption{Five examples of $\mathbb{F}_K^{(\kappa)}$: Each is a set of
	surjective functions, relevant for constructing polyspectra. The two
	sets $\mathbb{F}_1^{(0)}$ and $\mathbb{F}_1^{(1)}$ are needed to construct
	general $(g_0, g_1)$-polyspectra $S_{(g_0, g_1)}(\omega_1)$. The three sets
	$\mathbb{F}_2^{(0)}$, $\mathbb{F}_2^{(1)}$, and $\mathbb{F}_2^{(2)}$ are
	needed to construct general $(g_0, g_1, g_2)$-polyspectra $S_{(g_0, g_1,
	g_2)}(\omega_1, \omega_2)$.
	}
\label{fig:FunctionSets}
\end{figure}

First consider the $(\overline{X}, X)$-polyspectrum, $S_{\overline{X},
X}(\omega_1)$, which is simply the power spectrum $P(\omega_1)$. In this case,
$K=1$. So, we consider the functions contained in $\mathbb{F}_1^{(0)} = $
\Fonezeroapp and $\mathbb{F}_1^{(1)} = $ \Foneoneapp. For the compressive
function $\alpha = $ \FonezeroAapp, we obtain $\alpha^{-1}(0) = \{ 0, 1 \}$,
yielding:
\begin{align*}
\Omega_{g_{\alpha^{-1}(0) }} & = \Omega_{g_{ \{ 0, 1 \} }} \\
  & = \Omega_{|X|^2} \\
  &  = \sum_{s \in \SSet} \braket{|X|^2}_{\pdf( X | s )} \ket{ s } \bra{ s }
  ~.
\end{align*}
The $(\kappa=0)$-contribution to the power spectrum is thus:
\begin{align*}
\bra{\stationary} \Omega_{|X|^2} \ket{\one} = \sum_{s \in \SSet}
\braket{|X|^2}_{\pdf( X | s )} \braket{ \stationary | s } = \braket{ | x |^2
}
  ~,
\end{align*}
which is indeed the first term in Eq.~\eqref{eq:PcwFromResolvent}. The
$(\kappa=1)$-contribution to the power spectrum is:
\begin{align*}
\sum_{\alpha \in \mathbb{F}_1^{(1)}} \!
     \bra{\stationary} \Omega_{ g_{\alpha^{-1}(0)}}    T
   \bigl( e^{i \omega_{\alpha^{-1}(1)}} I  - T  \bigr)^{-1} 
   \Omega_{g_{\alpha^{-1}(1)}} \!
\ket{\one}
  ~,
\end{align*}
where it should be recalled that $\omega_0 = -\omega_1$.
Plugging in the identity and swap functions of $\mathbb{F}_1^{(1)}$,
this becomes:
\begin{align*}
\bra{\stationary} \Omega_{ g_{0}}    T
   \bigl( e^{i \omega_{1}} I  - T  \bigr)^{-1} 
   \Omega_{g_{1}} \!  \ket{\one}
  & + 
\bra{\stationary} \Omega_{ g_{1}}    T
   \bigl( e^{i \omega_{0}} I  - T  \bigr)^{-1} 
   \Omega_{g_{0}} \! \ket{\one} 
   \\
   & \quad = \bra{\stationary} \Omega_{ \overline{X}}    T
   \bigl( e^{i \omega_{1}} I  - T  \bigr)^{-1} 
   \Omega_{X} \!  \ket{\one} + 
\bra{\stationary} \Omega_{ X}    T
   \bigl( e^{ - i \omega_{1}} I  - T  \bigr)^{-1} 
   \Omega_{\overline{X}} \! \ket{\one} 
   \\
   & \quad = 2 \text{Re} 
\bra{\stationary} \Omega_{ \overline{X}}    T
   \bigl( e^{i \omega_{1}} I  - T  \bigr)^{-1} 
   \Omega_{X} \! \ket{\one} 
  ~,
\end{align*}
which is indeed the last term of Eq.~\eqref{eq:PcwFromResolvent}.

To see the general structure of other polyspectra, it is helpful to expand the first few $\kappa$ terms of the general polyspectra analytic expression Eq.~\eqref{eq:Polyspectra_closed_form_app}. Explicitly expanding the $\kappa$ terms from $0$ to $2$ yields:
\begin{align}
S_{g_0 , \dots , g_K}  (\omega_1, \dots , \omega_K) 
&=
  \sum_{\kappa = 0}^{K}
  \sum_{\alpha \in \mathbb{F}_K^{(\kappa)}}
     \bra{\pi} \Omega_{ g_{\alpha^{-1}(0)}} 
\Bigl( 
\prod_{\ell = 1}^{\kappa}    T
   \bigl( I /  \mathfrak{z}_{\ell:\kappa}^{(\alpha)}  - T  \bigr)^{-1} 
   \Omega_{g_{\alpha^{-1}(\ell)}}
\Bigr)
\ket{\one}           
\nonumber \\
& = 
     \bra{\pi} \Omega_{ g_{ \{ 0, 1, \dots , K \} }} 
\ket{\one}  
+ 
\biggl(
  \sum_{\alpha \in \mathbb{F}_K^{(1)}}
     \bra{\pi} \Omega_{ g_{\alpha^{-1}(0)}} 
  T
   \bigl( e^{i \omega_{\alpha^{-1}(1)} } I  - T  \bigr)^{-1} 
   \Omega_{g_{\alpha^{-1}(1)}}
\ket{\one} 
\biggr)
 \nonumber \\
& \qquad + 
  \biggl(
  \sum_{\alpha \in \mathbb{F}_K^{(2)}} \!
     \bra{\pi} \Omega_{ g_{\alpha^{-1}(0)}} 
     T
   \bigl( e^{i (\omega_{\alpha^{-1}(1)} + \omega_{\alpha^{-1}(2)} ) } I  - T  \bigr)^{-1} 
   \Omega_{g_{\alpha^{-1}(1)}}
   T
   \bigl( e^{i \omega_{\alpha^{-1}(2)} } I  - T  \bigr)^{-1} 
   \Omega_{g_{\alpha^{-1}(2)}} \!
    \ket{\one}  \!  \biggr)
\nonumber \\
& \qquad + 
\sum_{\kappa = 3}^{K}
  \sum_{\alpha \in \mathbb{F}_K^{(\kappa)}}
     \bra{\pi} \Omega_{ g_{\alpha^{-1}(0)}} 
\Bigl( 
\prod_{\ell = 1}^{\kappa}    T
   \bigl( I /  \mathfrak{z}_{\ell:\kappa}^{(\alpha)}  - T  \bigr)^{-1} 
   \Omega_{g_{\alpha^{-1}(\ell)}}
\Bigr)
\ket{\one}  
    ~.
\label{eq:Polyspectra_self_references}
\end{align}

From Eq.~\eqref{eq:Polyspectra_self_references}, it is now easy to specialize
to the $(\overline{X - \braket{X}},X - \braket{X},X -
\braket{X})$-polyspectrum denoted $S_{\overline{X - \braket{X}},X -
\braket{X},X - \braket{X}}(\omega_1, \omega_2)$. This is the \emph{cumulant
bispectrum}, $S_{\overline{X} , X, X}^\text{cumulant}  (\omega_1, \omega_2) $,
since the third-order cumulants of the original time series are the same as the
third-order moments of the modified time series with subtracted
mean~\cite{Coll98, Petr99}. It  is:
\begin{align} 
S_{\overline{X} , X , X}^{\text{cumulant}}  (\omega_1, \omega_2) 
&= 
  S_{\overline{X - \braket{X}},X - \braket{X},X - \braket{X}}(\omega_1,
  \omega_2) \nonumber \\
& = 
     \bra{\pi} \Omega_{  |X - \braket{x}|^2 (X - \braket{x}) } 
\ket{\one}  
+ 
\biggl(
  \sum_{\alpha \in \mathbb{F}_2^{(1)}}
     \bra{\pi} \Omega_{ g_{\alpha^{-1}(0)}} 
  T
   \bigl( e^{i \omega_{\alpha^{-1}(1)} } I  - T  \bigr)^{-1} 
   \Omega_{g_{\alpha^{-1}(1)}}
\ket{\one} 
\biggr)
 \nonumber \\
& \quad \, + \!
  \sum_{\alpha \in \mathbb{F}_2^{(2)}} \!\!
     \bra{\pi} \Omega_{ g_{\alpha^{-1}(0)}} 
     T
   \bigl( e^{i (\omega_{\alpha^{-1}(1)} + \omega_{\alpha^{-1}(2)} ) } I  - T  \bigr)^{-1} 
   \Omega_{g_{\alpha^{-1}(1)}}
   T
   \bigl( e^{i \omega_{\alpha^{-1}(2)} } I  - T  \bigr)^{-1} 
   \Omega_{g_{\alpha^{-1}(2)}} \!
    \ket{\one} 
    ~.
\label{eq:CumBispectrum}
\end{align}

This leads to a fraudulent white noise theorem for the cumulant bispectrum,
reminiscent of Cor.~\ref{cor:PSD_from_ZeroMean_pdfs}.

}
\end{widetext}

{ \color{blue} 
{\The \label{pro:Bispectrum_from_ZeroMean_pdfs} 
Any hidden Markov chain with any arbitrary state-paired collection of
equal-mean distributions, i.e.: 
\begin{align*}
\mathcal{P} \in \bigl\{ \{ \pdf( X | s) \}_{s \in \SSet} : \braket{X}_{\pdf( X |s)} = \braket{x}  \text{ for all } s \in \SSet \bigr\}
  ~,
\end{align*}
generates a flat bispectrum that is constant over all frequencies $( \omega_1,
\omega_2)$.
}

{\ProThe
Equation~\eqref{eq:CumBispectrum} shows that the cumulant bispectrum consists
of contributions from $\mathbb{F}_2^{(0)}$, $\mathbb{F}_2^{(1)}$, and
$\mathbb{F}_2^{(2)}$. The only $\mathbb{F}_2^{(0)}$ contribution is $ \bra{\pi}
\Omega_{  |X - \braket{x}|^2 (X - \braket{x}) } \ket{\one} $, which is a
constant independent of frequency. Whereas, we show that each contribution from
$\mathbb{F}_2^{(1)}$ and $\mathbb{F}_2^{(2)}$ is identically zero if the
stochastic process can be generated by a HMM with equal-mean PDFs associated
with each state.  For such processes, $\braket{X}_{\pdf(X|s)} = \braket{x}$,
where $\braket{x}$ is independent of the latent state $s$.

With the aid of Fig.~\ref{fig:FunctionSets}, it is easy to verify that, for
each $\alpha \in \mathbb{F}_2^{(1)}$, either $\Omega_{ g_{\alpha^{-1}(0)}} $ or
$\Omega_{ g_{\alpha^{-1}(1)}}$ equals either $\Omega_{ X - \braket{x}}$ or $\Omega_{ \overline{X - \braket{x}}}$. These latter two operators both
equal the zero operator $\boldsymbol{0}$ since:
\begin{align*}
\Omega_{ X - \braket{x}} 
  & =  \sum_{s \in \SSet} \braket{ X - \braket{x} }_{p(X | s)} \ket{s} \bra{s}
  \\
  & =  \sum_{s \in \SSet} \bigl( \braket{ X }_{p(X | s)}  - \braket{x} \bigr)
  \ket{s} \bra{s} \\
  & = \boldsymbol{0}
\end{align*}
and:
\begin{align*}
\Omega_{ \overline{X - \braket{x}}} 
  & = \sum_{s \in \SSet} \braket{ \overline{X} - \overline{\braket{x}} }_{p(X |
  s)} \ket{s} \bra{s} \\
  & =  \sum_{s \in \SSet} \bigl( \overline{\braket{ X }_{p(X | s)} } -
  \overline{\braket{x} } \bigr) \ket{s} \bra{s}  \\
  & = \boldsymbol{0}
  ~.
\end{align*}
Each potential contribution from $\alpha \in \mathbb{F}_2^{(1)}$ is therefore a
product of zero and thus vanishes.

Again with the aid of Fig.~\ref{fig:FunctionSets}, it is easy to verify that
for each $\alpha \in \mathbb{F}_2^{(2)}$, the operators $\Omega_{
g_{\alpha^{-1}(0)}} $, $\Omega_{ g_{\alpha^{-1}(1)}} $, and $\Omega_{
g_{\alpha^{-1}(2)}} $ are equal to either $\Omega_{ X - \braket{x}} $ or
$\Omega_{ \overline{X - \braket{x}}} $ which---as we showed---are all
zero. Each potential contribution from $\alpha \in \mathbb{F}_2^{(2)}$ is
therefore a product of zero and so vanishes.

For such processes, this establishes that the only nonzero contribution to the
cumulant bispectrum is independent of frequency. The corresponding cumulant
bispectrum is thus flat with uniform height $\bigl\langle  | x - \braket{x} \!
|^2 (x - \braket{x})  \bigr\rangle$.
}

}

\bibliography{chaos}

\end{document}

%% file: cmechabbrev.tex
\newcommand{\MeasAlphabet}  {\mathcal{A}}
\newcommand{\MeasSymbol}   { {X} }
\newcommand{\meassymbol}   { {x} }

\newcommand{\CausalState}   { \mathcal{S} }

\newcommand{\CausalStateSet}    { \boldsymbol{\CausalState} }

\newcommand{\Cmu}       {C_\mu}

\newcommand{\EE}        {{\bf E}}

\newcommand{\ProcessAlphabet}   {\MeasAlphabet}

\newcommand{\forward}{+}
\newcommand{\reverse}{-}
\newcommand{\forwardreverse}{\pm} 

\newcommand{\FutureCausalState} { {\CausalState}^{\forward} }

\newcommand{\PastCausalState}   { {\CausalState}^{\reverse} }

\newcommand{\one}{\mathbf{1}}

\newcommand{\lastindex}[2]{
  \edef\tempa{0}
  \edef\tempb{#2}
  \ifx\tempa\tempb
    \edef\tempc{#1}
  \else
    \edef\tempa{0}
    \edef\tempb{#1}
    \ifx\tempa\tempb
      \edef\tempc{#2}
    \else
      \edef\tempc{#1+#2}
    \fi
  \fi
  \tempc
}

\newcommand{\I}{\mathbf{I}}

\newcommand{\CSjoint}[1][,]{
   \edef\tempa{:}
   \edef\tempb{#1}
   \ifx\tempa\tempb
      \ensuremath{\FutureCausalState\!#1\PastCausalState}
   \else
      \ensuremath{\FutureCausalState#1\PastCausalState}
   \fi
}

\newif\ifpm
\edef\tempa{\forwardreverse}
\edef\tempb{\pm}
\ifx\tempa\tempb
   \pmfalse
\else
   \pmtrue
\fi